\documentclass[en, lightstyle]{unirostock}

\usepackage[backend=biber,style=ieee,maxbibnames=99,backref=false,citestyle=ieee]{biblatex}
\author{Arjun Haridas Pallath}
\enrolmentnumber{217205438} 

\title{Comparison of Convolutional neural network training parameters for detecting Alzheimer's disease and effect on visualization}
\type{Master Thesis}

\course{Master of Science Electrical Engineering}
\workperiod{07. Jan 2020 -- 20. July 2020}
\supervisor{Dr. rer. hum. Martin Dyrba}
\primaryreviewer{Prof. Dr.-Ing. Thomas Kirste}

\faculty{Faculty of Computer Science and Electrical Engineering}
\institute{Institute of Visual and Analytic Computing}
\workinggroup{Mobile Multimedia Information Systems}
\usepackage{pdflscape}
\usepackage{ragged2e}
\usepackage{multirow}
\usepackage{array}
\usepackage{titlesec}
\usepackage{placeins}
\titlespacing\section{0pt}{10pt plus 4pt minus 2pt}{0pt plus 2pt minus 2pt}
\begin{document}
\maketitle

\pagenumbering{Roman}

\tableofcontents 
\clearpage

\setstretch{1.263}

\listoffigures 
\clearpage

\listoftables
\clearpage
\pagenumbering{gobble}
\clearpage
{\normalfont
\color{uniblau}
\huge\sffamily\itshape
\chapter*{Abstract}
}

Convolutional neural networks (CNN) have become a powerful tool for detecting patterns in image data. Recent papers report promising results in the domain of disease detection using brain MRI data. Despite the high accuracy obtained from CNN models for MRI data so far, almost no papers provided information on the features or image regions driving this accuracy as adequate methods were missing or challenging to apply. Recently, the toolbox iNNvestigate has become available, implementing various state of the art methods for deep learning visualizations. Currently, there is a great demand for a comparison of visualization algorithms to provide an overview of the practical usefulness and capability of these algorithms.

Therefore, this thesis has two goals:
\begin{enumerate}
\item To systematically evaluate the influence of CNN hyper-parameters on model accuracy.
\item To compare various visualization methods with respect to the quality (i.e. randomness/focus, soundness).
\end{enumerate}

\vfill
\clearpage

\pagenumbering{arabic} 

\chapter{Introduction}

\section{Motivation}
Alzheimer's disease is an irreversible, progressive neurodegenerative disease and one of the world's most growing health issues. In the 2019 report by the Alzheimer's Disease International (ADI), it was estimated that over 50 million people are living with dementia globally. This number was projected to rise to 152 million by the year 2050. In Germany's case, it has been estimated that around 1.7 million people are affected by dementia \cite{ref1}.
It is the most common cause of dementia in older people above 65 years of age, accounting for around 60\% - 80\% of cases \cite{ref2}. The disease causes the death of nerve cells, which cause irreparable damage or atrophy in the brain, leading to memory loss, behavioral changes, speech impairment, and difficulty while eating and walking. Since all the brain regions are affected by atrophy in older people, it is difficult to diagnose in its early stages due to the slow progress of the disease and the difficulty in discriminating the brain regions affected due to Alzheimer's and normal age-related atrophy.

In clinical studies, the progression of the disease happens in three stages \cite{ref3}.
\begin{enumerate}
\item Normal control (NC) - Patients having minor age-related short-term memory losses.
\item Mild cognitive impairment (MCI) - It is the transitional stage between normal age-related memory losses and Dementia. This stage is further categorized into early and late MCI. During the early MCI (EMCI), the patient does not recall recent events, have difficulty in speaking, misplaces belongings and have trouble finding their way around , late MCI (LMCI) subjects have long term memory loss, may require assistance for performing daily routine activities and significant changes in their personality and behavior can be observed.
\item Alzheimer's dementia (AD) – It is the final stage of the disease progression, the cognitive ability continues to decline, and the patients require daily attention for performing daily activities, loses the ability to speak properly, and finds it difficult to eat and walk.
\end{enumerate} 
Imaging methods such as MRI and PET scans are used by researchers to diagnose Alzheimer's disease and assist in the search for more effective treatments. \\In recent years several papers have been published on AD classification using MRI images. However, most of these have not explained the selection of parameters used to train the model and the effects of varying the different hyperparameters of the convolutional neural network (CNN). \\This study tries to address the impact of varying a few of the hyperparameters that are more relevant to the CNN structure and evaluate the impact of these on its performance with respect to accuracy and Area under the curve and also on relevance heatmap visualizations produced on the various models. 

\section{Related Work}
Several papers have been published over the years for the classification of AD using MRI scans as input using convolutional neural networks. Some papers \cite{2dslice-1,2dslice-2} used 2D CNNs with input composed of the set of 2D slices extracted from the 3D MRI scans. The limitation of the 2D slice-level approach is that the 2D convolutional filters analyze individual slices of a 3D MRI subject independently and this way we cannot get a subject level accuracy.\\ To compensate for the absence of 3D information in the 2D slice-level approach few 2D slices covering a particular region of interest is fed in to the 2D CNN model as separate channels \cite{marzban}. 3D patch-level/Region of Interest (ROI) methods \cite{3d-patch1} uses the whole MRI by slicing it into smaller inputs by focusing on regions which are known to be informative. In this way, the complexity of the framework can be decreased as fewer inputs are used to train the networks. The drawback of this methodology is that it studies only one or few regions while AD
atrophy span over multiple brain areas. However, it may reduce the risk of overfitting because the trainable parameters are lower as it had to consider only a small portion of the input, thus a small input size.\\
Recently with the boost of high-performance computing resources, more studies used a 3D subject-level \cite{bohle,rieke,CNN-analysis-overview} approach. In this approach, the whole MRI was used at once and the classification was performed at the subject level. The advantage is that the spatial information was fully integrated. In the 3D-subject level approach, the number of samples was small compared to the number of parameters to optimize and the dataset will have only a few hundreds or thousands of subjects, thus increasing the risk of overfitting.

Most of the papers perform the classification by a certain model architecture and few papers explain the model by visualization methods \cite{bohle,rieke} and even fewer \cite{CNN-analysis-overview} have attempted to explain how the model is created by choosing certain hyperparameters and its impact on the accuracy. 

The goal of this Thesis are

\begin{enumerate}
    \item The hyper paramaters are tuned by using a ROI based input scheme due to the lack of computation power which restricts the number of parameters to be varied and attempts to study the impact of accuracy and visualization by varying the hyperparameters in the CNN model.
    \item The visulaizations obtained on the different models are evaluated for similarity and correlation with respect to hippocampal volume
\end{enumerate}

\chapter{Convolutional neural networks}

A Convolutional neural network (CNN) takes an input image which passes through several hidden layers which include convolutional layers, pooling layers, activation functions and fully connected (FC) layers to produce an output layer. This section describes the various building blocks used to build a CNN.  

\section{Convolutional Layer}
\begin{figure}[!htbp]
    \centering
    \includegraphics[]{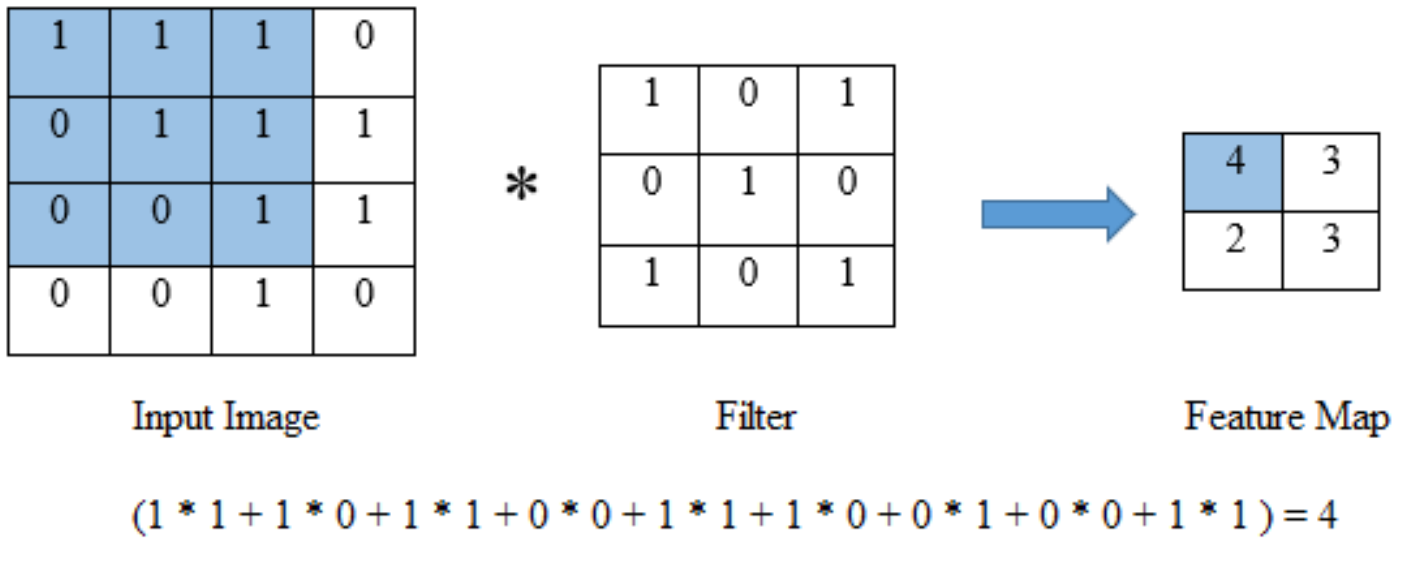}
    \label{fig:conv-op}
    \caption{Convolution operation: Using a 3 x 3 filter with stride of 1 }
\end{figure}

The convolutional layer is the main building block of a CNN. Convolutional layers extract features from an image to produce feature maps by a convolution operation, which is a dot product operation of the input image with a filter. The filter scans through the entire input image to produce feature maps. An example of a convolution operation is shown in Figure \ref{fig:conv-op} along with an illustration on how to calculate the element in the feature map highlighted in the image. The input image is usually zero-padded along its dimensions to maintain the output size the same as that of the input.\\
The outputs of a convolution layer are passed through a nonlinear activation function for the network to learn complex functions that map the input data to the classification output. The most commonly used activation function is a rectified linear unit (ReLu) which gives a sparse and nonlinear representation by setting all the negative values to zero. Convolution layers also increase the model efficiency, by reducing the number of parameters to learn compared to fully connected neural networks \cite{ref9}.

\section{Pooling Layer}
\begin{figure}[!htbp]
    \centering
    \includegraphics[]{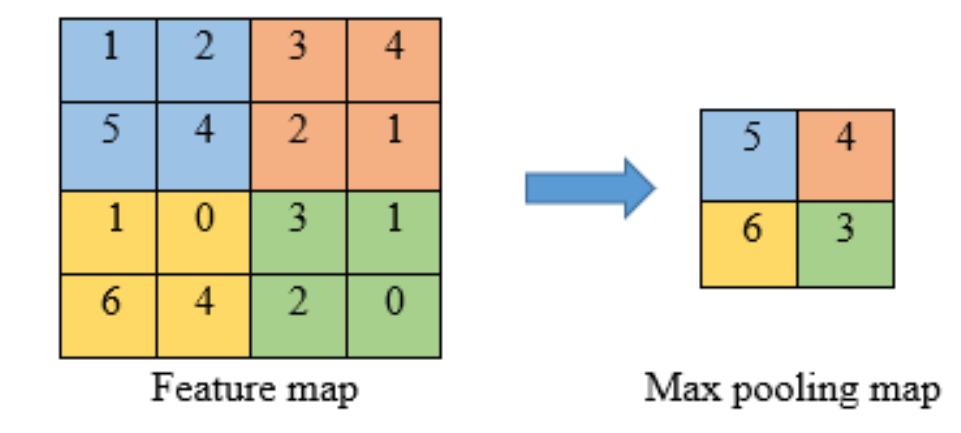}
    \label{fig:pooling-op}
    \caption{Max pooling: Using a 2 x 2 filter with stride of 2 }
\end{figure}

The pooling layer is applied after the convolution layer. It reduces the dimensionality of the feature map while retaining the information of the image by either taking the average or the largest value in the feature map. This reduces the number of parameters the network must learn thereby reducing overfitting and leads to shorter training time. Max pooling is one of the widely used pooling method \cite{ref9}. Unlike the filter used in the convolution layer for multiplication, the filter used in pooling operation extracts the maximum value within the filter region without overlapping, as it scans through the image.\\ An example of max pooling operation is shown in Figure \ref{fig:pooling-op}. A stride of 2 is used to prevent overlapping of the feature map.

\section{Fully connected layer}
\begin{figure}
    \centering
    \includegraphics[height=2.5in, width = 4.3in]{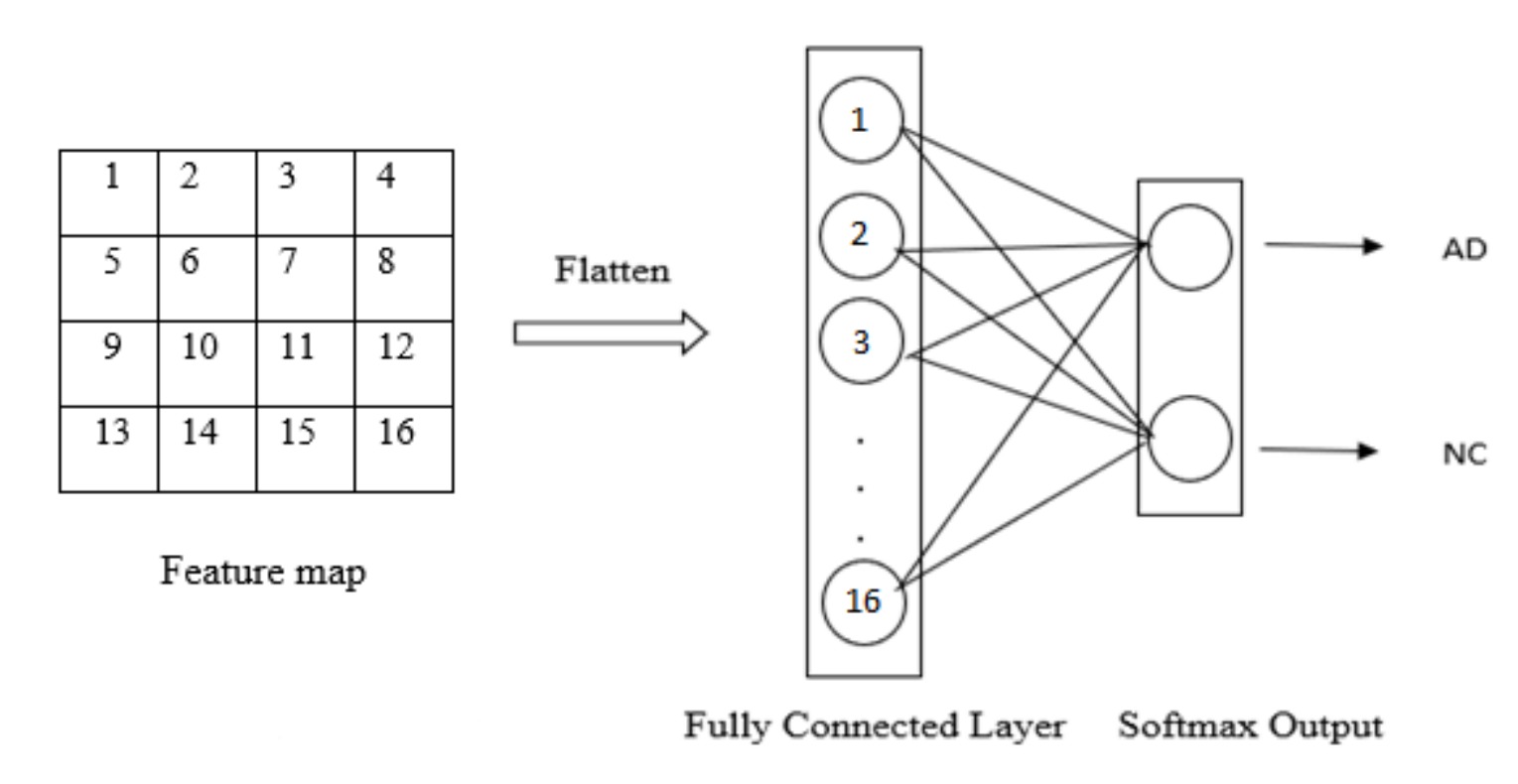}
    \label{fig:fc-op}
    \caption{Fully connected layer and output with Softmax activation for classes 'AD' and 'NC' }
\end{figure}

A fully connected layer uses the features extracted from the final convolution or pooling layer and uses it to classify the image into the respective classes. The output feature maps of the final convolution or pooling layer is converted to a one-dimensional (1D) array of numbers through a flattening operation (i.e., “reshaped to a vector”) and is connected to one or more fully connected layers known as dense layers, in which every input is connected to every output. The final fully connected layer typically has the same number of outputs as the number of classes, i.e. each output neuron encodes the presence of this class or object being recognized in the input image.\\ A Softmax activation layer is applied to the final fully connected layer which scales the output values in the range between 0 to 1 as a probability function of the classes.\\ An illustration of the fully connected layer along with the output using Softmax activation is shown in Figure \ref{fig:fc-op}.

\section{Training of the CNN}
The network learns to classify the data by modifying the weights of the filters used in the network in every iteration. It does this by considering the difference or the loss between the predicted class and the actual ground truth class and tries to minimize this loss. Some of the popular loss functions are Mean squared error, Absolute error, and Cross entropy.\\ The commonly used loss function for multiclass classification is cross entropy, whereas the mean squared error is typically applied for regression models \cite{ref9}. An optimization algorithm is used that iteratively updates the weights and biases, of the network to minimize the loss. Some of the widely used optimization algorithms are Stochastic Gradient Descent (SGD), Root mean square prop (RMSprop), AdaGrad and Adam.\\ Experiments by Kingma et.al., \cite{ref10} have reported Adam optimizer to be robust and well-suited for a wide range of optimization problems in the field of machine learning, since the loss function is observed to converge faster than the other optimizers.

\section{Problem of overfitting}
During the training of CNN, sometimes the network tries to memorize all the data available in the training data. This makes it not able to generalize well to make predictions on new and unseen data, leading to low accuracy. It could be because the network tries to learn too many features in the data or when there is not enough data for the network to learn.
This is referred to as overfitting. Batch normalization and dropout layers are used to tackle this problem. \\\textbf{Batch normalization} is a method for improving the speed, performance, and stability of neural networks \cite{ref11}. It is used between two consecutive convolution or dense layers. It normalizes the activations of the previous layer by transforming the activations to zero mean and unit standard deviation, which mitigates the risk of overfitting, improves gradient flow through the network, reduces the number of training epochs required to train the network and reduces initialization dependence. \\\textbf{Dropout} is a regularization method, in which randomly chosen activations are set to 0 during training, making the model less susceptible to learn specific weights in the network \cite{ref12}.

\chapter{Interpretable AI}
\section{Need for interpretability of neural networks}
Deep convolution neural networks have become the state-of-the-art technique for various image classification tasks during recent times. The performance of these systems has been reported to be at par with humans. These networks are actively being researched and developed in areas related to computer vision in many fields, such as self-driving cars \cite{selfdriving}, face recognition \cite{facedet}, object detection \cite{objdet}, and medical imaging \cite{medimag}. These systems could aid physicians in the early diagnosis of diseases and highlight the concerning areas in medical scans. However, there is a lack of transparency in the accuracy results derived from these networks, as there is no direct way to identify, on what basis the network performs the classification.\\ This brings us to a need to build CNN models that can be interpretable and explainable and are class-discriminative, in the sense that it should highlight the critical region of the input for a specific class prediction while the model is built for classifying different classes.\\
Some of the main reasons for the need for explainable AI are reported in \cite{ref13} and these are briefly explained in the following sections.

\subsection{Verification of the system} 
It will be helpful when visual insights are obtained that highlight the regions in the input images that the model predicts to contribute to a particular class's prediction. These regions can then be compared against the actual ground truth regions to check if they correspond well. For example, the regions highlighted by the network predicted for a particular disease, and the regions found in the medical literature corresponding to the disease can be compared to check whether they match. Such systems can then be trusted to be used by clinical practitioners.

\subsection{Improvement of the system}
Model interpretability can help detect where the model is making a wrong prediction and figure out the biases in the model or the dataset \cite{ref13}. It can also be used to compare different models or architectures. For instance, the authors in \cite{FV-bias} compared the prediction of an image for a particular class for two different models, Fischer vector (FV) and Deep Neural network (DNN). It was observed that both the models had similar prediction accuracies. However, FV showed visualization in the area of a copyright tag in the image rather than parts of the class object, which was rightly highlighted by the DNN model. Removing the copyright tag changed the FV visualization showing parts of the class object.

\subsection{Learning from the system}
As the AI systems learn by analyzing millions of data, it infers its prediction by observing for patterns in the data. Since it is practically impossible for humans to analyze this vast amount of data, we can try to use the knowledge gained from the AI system to acquire new insights that were previously unknown to us. For example, the moves used by Google DeepMind's AlphaGo system, an AI system that defeated professional players in the board game of Go, can be analyzed by players to play the game better. Another example is the AI system used in a food industry based company, NotCo \footnote{https://www.notco.com/}, where they try to reproduce a meat-based food from plant-based food with the same perception and taste as that of meat product. The AI algorithm analyses the molecular components in meat-based food with a combination of similar molecular components present in the plant-based foods and matches it with the human perception of taste and texture.  

\subsection{Compliance to legislation}
In the future, when AI systems are used in critical industries such as banking, automation, and medical, the decisions made by these systems need to governed with legal aspects, especially in the case when the system makes a wrong decision. The system must be able to explain the reasoning behind its decision. Due to these concerns, the European Union (EU) has added the law to “right to explanation” in the EU's General Data Protection Regulation, whereby the AI system must give out the reasoning behind its decision. \cite{AI-regulation}.

\section{Overview of visualization method to interpret neural networks}
Several papers have been produced over the last couple of years for various visualization methods that can enable us to visualize the deep learning models' decision-making process. 
\newpage
They are categorized into three main categories.
\begin{enumerate}
    \item Perturbation methods - Occlusion Sensitivity Analysis and Local Interpretable Model-agnostic Explanations (LIME)
    \item Gradient based methods - Saliency map, DeconvNet,Guided backpropagation and Class activation maps (Grad-CAM)
    \item Decomposition based methods - Layer-Wise Relevance Propagation (LRP) and its variants
\end{enumerate}

In the first method, the input image gets occluded, and the network output was recalculated. The intuitive reason is that, if the probability of the target class decreases compared to the original image, the image region which had been occluded is more relevant. Though these methods produced class-discriminative visualizations, they are computationally intensive and take more time for processing, as they need to run multiple feedforwards to generate the heatmap for a given input image. For instance, in order to obtain the relevance heatmaps, the image patches in the form of regular grids in the case of occlusion sensitivity analysis \cite{deconvnet}, and a group of adjacent pixels of similar intensities and such regions are occluded in the case of LIME \cite{LIME}, across the input image iteratively, and the difference between unoccluded and occluded probability is generated.

For the methods in the second category, the relevance maps are generated by using the gradient (partial derivative) of the output or the internal units such as the fully connected or the convolution/pooling layers. For instance, saliency maps \cite{sensitivity} are generated by projecting back the fully connected layers of the network to the input image through backpropagation by taking the gradients of networks output probability with respect to the input image pixel intensities. Whereas, in the case of DeconvNet \cite{deconvnet} and  Guided Backpropagation \cite{guidedbackprop}, interpret the network by projecting the features extracted by the convolution/pooling layers back to the input through different ways of backward pass. However, these methods do not produce class-discriminative visualizations and require the gradients to backpropagate through the entire network.\\ In contrast, Grad-CAM \cite{GradCam} produces class-discriminative visualizations and uses only the activation maps of the final convolution or pooling layer in the CNN model, as it contains most of the spatial information of the image to generate heatmaps. \\One of the drawbacks of this method is that the resulting heatmaps are not of high resolution and do not show fine-grained details as it visualizes the linear combination of activations and class-specific weights from the last convolutional layer which is of low resolution, and these are up-scaled back to the size of the input image \cite{ref4}.

LRP is a decomposition-based visualization method where the classification decision is decomposed into pixel-wise relevance indicating, the contributions of pixels to the overall classification score. This approach uses a layer-wise conservation principle, which forces the evidence for a predicted class to be preserved between neurons of two adjacent layers. The visualizations in this method are of high resolution with fine-grained details and are class-discriminative \cite{lrpfig}.

Both gradient and decomposition-based methods were compared for the classification between AD and CN in a recent study, \cite{ref4}, and it was reported that LRP with $\alpha=1$, $\beta=0$ rule produced high-resolution activation patterns with fine-grained details. The working of the LRP visualization method is briefly explained in the next section and this method is used in this study to visualize the different models.

\section{Layer-Wise Relevance Propagation (LRP)}
\begin{figure}[!htbp]
    \centering
    \includegraphics[height=3in, width = 4in]{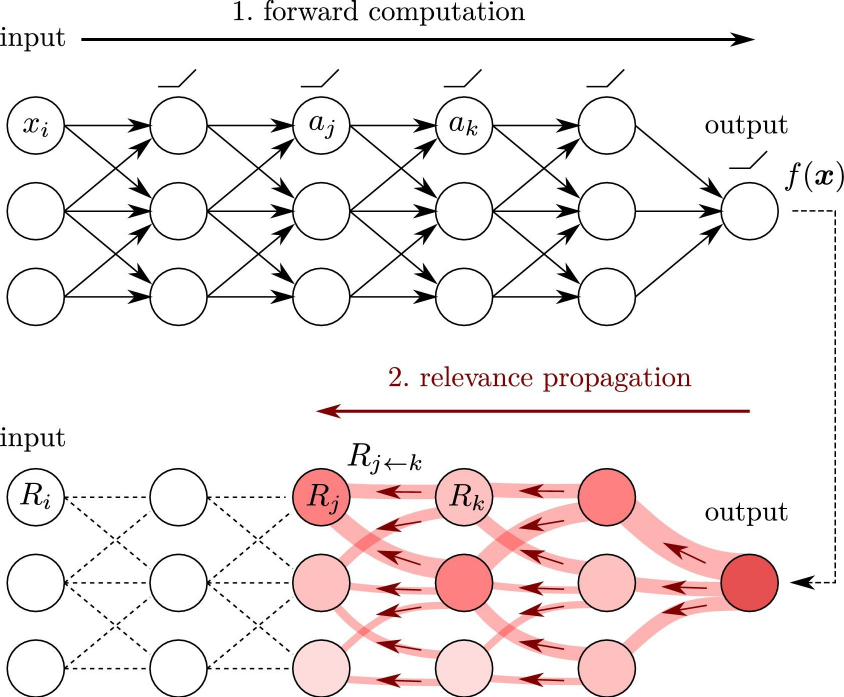}
    \caption{Representation of LRP procedure. Red arrows indicate the relevance propagation flow. Image adapted from \cite{lrpfig}}
    \label{fig:lrpfig}
\end{figure}

LRP uses the prediction output of the model, i.e., the network weights and the neural activations created by the forward-pass of the network, to be redistributed back through the network up until the input layer in a backward pass in order to produce the pixel-wise decomposition as represented in Figure \ref{fig:lrpfig}. Through this process we can visualize the heatmap indicating the pixels that contribute to the classification decision. The magnitude of the contribution of each pixel is referred to as \textit{relevance}. 
The procedure of the "alpha-beta" rule of LRP decomposition which is used for the visualization purpose in this study is illustrated in eq \ref{eqn:lrp-eq} wherein two consecutive layers $l$ and $l+1$ are considered and shows how the relevance is redistributed from layer $l + 1$ to layer $l$

\begin{center}
\begin{equation}
    R_i=\sum_{j} (\alpha \cdot \frac{(x_i\cdot w_{ij})^+}{\sum_{i}(x_i\cdot w_{ij})^+}-\beta \cdot \frac{(x_i\cdot w_{ij})^-}{\sum_{i}(x_i\cdot w_{ij})^-})R_j
    \label{eqn:lrp-eq}
\end{equation}
\end{center}

 In the eq 3.1 $(x_i)_i$ represent the neuron activations at layer $l$. $(R_j)_j$ is the relevance scores associated to the neurons at layer $l+1$, $w_{ij}$ represents the weight connecting neuron $i$ to neuron $j$ and $()^+$ and $()^-$ denote the positive and negative relevance scores respectively.

LRP is a conservative technique, meaning the magnitude of the output prediction $f(x)$ is redistributed to each input pixel in the image and a relevance score $R_i$ is assigned to each each input pixel $i$. This can be represented as $\sum_{i}(R_i)^l$ = ... = $\sum_{j}(R_j)^l$ = ... = $f(x)$ which shows that the total relevance is conserved at each layer. \cite{lrpfig}.

\chapter{Experimental Setup and Methods}

This section discusses the steps taken in pre-processing the input data and the architectures of the 3D CNN models used.

\section{Datasets used for AD classification}
A supervised learning approach was used in this study, which requires a set of training image data and its corresponding label to be trained. The MRI dataset was obtained from two publicly available datasets have been used for this study, the Alzheimer's Disease Neuroimaging Initiative (ADNI) \footnote{"ADNI database” http://adni.loni.usc.edu} and the Australian Imaging Biomarkers and Lifestyle (AIBL) \footnote{"AIBL database" https://aibl.csiro.au/adni/imaging.html}. These are two independent imaging data with different diagnostic criteria and there is no strict equivalence between the labels of ADNI and AIBL. A brief description of these datasets along with the diagnostic labels provided are explained below.

\subsection{Training Data}
\begin{figure}
    \centering
    \includegraphics[height=4.2in, width = 2in]{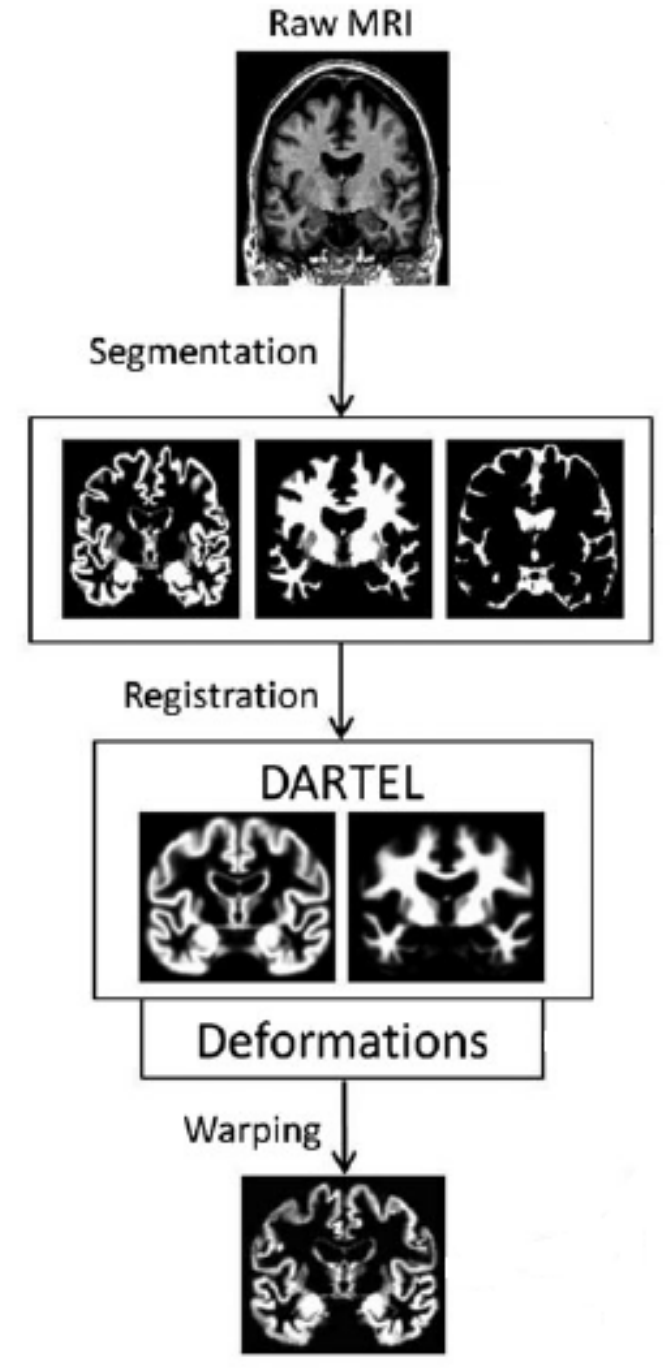}
    \caption{Data processing of raw MRI image Image adapted from \cite{ref6}}
    \label{fig:spm8}
\end{figure}

For training the model and evaluating the different hyper-parameters, the ADNI dataset was used. It is a study that aims to develop clinical, imaging, genetic, and biochemical biomarkers for the early detection and tracking of Alzheimer's disease (AD). For this study, T1-weighted 3D MPRAGE sequences of ADNI-GO/-2 MRI data acquired on 3 Tesla MRI scanners were used.\\ The raw MRI data then underwent a preprocessing step to increase signal uniformity across the multicenter scanner platforms \cite{ref6}. MRI scans were segmented into the grey matter and white matter using reference brain template specific to aging/AD by the SPM8/VBM8 software \footnote{ "SPM8/VBM8 software" http://dbm.neuro.uni-jena.de/vbm} which are spatially normalized and modulated using DARTEL \cite{ref7} algorithm to preserve the total amount of grey matter in the scans, as shown in Figure \ref{fig:spm8}. These preprocessed images were taken as the dataset used for training the Convolution network to classify MRI scans from the Alzheimer's disease subjects (AD) and the healthy control subjects (NC). Diagnostic labels are given by the physician using the MMSE \cite{ref8} test result, which is an exam used in clinical research to measure cognitive impairment.\\ A total of 662 3D MRI scans were used in this study. The subjects were categorized into AD, MCI and NC based on extensive neuropsychological testing including the MMSE  test. Details regarding the gender, age, years of education and MMSE scores are given in Table \ref{tab:ADNI}.\\ For training and validation, a total of 662 scans, 189 scans from AD, 254 scans from NC and 219 scans from LMCI were used, both AD and LMCI scans were combined to be labeled as AD for binary classification. Each MRI scan is a 3D volume of grayscale intensities consisting of 111 coronal slices with a slice dimension of 88 x 94. 

\begin{figure}
    \centering
    \includegraphics[height=1.5in,width=5in]{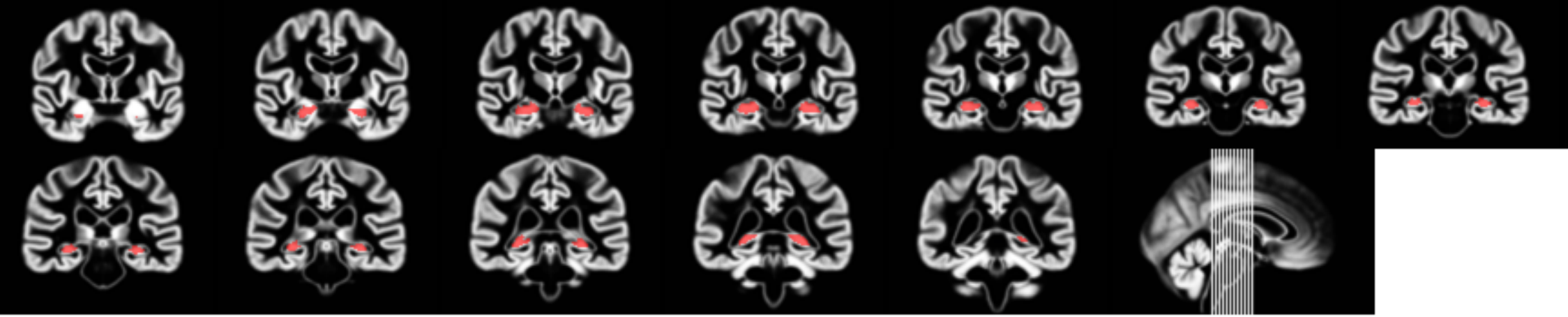}
    \caption{Segmentation of left and right Hippocampus region based on harmonized hippocampus segmentation protocol. Overlay map obtained from DZNE, Rostock based on \cite{ref-coronal-slices}}
    \label{fig:coronal-slices}
\end{figure}

\begin{center}
\begin{table}[]
    \centering
     \caption{Demographic data for ADNI dataset}
\begin{tabular}{l l l l l} \hline
& Male/Female    & Age & Years of Education & MMSE \\ \hline
 \multirow{1}{10em}{ADNI(n=662)} & & & \\ \hline
 \multirow{1}{10em}{NC(n=254)}& 124/130(51.2\%) & $75.4 \pm 6.6$ & $16.4 \pm 2.7$ & $29.1 \pm 1.2$ \\
 \multirow{1}{10em}{MCI(n=219)} & 126/93(42.5\%) & $74.1 \pm 8.1$ & $16.2 \pm 2.8$ & $27.6 \pm 1.9$ \\
 \multirow{1}{10em}{AD(n=189)} & 99/80(42.3\%) & $75.0 \pm 8.0$ & $15.9 \pm 2.7$ & $22.6 \pm 3.2$ \\ \hline
 \end{tabular}
NC: normal controls, MCI: mild cognitive impairment, AD: Alzheimer’s dementia, MMSE: mini-mental state exam. Values are Denoted as mean $\pm$ SD
\label{tab:ADNI}
\end{table}
\end{center}
\subsubsection{32 slice model}
For tuning the model by changing the model’s Hyper parameters, only a subvolume of the MRI scan is considered as input to the convolution network.  This is due to the lack of computational power which limits the flexibility to vary the different parameters. The Hippocampus region in the MRI scan which is observed to be one of the affected areas due to Alzheimer’s is only considered and out of the 111 coronal slices, only 32 consecutive coronal slices that contain the hippocampal region were taken as input with an input dimension of 89 * 32 * 94. An example of how the activations look on the coronal slices are shown in Figure \ref{fig:coronal-slices}. 

\subsubsection{Whole brain model}
The best model gained by adjusting the parameters in the 32 slices model is applied to the whole brain and trained, with an input dimension of 89 * 111 * 94.

\subsection{Testing Data}
The best model trained with the ADNI dataset is then applied to the AIBL dataset for testing. Like ADNI, the Australian Imaging Biomarkers and Lifestyle Flagship Study of Ageing seeks to determine the biomarkers, cognitive characteristics, and lifestyle factors contributing to the development of AD. Similar to ADNI data, T1-weighted 3D MPRAGE MRI sequences were used which underwent the preprocessing pipeline using the SPM8/VBM8 software. The diagnosis is given according to a series of clinical tests \cite{ref14} and the subjects were categorized into AD, MCI and NC. A total of 621 MRI scans, 67 scans from AD, 455 scans from NC and 99 scans from MCI were used for testing. Details regarding the gender, age and MMSE scores are given in Table \ref{tab:AIBL}.

\begin{center}
\begin{table}[]
    \centering
     \caption{Demographic data for AIBL dataset}
\begin{tabular}{l l l l} \hline
& Male/Female    & Age & MMSE \\ \hline
 \multirow{1}{10em}{AIBL(n=621)} & & & \\ \hline
 \multirow{1}{10em}{NC(n=455)}& 190/265(58.2\%) & $72.4 \pm 6.1$ & $28.7 \pm 1.2$\\
 \multirow{1}{10em}{MCI(n=99)} & 52/47(47.5\%) & $74.3 \pm 6.8$ & $27.1 \pm 2.2$ \\
 \multirow{1}{10em}{AD(n=67)} & 28/39(58.2\%) & $73.7 \pm 7.6$ & $21.0 \pm 5.3$ \\ \hline
 \end{tabular}
    \label{tab:AIBL}
\\ NC: normal controls, MCI: mild cognitive impairment, AD: Alzheimer’s dementia, MMSE: mini-mental state exam. Values are Denoted as mean $\pm$ SD \par
\end{table}
\end{center}

\section{Baseline CNN architecture}
\begin{figure}[!htbp]
    \centering
    \includegraphics[height=2.2in, width = 5.5in]{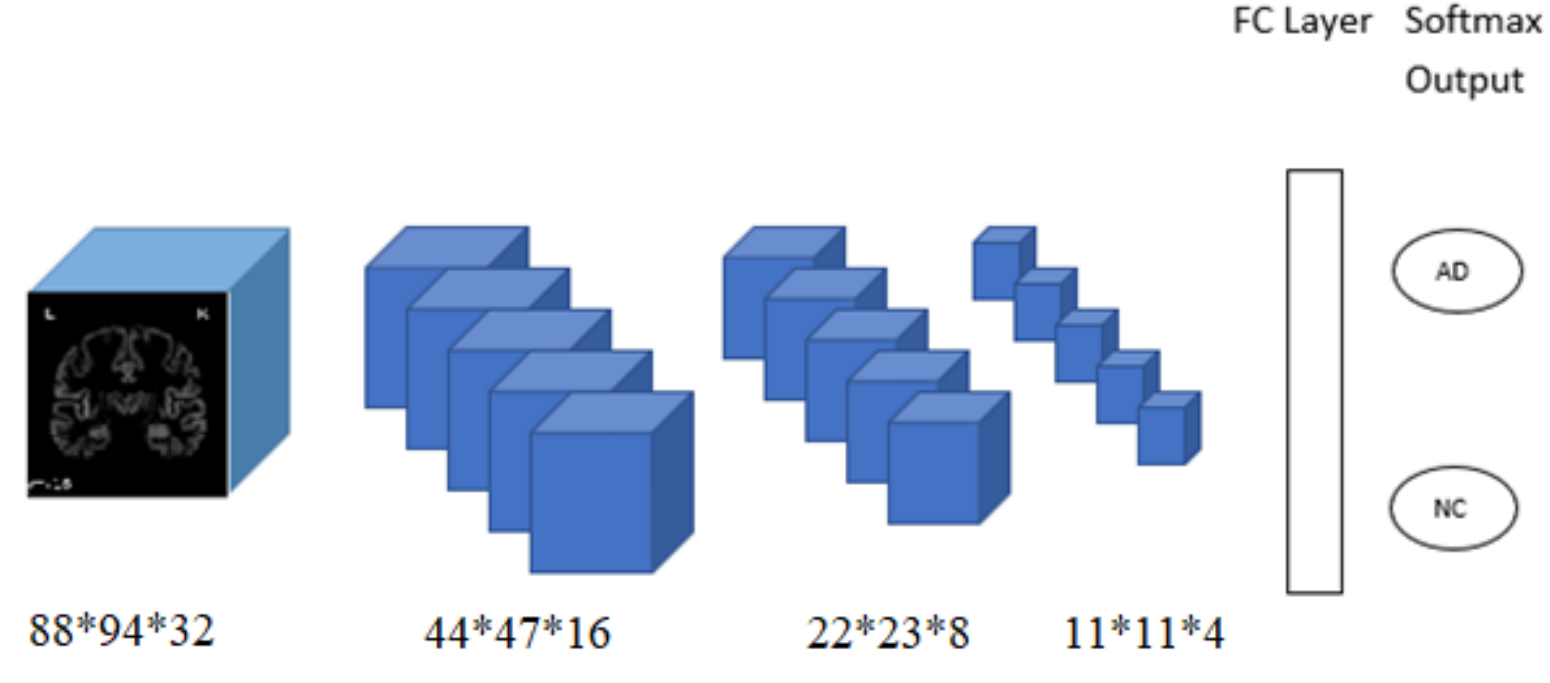}
    \caption{3D Convolution network using 3 convolutional blocks. Each block is made up of 5 layers, 3D convolution, batch normalization, 3D max pooling, ReLu activation and a dropout layer}
    \label{fig:base-arch}
\end{figure}

The architecture consists of three convolution blocks, followed by a fully connected layer for classification. Each of the convolution blocks consists of a 3D convolution, batch normalization, 3D max pooling, rectified linear unit, and a dropout layer. \\
The convolution layers use 5 filters each of dimension 3 x 3 x 3 with a stride of 1 and the max-pooling layers use 5 filters each of dimension 2 x 2 x 2 with a stride of 2 to obtain the feature maps that contains the spatial information from the MRI. A dropout of 50\% is used after each max pooling layer and the final layer uses Softmax activation for classification. \\
A 3D volume of MRI scans with 32 consecutive coronal slices per subject with a dimension of 88 x 32 x 94 (Axial, Coronal, Sagittal) was taken as input in this model. The model can learn 3D subject-wise spatial information from adjacent slides. The architecture of the model is shown in Figure \ref{fig:base-arch}. 

The models have been implemented in Python 3.6 version and were run on Google Colab which is a cloud hosted environment for running python code supporting 12 GB Tesla K80 GPU. The deep learning framework used is Keras \footnote{https://keras.io/} with TensorFlow \footnote{https://www.tensorflow.org/} 1.15 as a backend.

\chapter{Hyper Parameter tuning for Convolutional neural networks}

This chapter explores the different experiments done on the baseline CNN model by varying the different hyperparameters related to the training (e.g., batch size, learning rate, decay) and the architecture (e.g., number of layers, dropout). The model selection procedure, including the model architecture selection and fine-tuning the hyperparameter, was performed using a 10-fold cross-validation, where one fold (10\%) of the data was used for validation while making sure that each train and validation dataset have an equal number of both AD and NC labeled data as to avoid bias and the rest is used for training. The 10-fold data split was performed only once for all the experiments with a fixed seed number (random\_state = 1), thus guaranteeing that all the experiments used the same subjects during cross-validation per each fold.\\ There are several hyperparameters that can be varied in a CNN structure. Therefore, for the purpose of experimentation, the range of parameters to vary in each hyperparameter was limited since a grid search of all the parameters is extremely time-consuming and computationally expensive. \\
The performance of the parameters were evaluated by comparing the average accuracy and standard deviation on the validation set obtained after the 10 folds of cross-validation. Hyperparameters were varied one at a time and the subsequent experiments were done by considering the parameter with the highest validation accuracy obtained in the previous experiment. \\The models were run initially for 100 epochs using categorical cross-entropy as the loss function with a batch size of 64 and a learning rate of 0.0001 and a softmax activation layer after the dense layer for classification. 

\section{Varying the input type of Data}
In this experiment, two types of input scans were used for training the model. The first was the scans that were obtained from the SPM8 preprocessing pipeline. \\For the second type of input scans, a multiple linear regression model was used to estimate the effect of the covariates based on the subsample of healthy controls. These included age, gender, total intracranial volume (TIV), and scanner magnetic field strength of each voxel in the training sample given by the eq. \ref{eq:1} where $v_{ij}$ being the intensity of voxel j for subject i, $\beta_{0j}$ to $\beta_{4j}$ being the estimated regression parameters for voxel j, age, gender, TIV and field strength being the covariates for subject i, and $\epsilon_{ij}$ being the error/residual for voxel j and subject i.

For the training dataset, the models were fitted for the healthy control subjects and the error (mean squared error loss) was minimized. For the remaining dataset the error/residual is calculated based on eq. 5.2 by subtracting the actual voxel intensity by the predicted intensity.

These residualized scans were taken as the second type of input for the model. This enables us to reduce the effects of the different covariates and to separate the atrophy due to aging from atrophy due to Alzheimer’s.

\begin{equation}
    v_{ij}=\beta_{0j} + \beta_{1j}\cdot age + \beta_{2j}\cdot gender + \beta_{3j}\cdot TIV + \beta_{4j}\cdot field strength + \epsilon_{ij} \label{eq:1} \\
\end{equation}
\begin{equation}
   \epsilon_{ij} =v_{ij} - (\beta_{0j} + \beta_{1j}\cdot age + \beta_{2j}\cdot gender + \beta_{3j}\cdot TIV + \beta_{4j}\cdot field strength) \label{eq:2}
\end{equation}

\begin{landscape}
\begin{figure}
    \includegraphics[height=4.2in,width=4.5in]{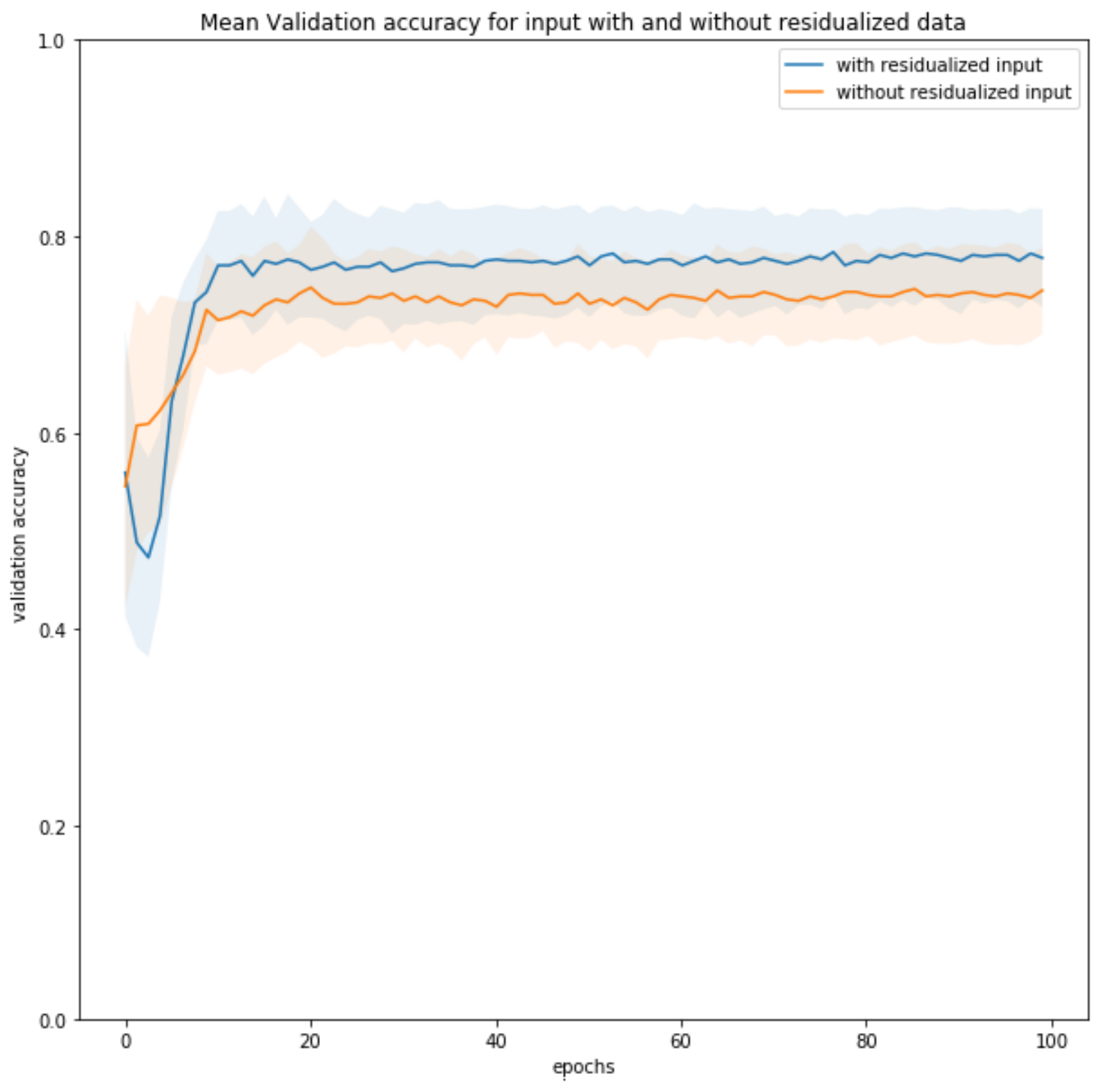}
    \includegraphics[height=4.2in,width=4.5in]{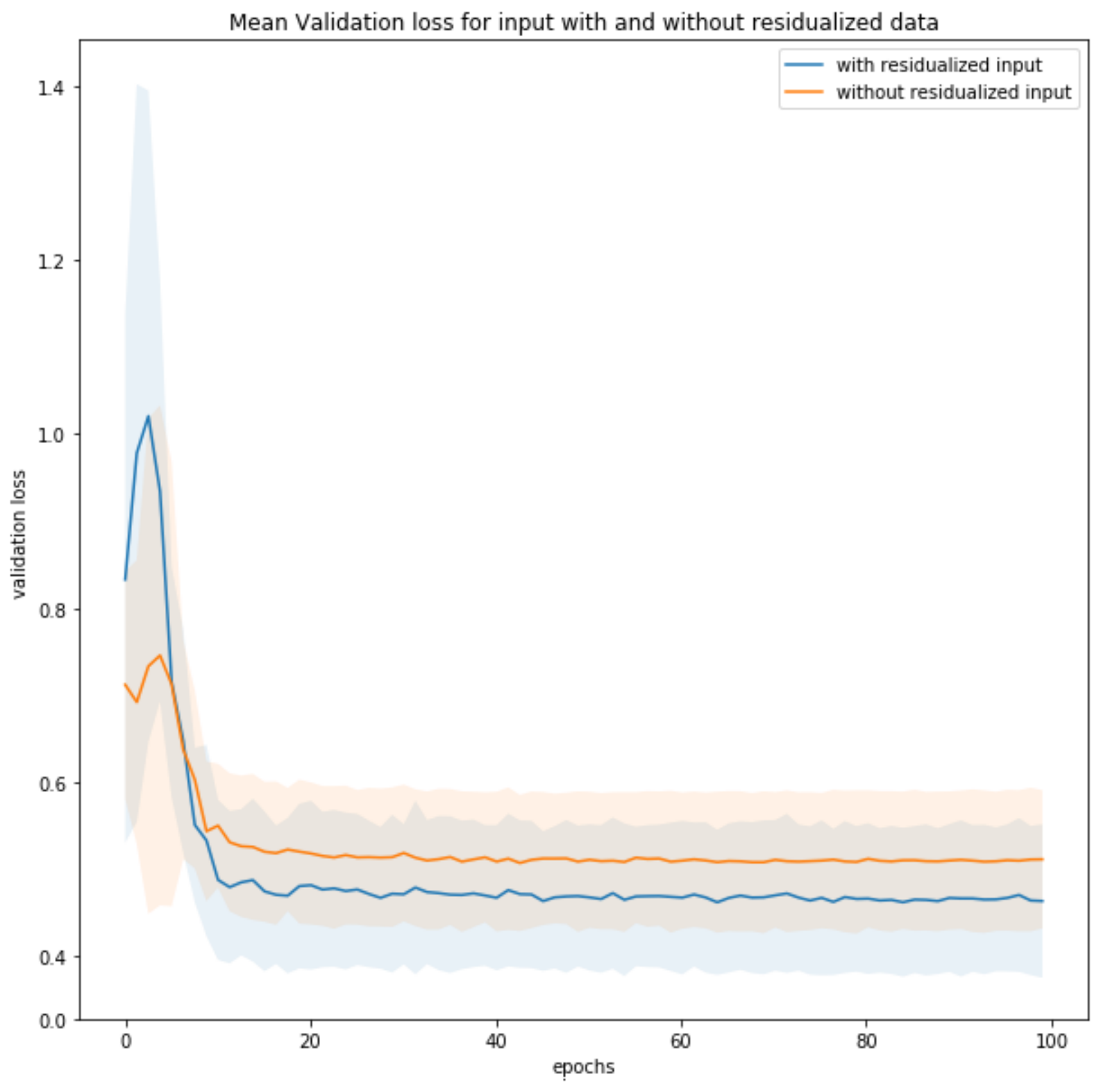}
    \label{fig:exp-residual}
    \caption{Mean Validation accuracy and validation loss across all the 10 folds for input scans with and without residualization; Standard deviation indicated by the shaded region}
\end{figure}
\end{landscape}

From the figure \ref{fig:exp-residual} it was observed that the accuracy of the model trained with the residualized scans as the input had a higher accuracy than that of the model trained without removing for the effects of the covariates. Further experiments were done by taking the residualized scans as the input to the model.

\section{Varying the amount of Data Augmentation}
\begin{landscape}
\begin{figure}
    \includegraphics[height=4.2in,width=4.5in]{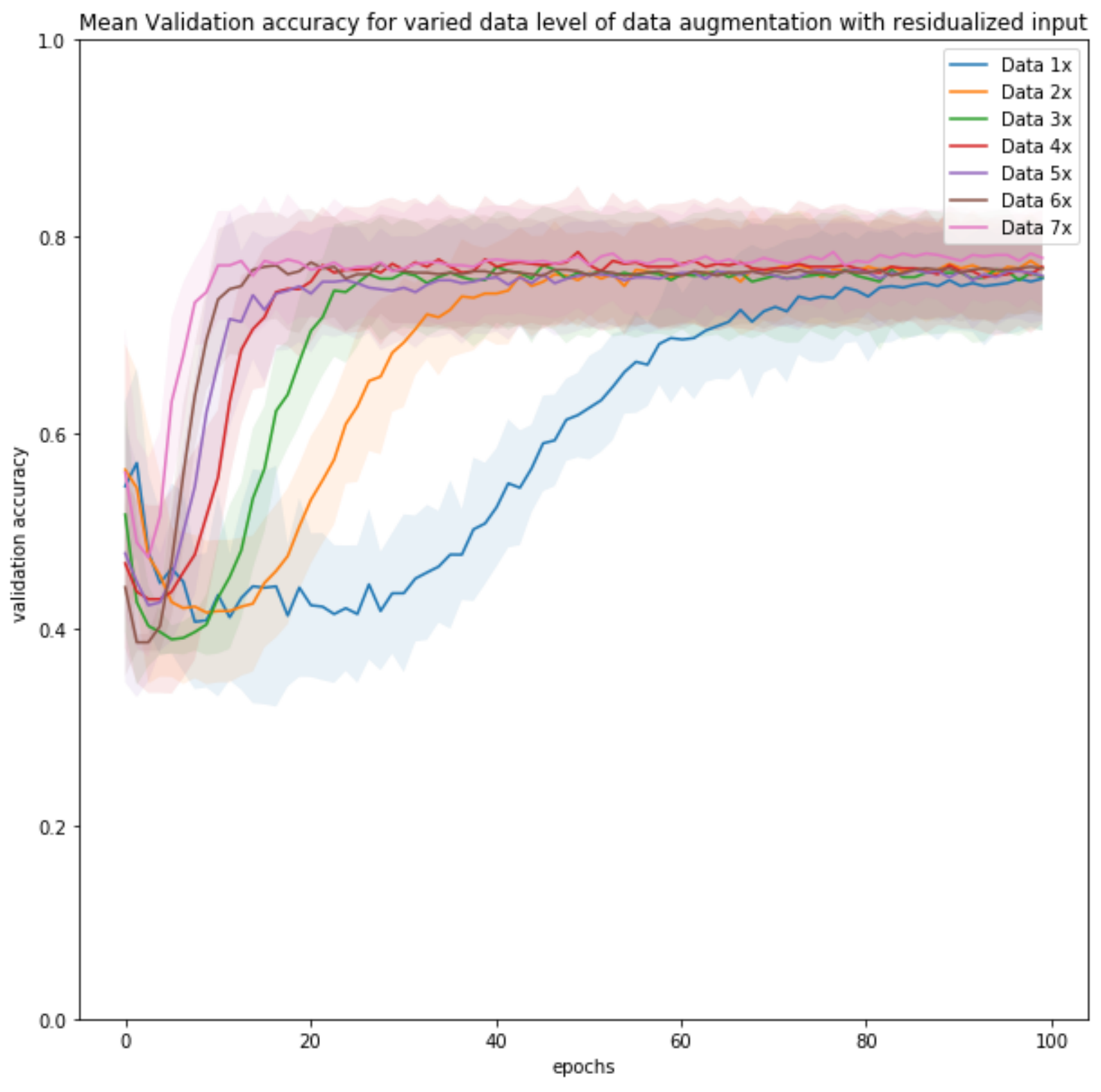}
    \includegraphics[height=4.2in,width=4.5in]{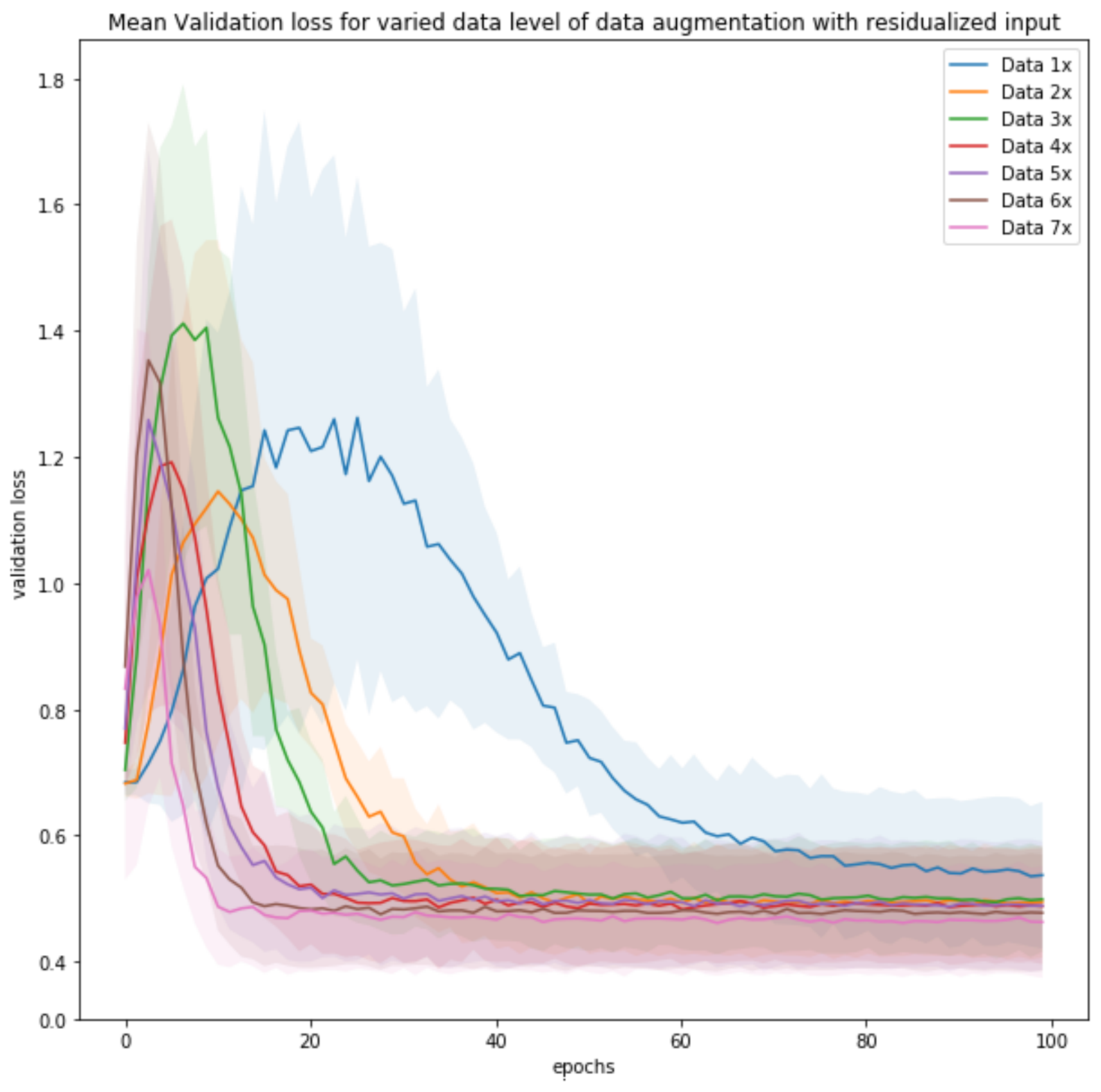}
    \label{fig:exp-DA}
    \caption{Mean Validation accuracy and validation loss across all the 10 folds for varied level of data augmentation; Standard deviation indicated by the shaded region}
\end{figure}
\end{landscape}

In this experiment, the amount of training data available for training was increased linearly by adding additional copies of the training scans shifted by $\pm2$ voxels in the x/y/z directions one at a time, as shown in Table \ref{tab:Table DA}. It was ensured that the augmentation was done only to the training data after keeping aside 10\% of the sample data for the validation set in the 10-fold cross-validation so that the training sample will not be present in the validation set for each fold. 
From Figure \ref{fig:exp-DA}, it was observed that all the models with varying data augmentation reached a stable mean accuracy between the range of  $75.7 \pm5.6$ and $77.8 \pm5.2$. However, the model having 7x data augmentation reaches the stable accuracy fastest. Thus when a model is presented with a large amount of training sample, the time taken for reaching a stable accuracy is faster. Hence for all further experiments were done with 7 times data augmentation.

\begin{table}[!htbp]
\caption{Mean accuracy of model trained with varying data augmentation}
\begin{tabular}{l l l} \hline
Training data & Augmentation & Mean accuracy \\ && $\pm$ SD\\  \hline
\multirow{1}{10em}{Data 1x} & Data & 75.7 $\pm5.6$\\
\multirow{1}{10em}{Data 2x}& Data + Data shifted by -2 voxels in x axis & 76.7 $\pm4.8$ \\
\multirow{1}{10em}{Data 3x}& Data + Data shifted by $\mp2$ voxels in x axis & $75.8 \pm5.7$ \\
\multirow{1}{10em}{Data 4x}& Data + Data shifted by $\mp2$ voxels in x axis & $76.9 \pm6.1$ \\ & + Data shifted by -2 voxels in x axis \\
\multirow{1}{10em}{Data 5x}& Data + Data shifted by $\mp2$ voxels in both x \& y axis & $76.0 \pm4.8$ \\
\multirow{1}{10em}{Data 6x}& Data + Data shifted by $\mp2$ voxels in both x \& y axis  & $76.7 \pm5.1$ \\ & + Data shifted by -2 voxels in z axis \\
\multirow{1}{10em}{Data 7x}& Data + Data shifted by $\mp2$ voxels in x,y \& z axis & $77.8 \pm5.2$ \\ \hline
 \end{tabular}
    \label{tab:Table DA}
\end{table}

\section{Varying the batch size} \label{batch size}
\begin{landscape}
\begin{figure}
    \includegraphics[height=4.2in,width=4.5in]{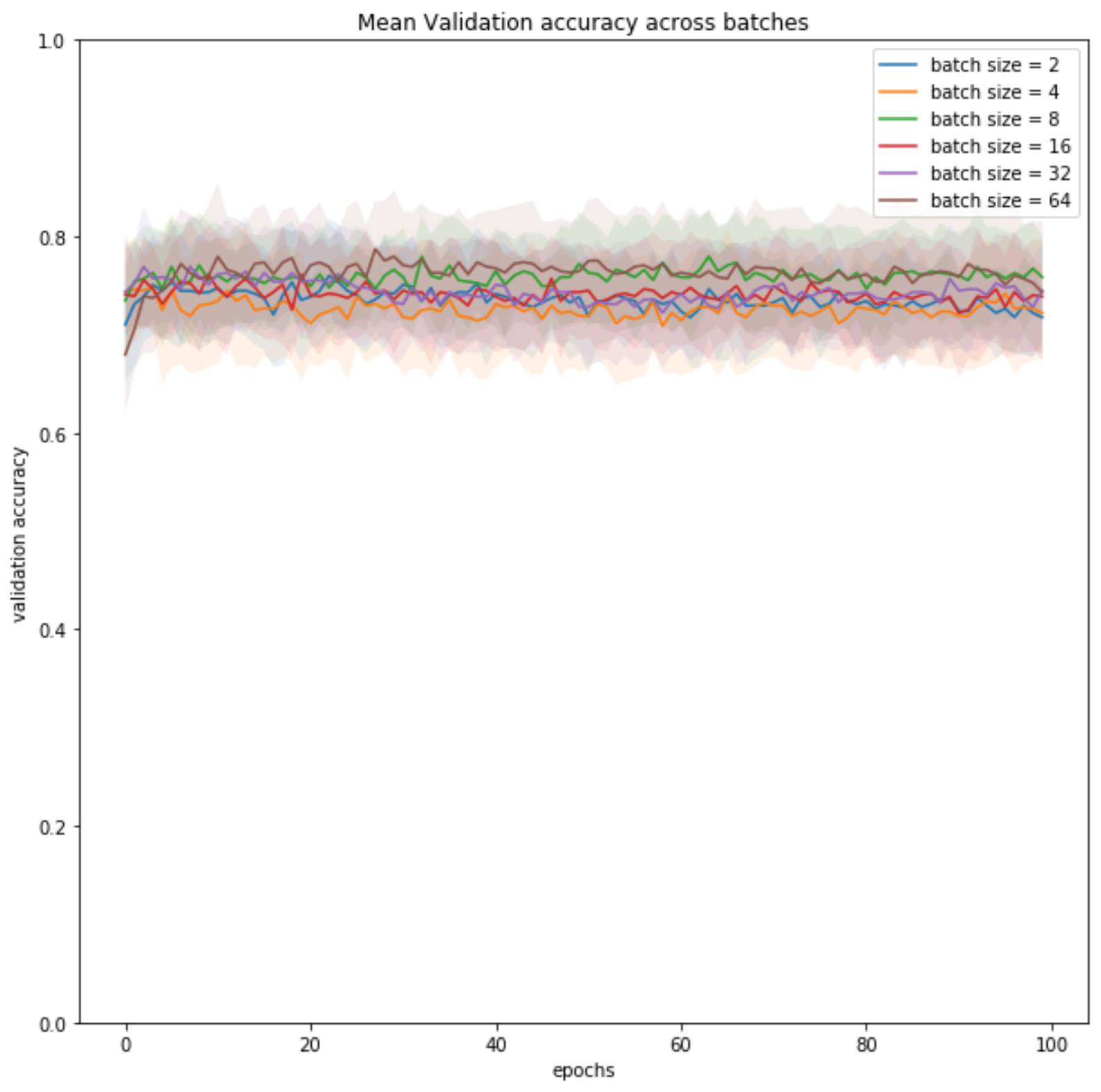}
    \includegraphics[height=4.2in,width=4.5in]{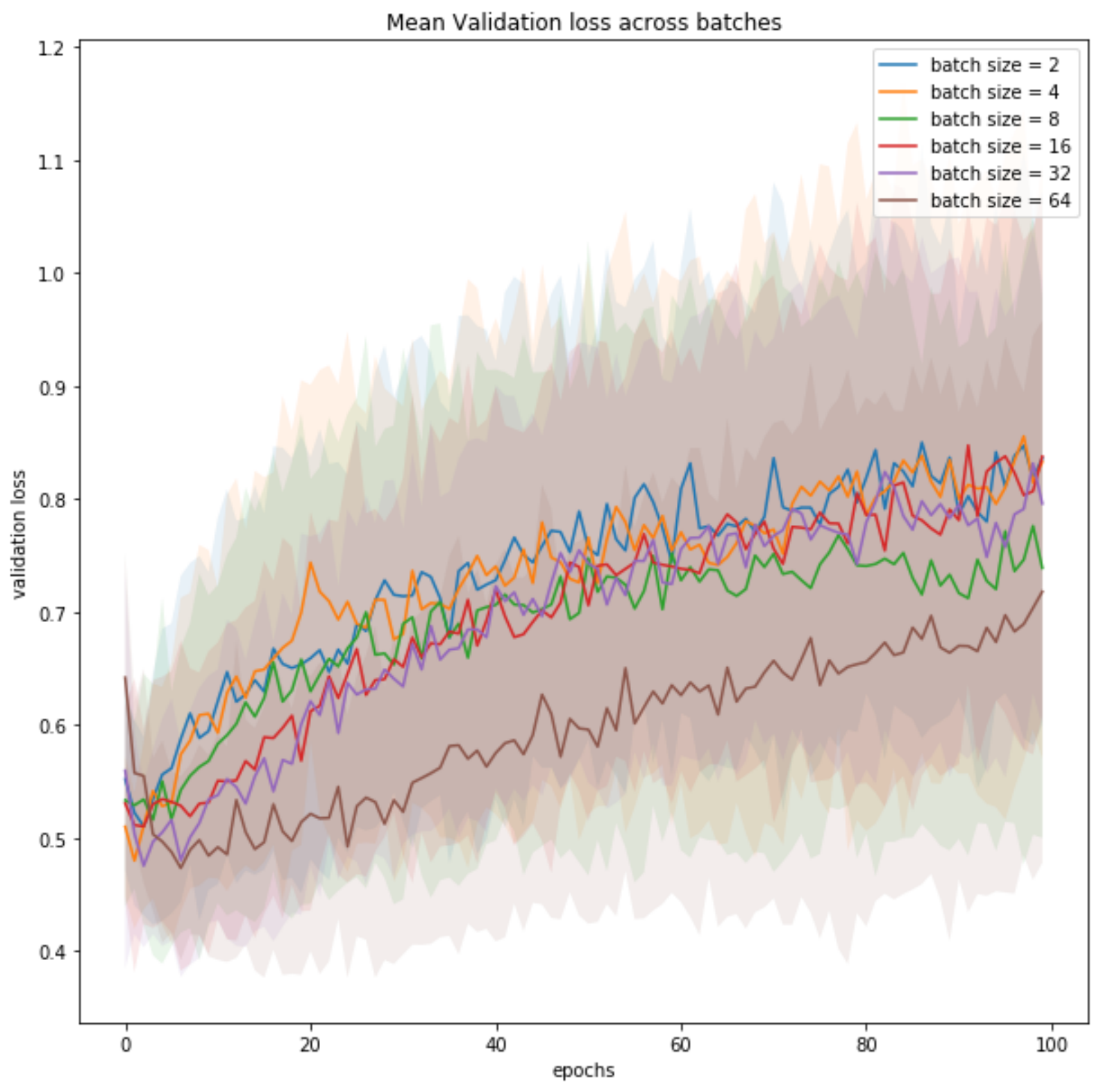}
    \label{fig:exp-bs}
    \caption{Mean Validation accuracy and validation loss across all the 10 folds for varied batch sizes; Standard deviation indicated by the shaded region}
\end{figure}
\end{landscape}

Deciding the batch size for training a deep CNN structure is essential as it controls the accuracy of the estimate of the error gradient when training neural networks. Batch size refers to the number of training examples considered simultaneously to estimate the error gradient while training the model. They have an impact on the training duration and the memory capacity. For higher batch size, higher is the memory required and training duration is short.
For example, a batch size of 64 means that 64 samples from the training dataset will be used to estimate the error gradient before the model weights are updated.\\
A study \cite{ref5} had reported that using small batch sizes achieved the best training stability and generalization performance, for a given computational cost. Best results in that study were obtained with batch sizes of 32 or smaller sizes of 2 or 4.\\
The model was trained for 100 epochs with varying batch sizes of 2,4,8,16,32 and 64, and the corresponding mean validation accuracies and validation losses across all 10 folds are plotted in Figure \ref{fig:exp-bs}.\\
From the plots, it was observed that the mean validation accuracy is found to be stable for batch sizes of 8 and 64. Hence a batch size of 8 is chosen to be the best choice for further experimentation after considering the memory limitation.

\section{Varying the learning rate and decay}
Deciding on the learning rate and decay is the next most crucial hyperparameter to be tuned. The learning rate decides the rate at which the weights are updated during the model training. Lower learning rates require more training epochs given the smaller changes made to the weights after each update, whereas higher learning rates result in rapid changes and require fewer training epochs. A balance needs to be maintained for choosing the optimal learning rate as the network may not learn if the learning rate is too low or the network can overfit if the learning rate is too high.\\
The learning rate can be adjusted after each batch update by using the decay parameter.
The relation between the learning rate and decay is summarised in eq. \ref{eq:3}.

\begin{equation}
Learning rate = initial learning rate * (1 / 1 + decay * iterations) \label{eq:3}
\end{equation}

\begin{landscape}
\begin{figure}
    \includegraphics[height=4.2in,width=4.5in]{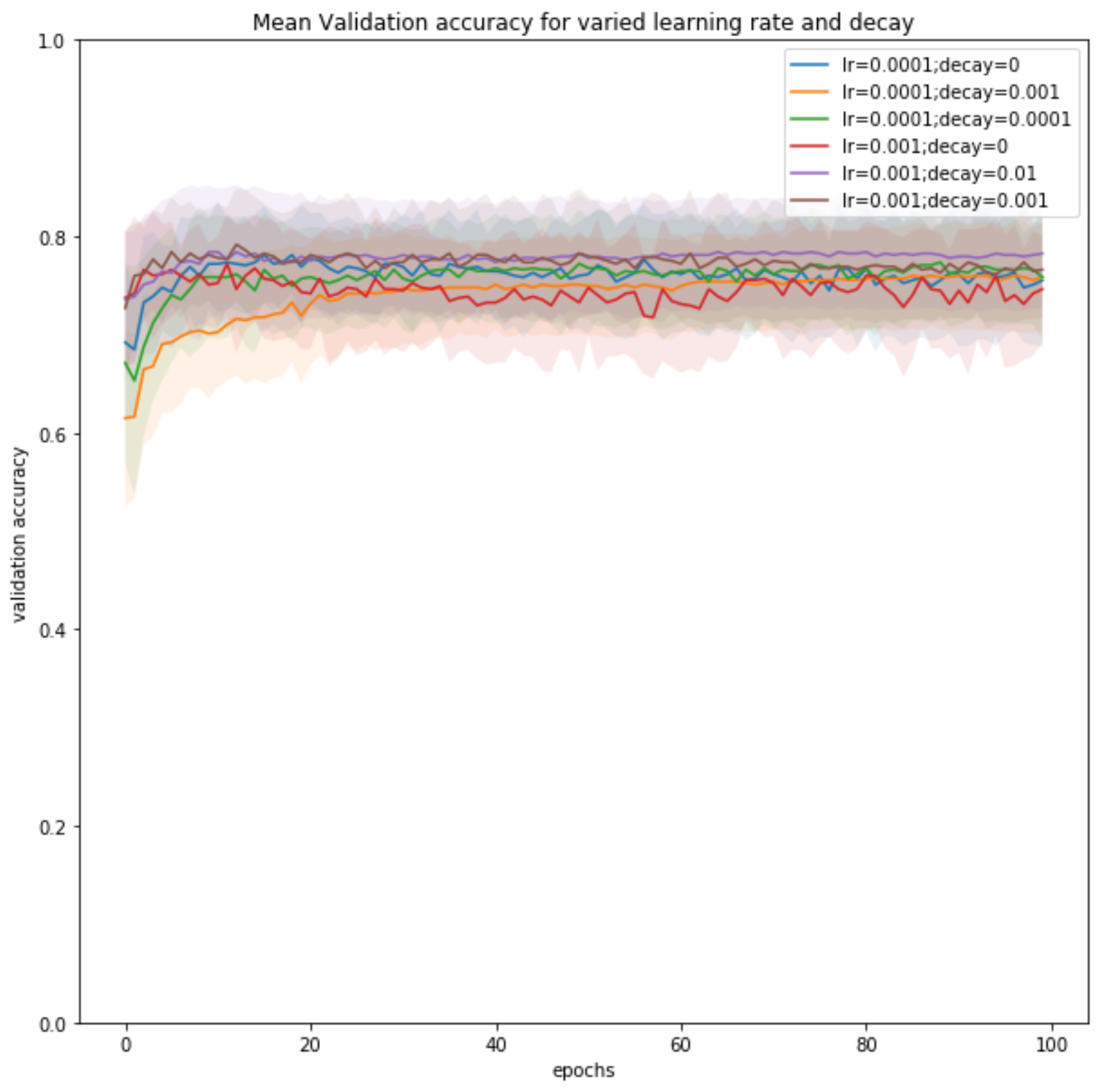}
    \includegraphics[height=4.2in,width=4.5in]{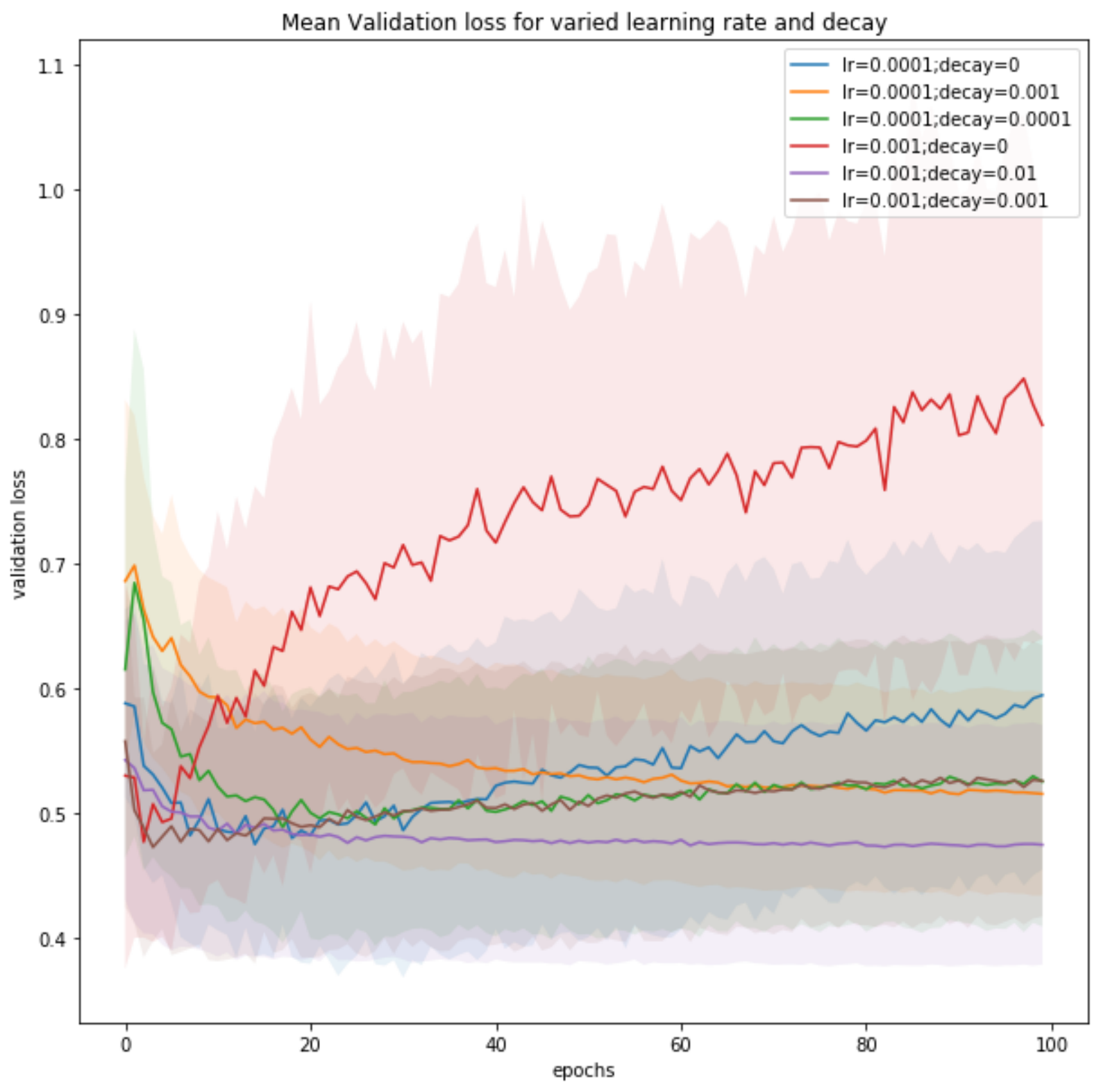}
    \label{fig:exp-lr}
    \caption{Mean Validation accuracy and validation loss across all the 10 folds for varied learning rate and decay; Standard deviation indicated by the shaded region}
\end{figure}
\end{landscape}

\begin{landscape}
\begin{figure}
    \includegraphics[height=4.2in,width=4.5in]{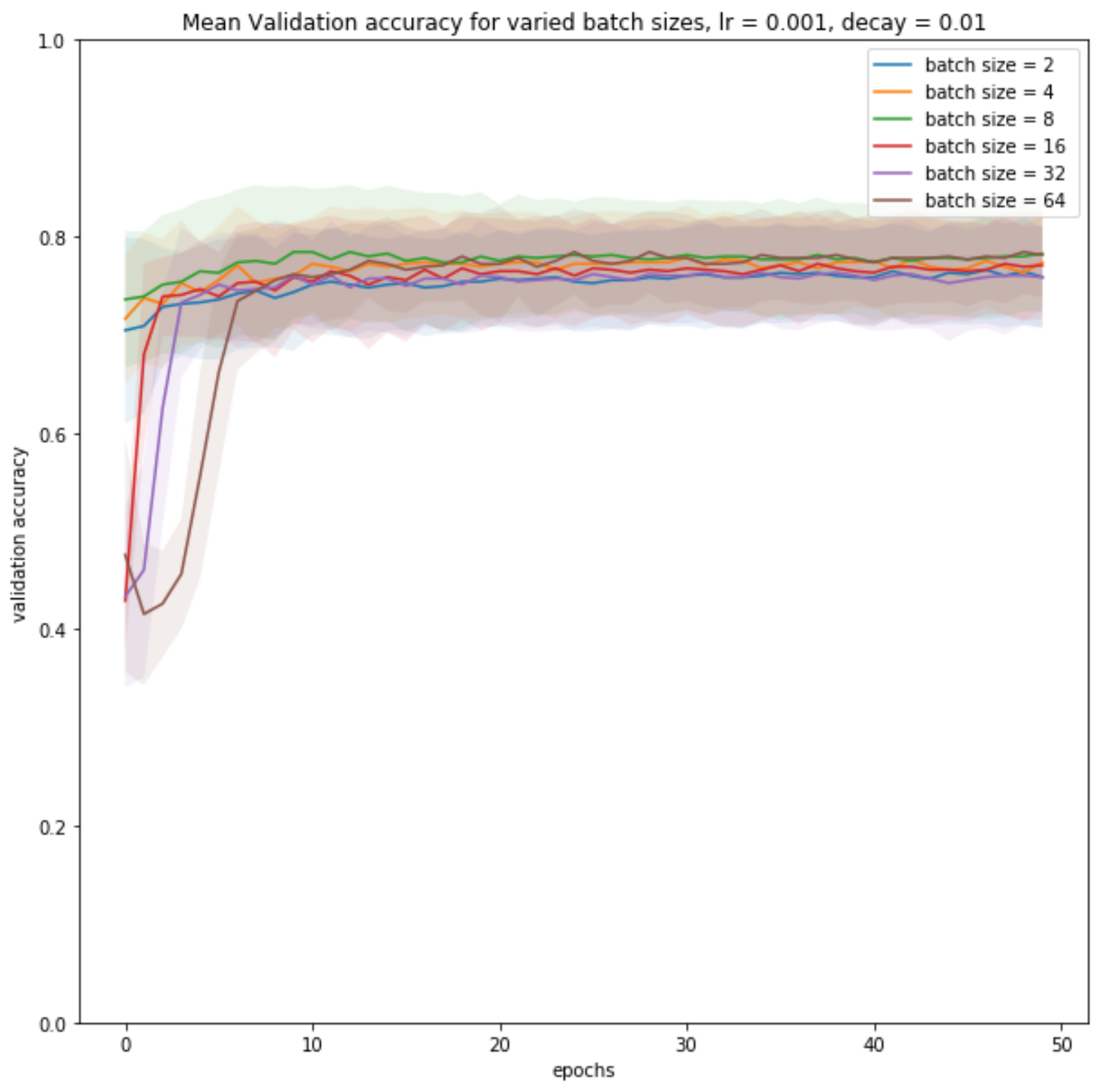}
    \includegraphics[height=4.2in,width=4.5in]{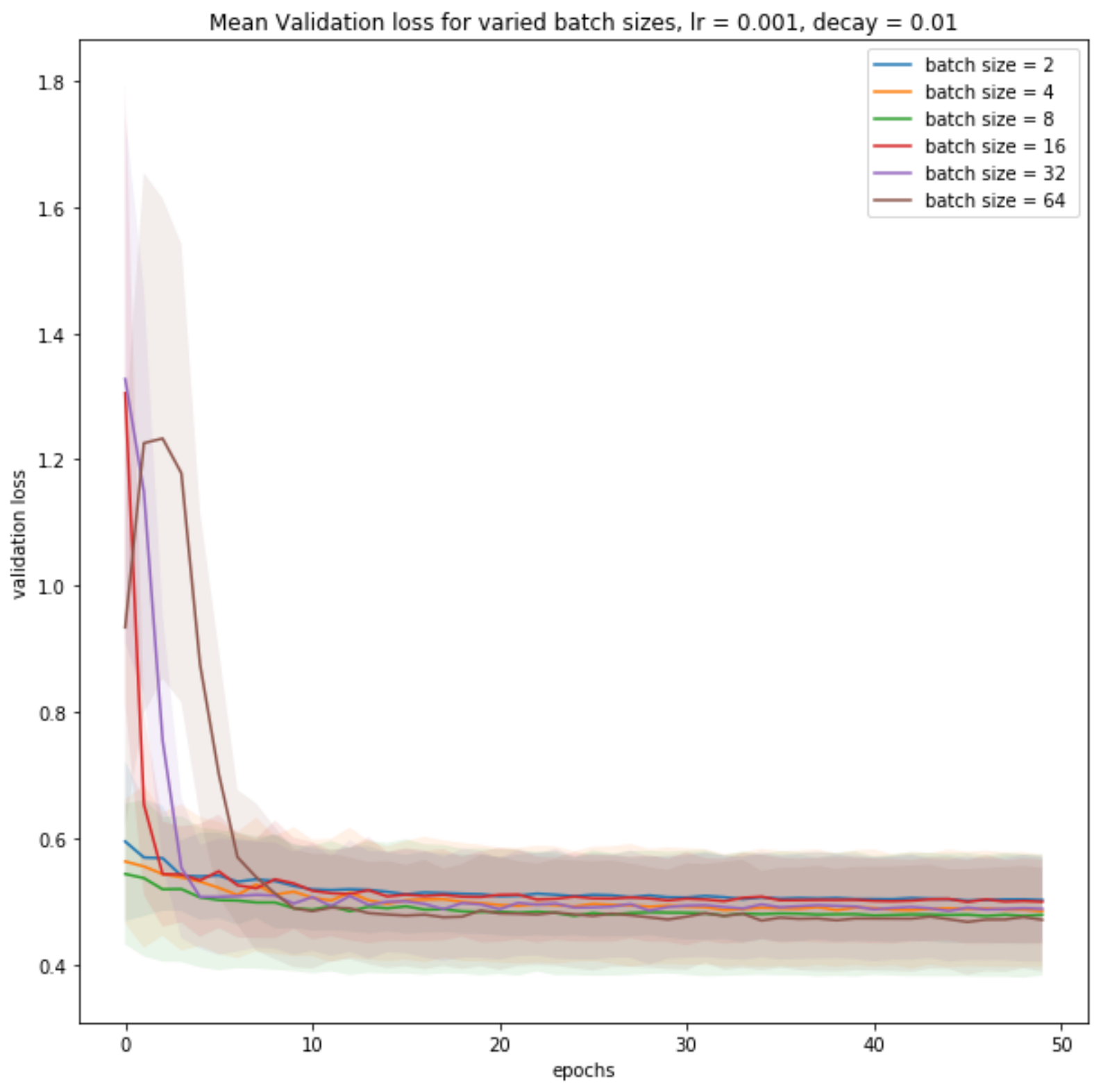}
    \caption{Experiment 5.3 repeated with new lr=0.001 and decay =0.01 to check for any change in the results due to the interaction between parameters. Mean Validation accuracy and validation loss across all the 10 folds for varied batch sizes; Standard deviation indicated by the shaded region}
    \label{fig:exp-bs-repeat}
\end{figure}
\end{landscape}

The model was trained for 100 epochs with a combination of different learning rates of 0.01 and 0.001 and decays of 0, 0.01, and 0.001 using the ADAM optimizer and the corresponding mean validation accuracies and validation losses across all 10 folds were plotted as shown in Figure \ref{fig:exp-lr}.
It was observed from the mean validation accuracy plot that the parameter of the learning rate of 0.001 and decay of 0.01 was found to have the least validation loss and the loss value decreases until it reaches a stable value after 20 epochs. 
Thus the model needs to be therefore trained only for 20 epochs.
The experiment done in section \ref{batch size} was repeated with the new learning rate and decay parameters chosen to check if there is any change in the results due to the interaction between parameters.
The results as shown in Figure \ref{fig:exp-bs-repeat} show that the results obtained remained the same with the batch sizes of 8 and 64 attaining the highest accuracy.

\section{Varying the placement of dropouts}
\begin{figure}[!htbp]
\centering
\includegraphics[width = 6in]{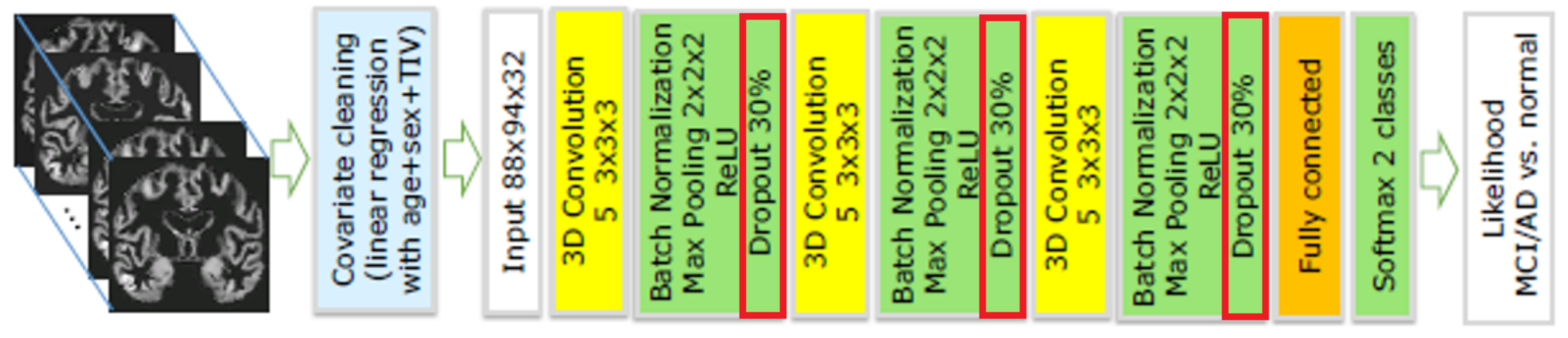}
\caption{Model with dropout layer after each CNN block}
\label{fig:drop-each}

\centering
\includegraphics[width = 6in]{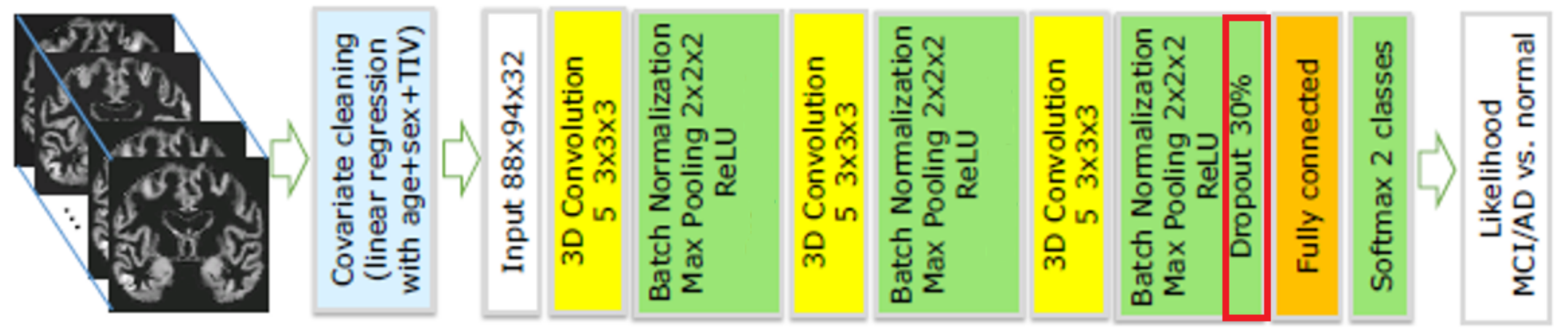}
\caption{Model with dropout layer all each CNN block}
\label{fig:drop-all}
\end{figure}

\begin{landscape}
\begin{figure}
    \includegraphics[height=4.2in,width=4.5in]{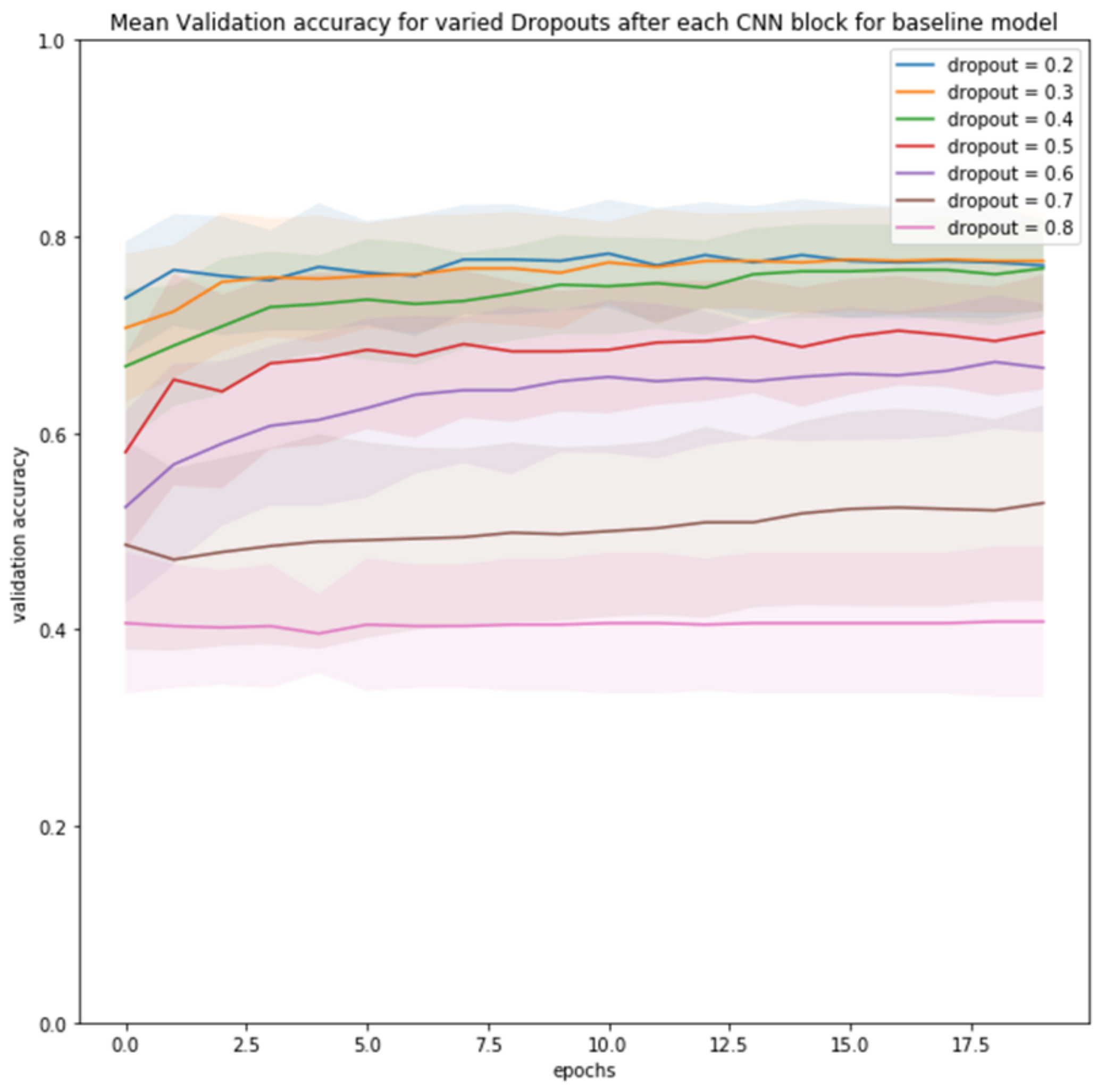}
    \includegraphics[height=4.2in,width=4.5in]{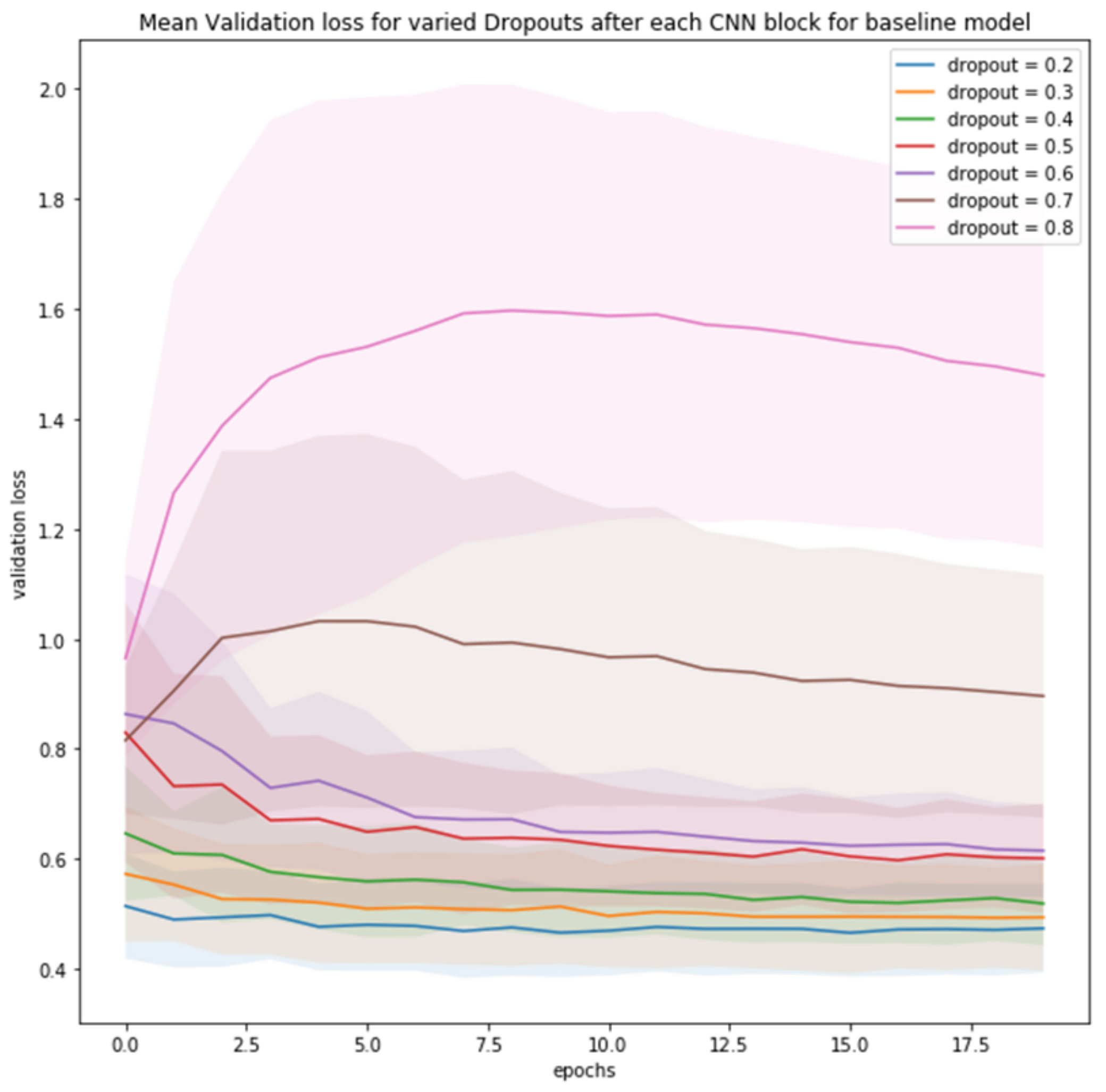}
    \caption{Mean Validation accuracy and validation loss across all the 10 folds for dropout layer after each convolution block; Standard deviation indicated by the shaded region}
    \label{fig:exp-drop-each}
\end{figure}
\end{landscape}

\begin{landscape}
\begin{figure}
    \includegraphics[height=4.2in,width=4.5in]{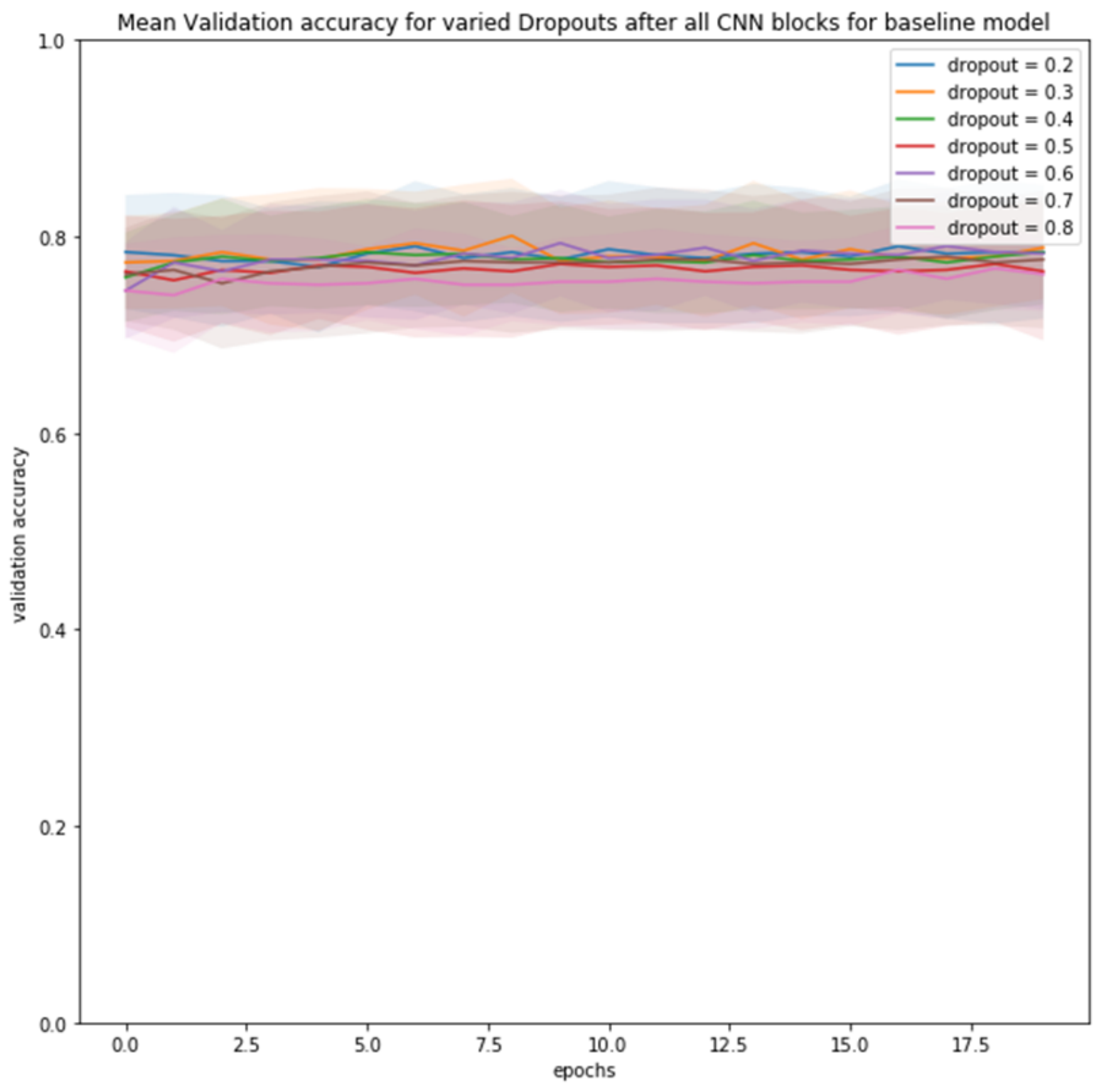}
    \includegraphics[height=4.2in,width=4.5in]{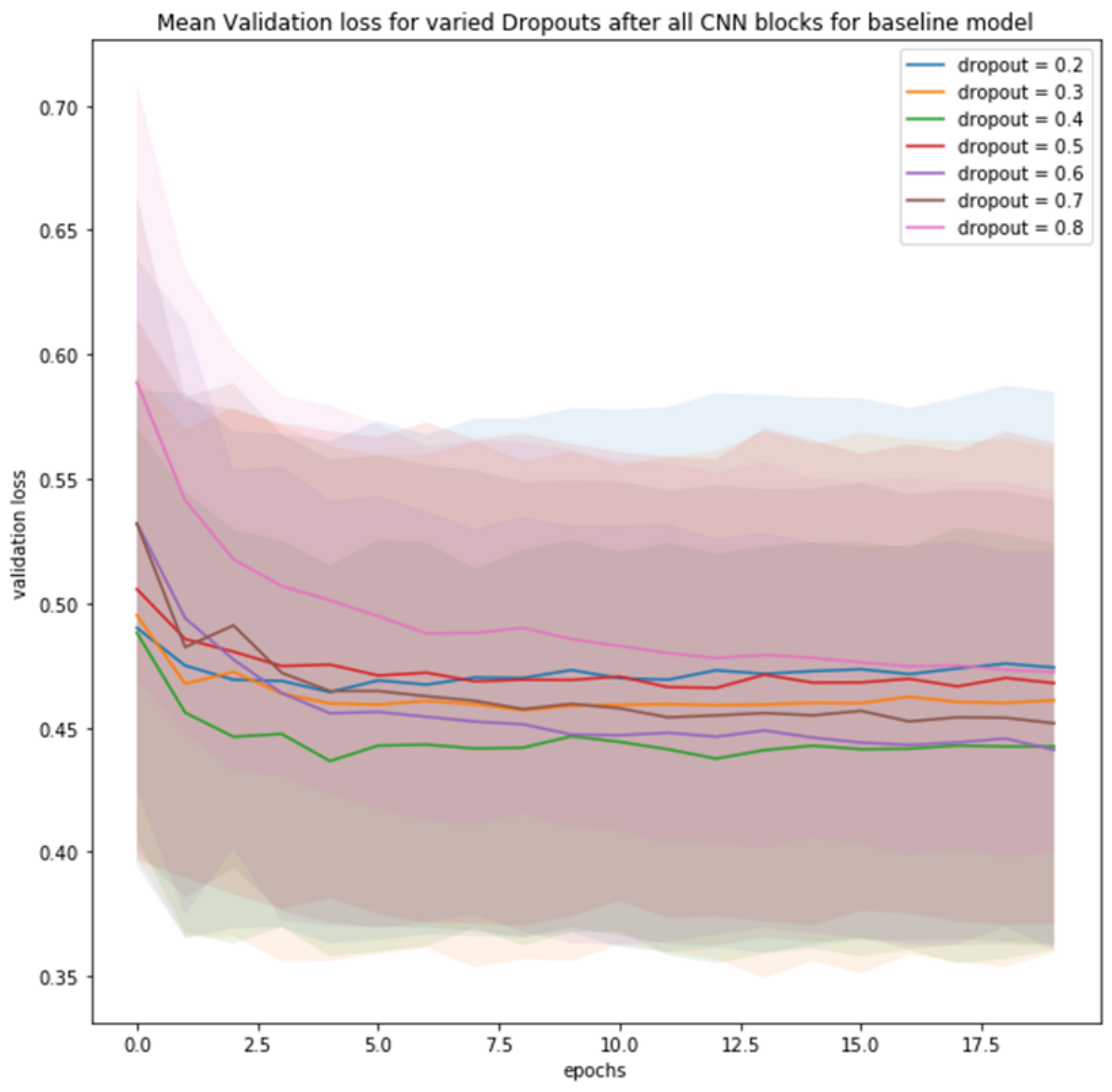}
    \caption{Mean Validation accuracy and validation loss across all the 10 folds for dropout layer after all convolution block; Standard deviation indicated by the shaded region}
    \label{fig:exp-drop-all}
\end{figure}
\end{landscape}

In this experiment, the dropout layer's placement in the CNN model and the percentage of dropout is varied.
The placement of the dropout layer is varied in two places in the CNN model, as illustrated in the Figures \ref{fig:drop-each} and \ref{fig:drop-all}. \\
The dropout value was varied in the range from 0.2 to 0.8. It was observed from the results as shown in tables \ref{tab:Table-Drop-each} and \ref{tab:Table-Drop-all} and the mean accuracy and loss plots in figures \ref{fig:exp-drop-each} and \ref{fig:exp-drop-all} that it is ideal to place the dropout layer after all the convolutional blocks with a dropout value of 0.4.

\begin{table}[!htbp]
\caption{Mean accuracy of model trained with dropout layer after each convolutional layer block}
\begin{tabular}{l l l l l l} \hline
\multicolumn{2}{c}{\multirow{2}{*}{}} & \multicolumn{4}{c}{Accuracy / AUC $\pm$ SD averaged across all 10 folds} \\
\cline{3-6}
\multicolumn{3}{c}{\multirow{3}{*}{}} & \multicolumn{3}{c}{AUC} \\
\cline{4-6}
Model & Dropout & Mean Acc & All test data & MCI vs CN & AD vs CN\\  \hline
\multirow{1}{10em}{CNN block + Dropout\\ + CNN block + Dropout \\+ CNN block + Dropout \\+ Flatten + 1 FC} & 0.2 & $77.0 \pm4.8$ & $\textbf{0.860} \pm0.053$  & $\textbf{0.780} \pm0.081$  & $0.953 \pm0.031$ \\
\multirow{1}{10em}&  0.3 & $\textbf{77.3} \pm5.0$ & $0.858 \pm0.055$ & $0.775 \pm0.085$  & $0.953 \pm0.037$ \\
\multirow{1}{10em}&  0.4 & $77.1 \pm5.0$ & $0.858 \pm0.052$  & $0.775 \pm0.084$  & $\textbf{0.955} \pm0.037$   \\
\multirow{1}{10em}&  0.5 & $75.4 \pm6.0$ & $0.855 \pm0.052$  & $0.772 \pm0.082$  & $0.951 \pm0.041$  \\ 
\multirow{1}{10em}&  0.6 & $66.6 \pm6.9$ & $0.831 \pm0.056$  & $0.746 \pm0.081$  & $0.929 \pm0.054$  \\
\multirow{1}{10em}&  0.7 & $59.7 \pm11.1$ & $0.825 \pm0.058$  & $0.737 \pm0.083$  & $0.928 \pm0.053$  \\ 
\multirow{1}{10em}&  0.8 & $53.4 \pm13.6$ & $0.809 \pm0.081$  & $0.726 \pm0.094$  & $0.905 \pm0.086$  \\ \hline
 \end{tabular}
    \label{tab:Table-Drop-each}

\vspace{1cm}

\caption{Mean accuracy of model trained with dropout layer after all convolutional layer block}
\begin{tabular}{l l l l l l} \hline
\multicolumn{2}{c}{\multirow{2}{*}{}} & \multicolumn{4}{c}{Accuracy / AUC $\pm$ SD averaged across all 10 folds} \\
\cline{3-6}
\multicolumn{3}{c}{\multirow{3}{*}{}} & \multicolumn{3}{c}{AUC} \\
\cline{4-6}
Model & Dropout & Mean Acc & All test data & MCI vs CN & AD vs CN\\  \hline
\multirow{1}{10em}{3 CNN blocks + Dropout \\ + Flatten + 1 FC} & 0.2 & $78.4 \pm7.1$ & $0.861 \pm0.058$  & $\textbf{0.785} \pm0.083$  & $0.950 \pm0.045$ \\
\multirow{1}{10em}&  0.3 & $78.6 \pm6.2$ & $0.860 \pm0.058$  & $0.780 \pm0.090$  & $0.953 \pm0.041$  \\
\multirow{1}{10em}&  0.4 & $\textbf{78.7} \pm5.3$ & $\textbf{0.864} \pm0.056$  & $0.784 \pm0.087$  & $\textbf{0.957} \pm0.039$  \\
\multirow{1}{10em}&  0.5 & $78.0 \pm6.3$ & $0.862 \pm0.056$  & $0.781 \pm0.085$  & $0.955 \pm0.040$  \\ 
\multirow{1}{10em}&  0.6 & $78.1 \pm6.2$ & $0.863 \pm0.055$  & $0.783 \pm0.083$  & $0.956 \pm0.045$  \\
\multirow{1}{10em}&  0.7 & $77.6 \pm7.4$ & $0.860 \pm0.059$  & $0.782 \pm0.086$  & $0.950 \pm0.051$  \\ 
\multirow{1}{10em}&  0.8 & $76.9 \pm5.8$ & $0.858 \pm0.055$  & $0.778 \pm0.082$  & $0.950 \pm0.045$  \\ \hline
 \end{tabular}
    \label{tab:Table-Drop-all}
\end{table}

\section{Varying the number of CNN blocks with 1 FC layer}
\begin{landscape}
\begin{figure}
    \includegraphics[height=4.2in,width=4.5in]{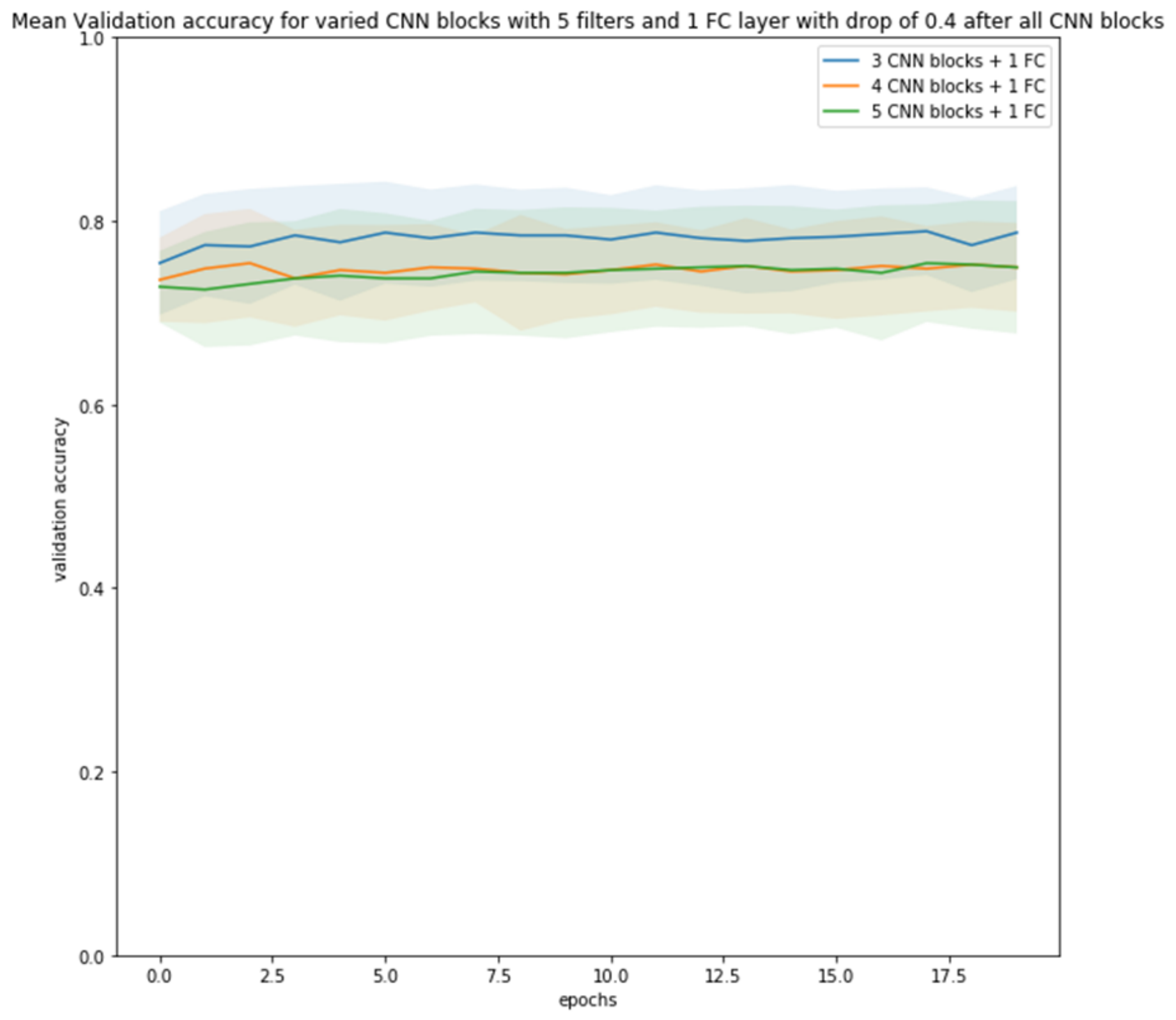}
    \includegraphics[height=4.2in,width=4.5in]{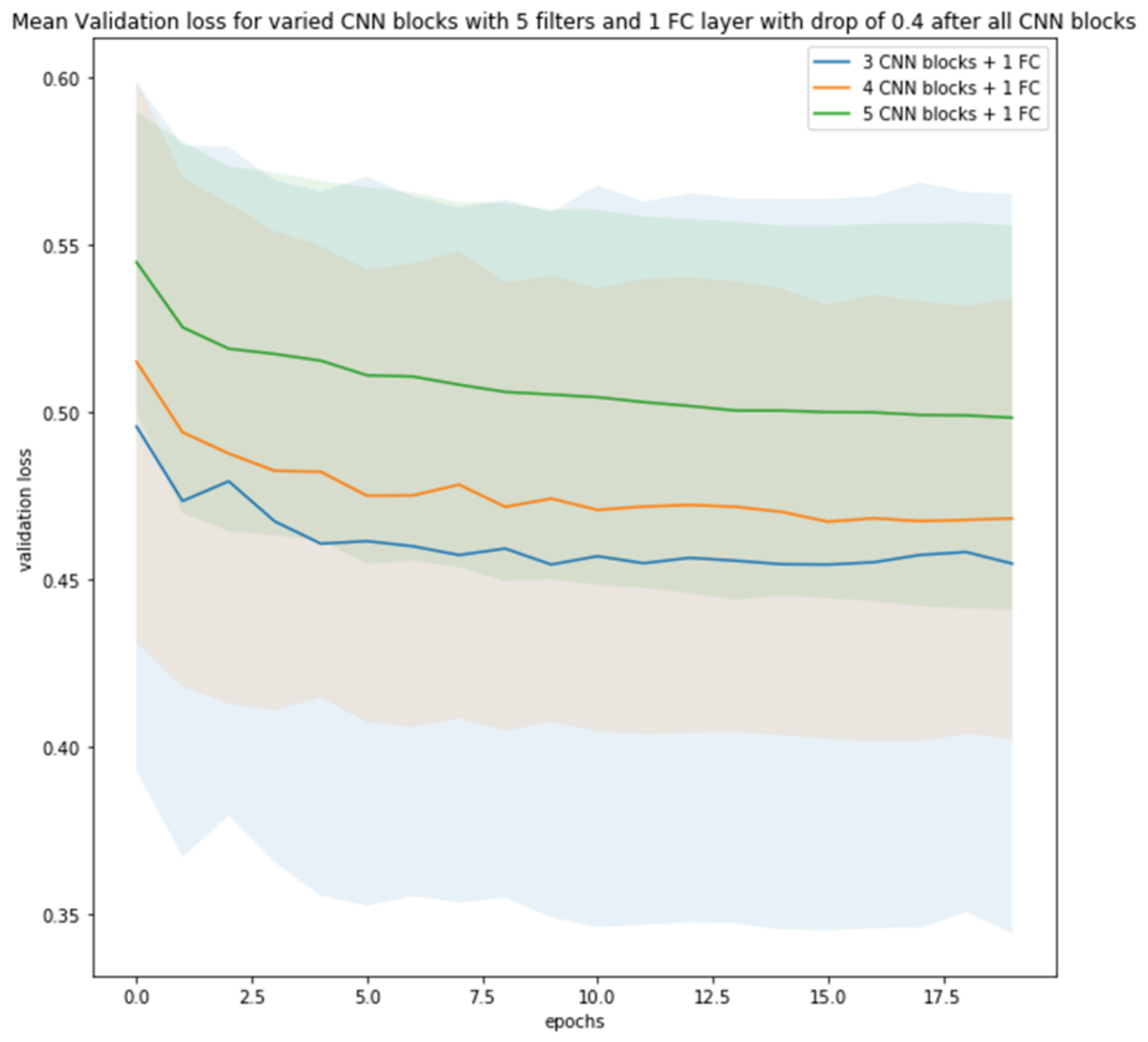}
    \label{fig:exp-cnn-blocks}
    \caption{Mean Validation accuracy and validation loss across all the 10 folds for varying CNN block sizes; Standard deviation indicated by the shaded region}
\end{figure}
\end{landscape}

The baseline model is modified from the simplest deep learning model\footnote{https://de.mathworks.com/help/deeplearning/ug/create-simple-deep-learning-network-for-classification.html} for the classification of MNIST \footnote{http://yann.lecun.com/exdb/mnist/} data. \\It uses 3 CNN blocks consisting of a CNN layer followed by a Batch normalisation layer and a Max pooling layer and finally a Fully connected layer with Softmax activation for classification.
By stacking several convolutional CNN blocks it is possible to extract more complex features of the input image and also since each CNN block is followed with a max pooling layer, the input size reduces as the number of CNN blocks increases which in turn reduces the number of parameters to be trained thereby reducing the possibility of overfitting.\\
The model was experimented for 3 different block sizes of depth 3,4 and 5 and adding a dropout of 0.4 after all the convolutional blocks. The model structure is summarised in Appendix A1. \\
Models are trained for 20 epochs with varying CNN block depth and the corresponding mean validation accuracies and validation losses across all 10 folds are plotted as shown in Figure \ref{fig:exp-cnn-blocks} and the model accuracies are shown in Table \ref{tab:Table-varying-cnn-block}.
It is observed that for the model with the simple 3-layer CNN block performs better than the models with more depth of CNN blocks.

\begin{table}[!htbp]
\caption{Mean accuracy of model trained with varying CNN block depths with 1 FC layer}
\begin{tabular}{l l l l l} \hline
\multicolumn{1}{c}{\multirow{1}{*}{}} & \multicolumn{4}{c}{Accuracy / AUC $\pm$ SD averaged across all 10 folds} \\
\cline{2-5}
\multicolumn{1}{c}{\multirow{1}{*}{}} & \multicolumn{4}{c}{AUC} \\
\cline{3-5}
Model & Mean Acc & All test data & MCI vs CN & AD vs CN\\  \hline
\multirow{1}{10em}{3 CNN blocks + 1 FC} & $\textbf{78.7} \pm5.3$ & $\textbf{0.865} \pm0.065$ & $\textbf{0.785} \pm0.097$ & $\textbf{0.958} \pm0.046$\\
\multirow{1}{10em}{4 CNN blocks + 1 FC} & $74.9 \pm5.1$  & $0.845 \pm0.050$ & $0.763 \pm0.070$ & $0.940 \pm0.043$\\
\multirow{1}{10em}{5 CNN blocks + 1 FC} & $74.9 \pm7.5$  & $0.836 \pm0.053$ & $0.756 \pm0.076$ & $0.930 \pm0.057$\\ \hline
 \end{tabular}
    \label{tab:Table-varying-cnn-block}
\end{table}

\section{Varying the number of filters in each Convolutional layer per CNN block}
\begin{landscape}
\begin{figure}
    \includegraphics[height=4.2in,width=4.5in]{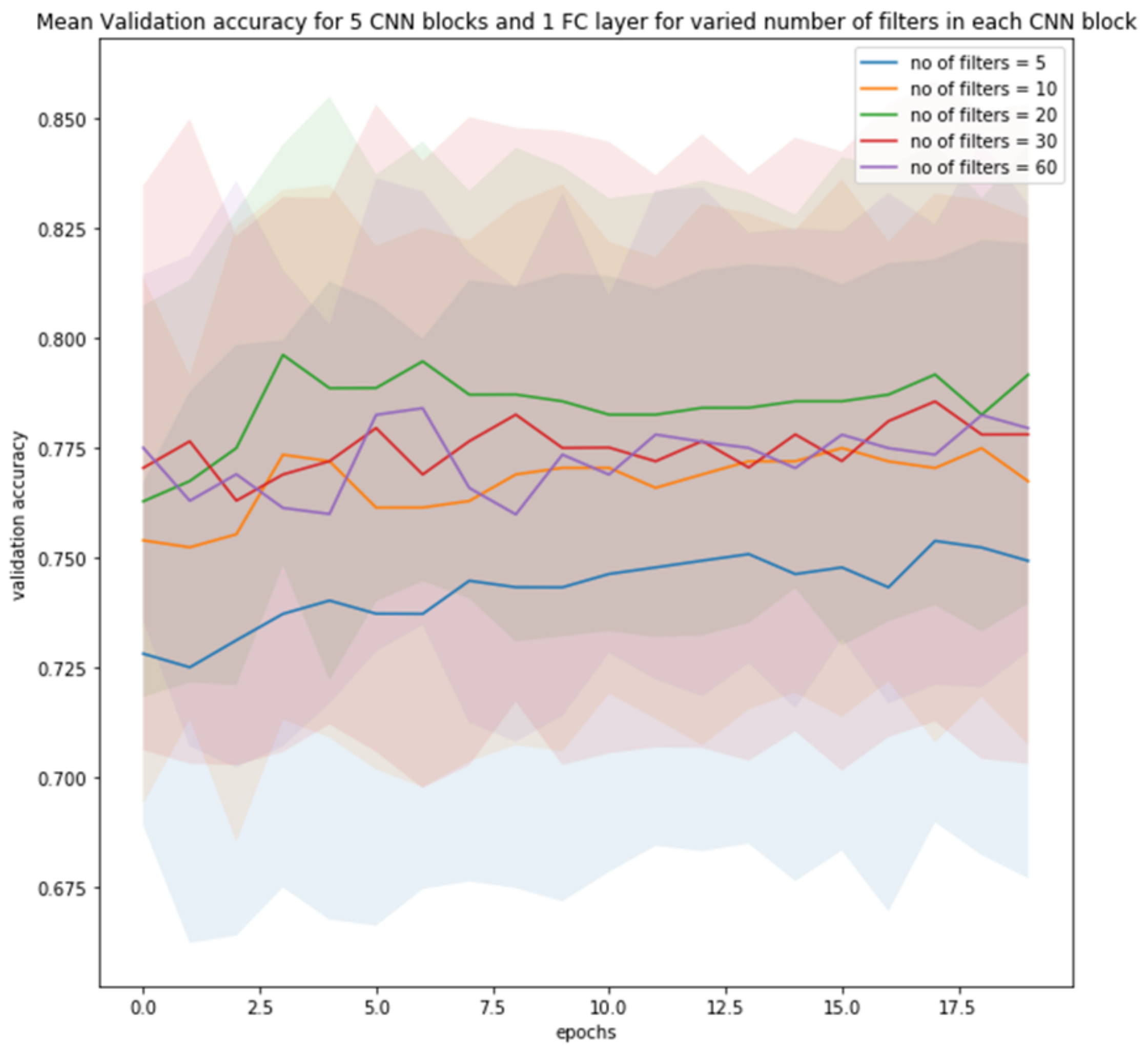}
    \includegraphics[height=4.2in,width=4.5in]{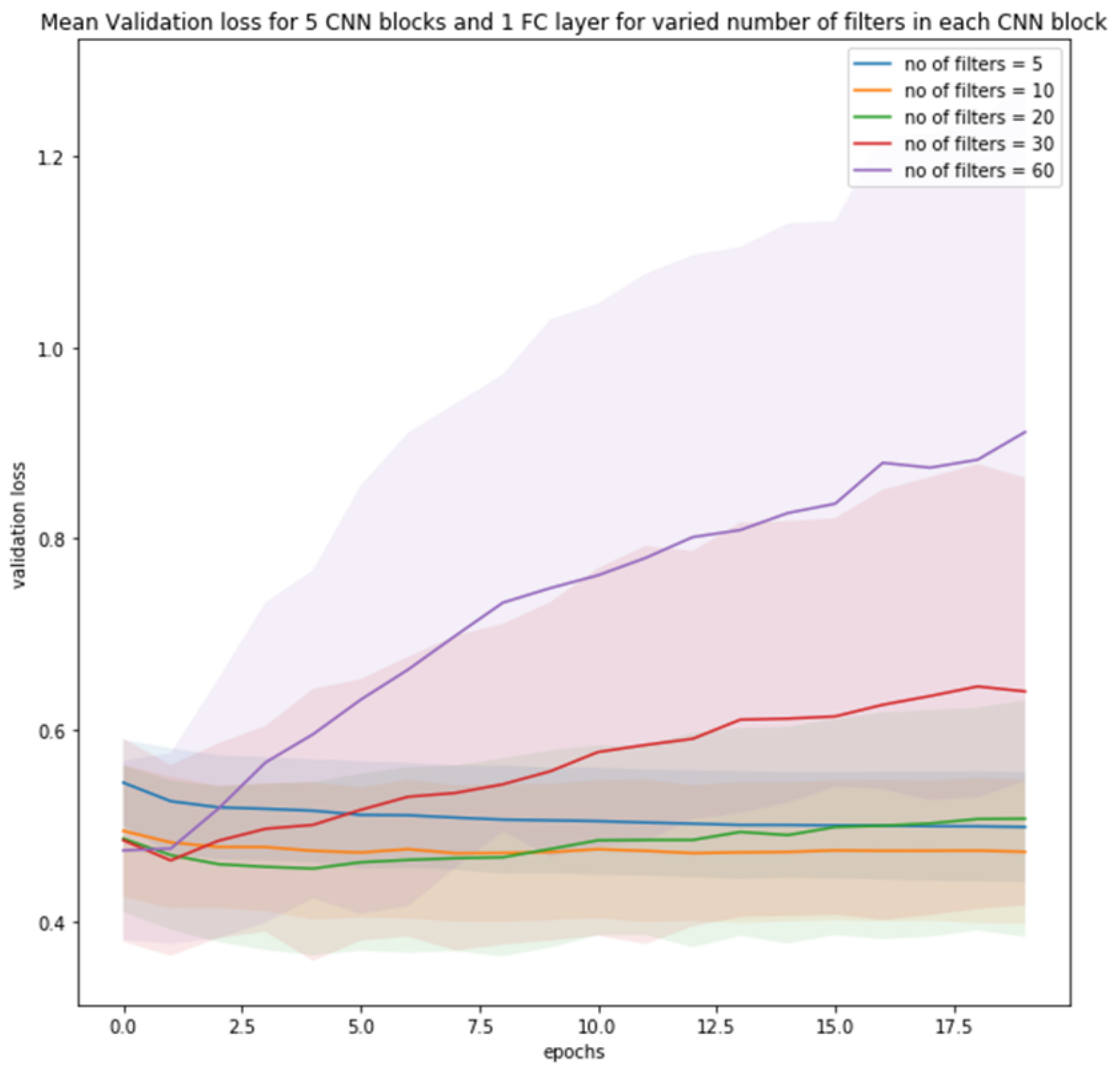}
    \label{fig:exp-cnn-filters}
    \caption{Mean Validation accuracy and validation loss across all the 10 folds for varying number of convolution layer filters; Standard deviation indicated by the shaded region}
\end{figure}
\end{landscape}

The convolutional filters are used to extract features from the input image and each filter captures different features. As we stack up the convolutional layers, the filters try to learn more complicated and fine-tuned features from the features extracted from the preceding convolutional layer filter. In the cases of models used for IMAGENET \cite{imgnet} the number of filter size is increased in the subsequent layers to capture as many combinations of features as possible. However, as the number of filters increases the number of trainable parameters also increases, leading to overfitting. In order to avoid this, the number of filters is increased in the convolutional layer, but in the subsequent layers, the number of filters is kept the same instead of increasing to avoid overfitting \cite{featureext}. \\
In the 3-layer CNN blocks, as the number of filters was increased per block, the models got overfitted as the number of training parameters increases. Thus, experiments are done for 5-layer CNN blocks with the number of convolutional filters varied from 5,10,20,30 and 60 per convolutional block with 1 FC layer and dropout of 0.4 after all convolutional blocks and trained for 20 epochs. \\The mean validation accuracies and validation losses across all 10 folds are plotted, as shown in Figure \ref{fig:exp-cnn-filters} and the model accuracies are shown in Table  \ref{tab:Table Varying cnn filters}. Model structure is summarised in Appendix A2. The best result was obtained for the 5 layer CNN block model using 20 convolutional filters per block. 

\begin{table}[!htbp]
\caption{Mean accuracy of model trained with varying number of convolutional filters on the 5 CNN block model}
\begin{tabular}{l l l l l l} \hline
\multicolumn{2}{c}{\multirow{2}{*}{}} & \multicolumn{4}{c}{Accuracy / AUC $\pm$ SD averaged across all 10 folds} \\
\cline{3-6}
\multicolumn{3}{c}{\multirow{3}{*}{}} & \multicolumn{3}{c}{AUC} \\
\cline{4-6}
Model & Filters & Mean Acc & All test data & MCI vs CN & AD vs CN\\  \hline
\multirow{1}{10em}{5 CNN blocks + 1 FC} & 5  & $74.9 \pm7.5$  & $0.836 \pm0.053$  & $0.756 \pm0.076$  & $0.930 \pm0.057$\\
\multirow{1}{10em}{5 CNN blocks + 1 FC}&  10 & $76.7 \pm6.2$   & $0.839 \pm0.057$  & $0.755 \pm0.085$  & $0.936 \pm0.049$\\
\multirow{1}{10em}{5 CNN blocks + 1 FC}&  20 & $\textbf{79.2} \pm5.5$   & $\textbf{0.866} \pm0.064$  & $\textbf{0.794} \pm0.089$  & $0.950 \pm0.046$\\
\multirow{1}{10em}{5 CNN blocks + 1 FC}&  30 & $77.8 \pm7.8$   & $0.862 \pm0.055$  & $0.790 \pm0.0855$  & $\textbf{0.959} \pm0.047$\\ 
\multirow{1}{10em}{5 CNN blocks + 1 FC}&  60 & $77.9 \pm5.4$   & $0.859 \pm0.062$  & $0.781 \pm0.0855$  & $0.950 \pm0.045$\\ \hline
 \end{tabular}
    \label{tab:Table Varying cnn filters}
\end{table}

\section{Varying the number of Fully connected layers}
\begin{landscape}
\begin{figure}
    \includegraphics[height=4.2in,width=4.5in]{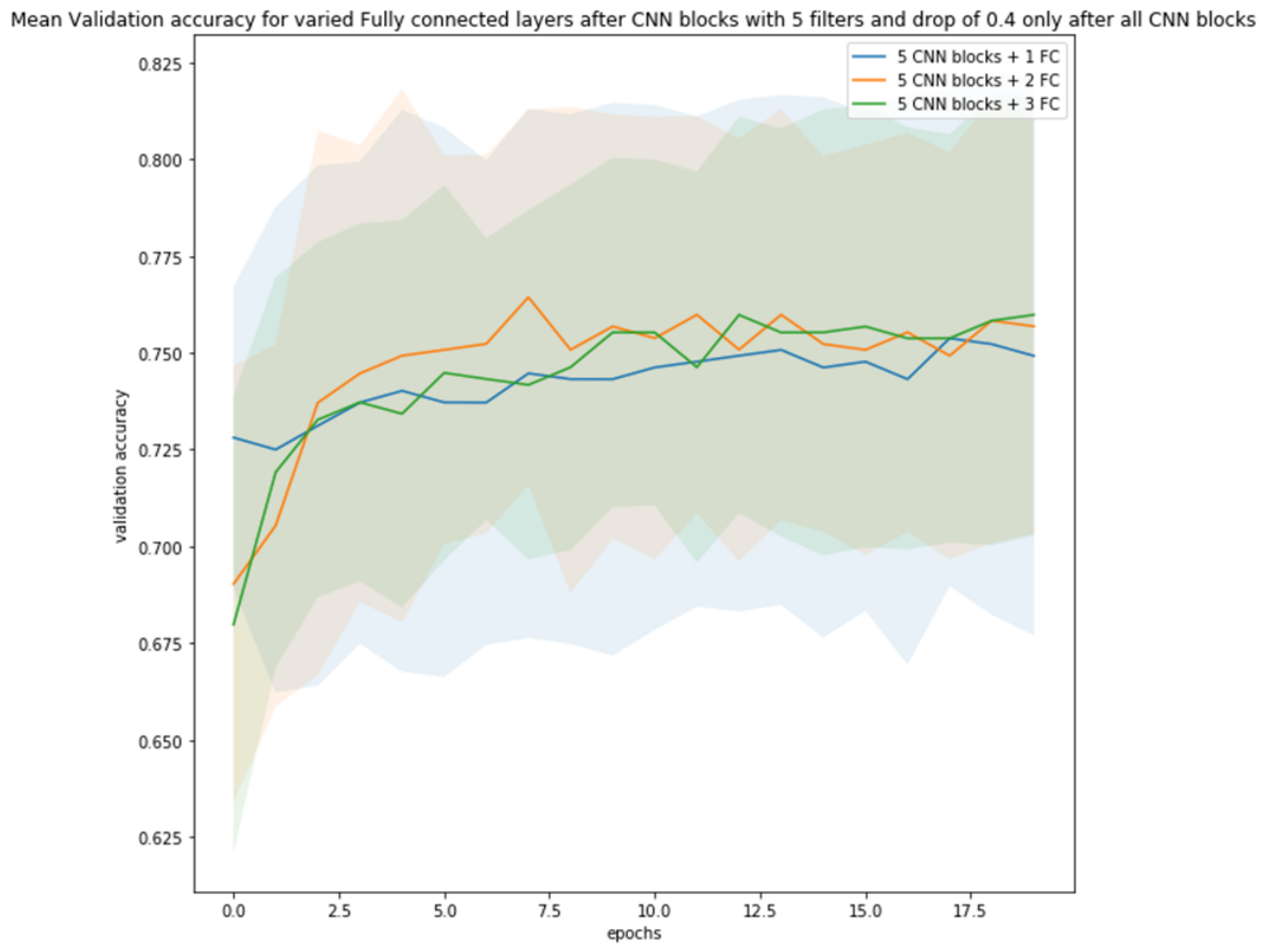}
    \includegraphics[height=4.2in,width=4.5in]{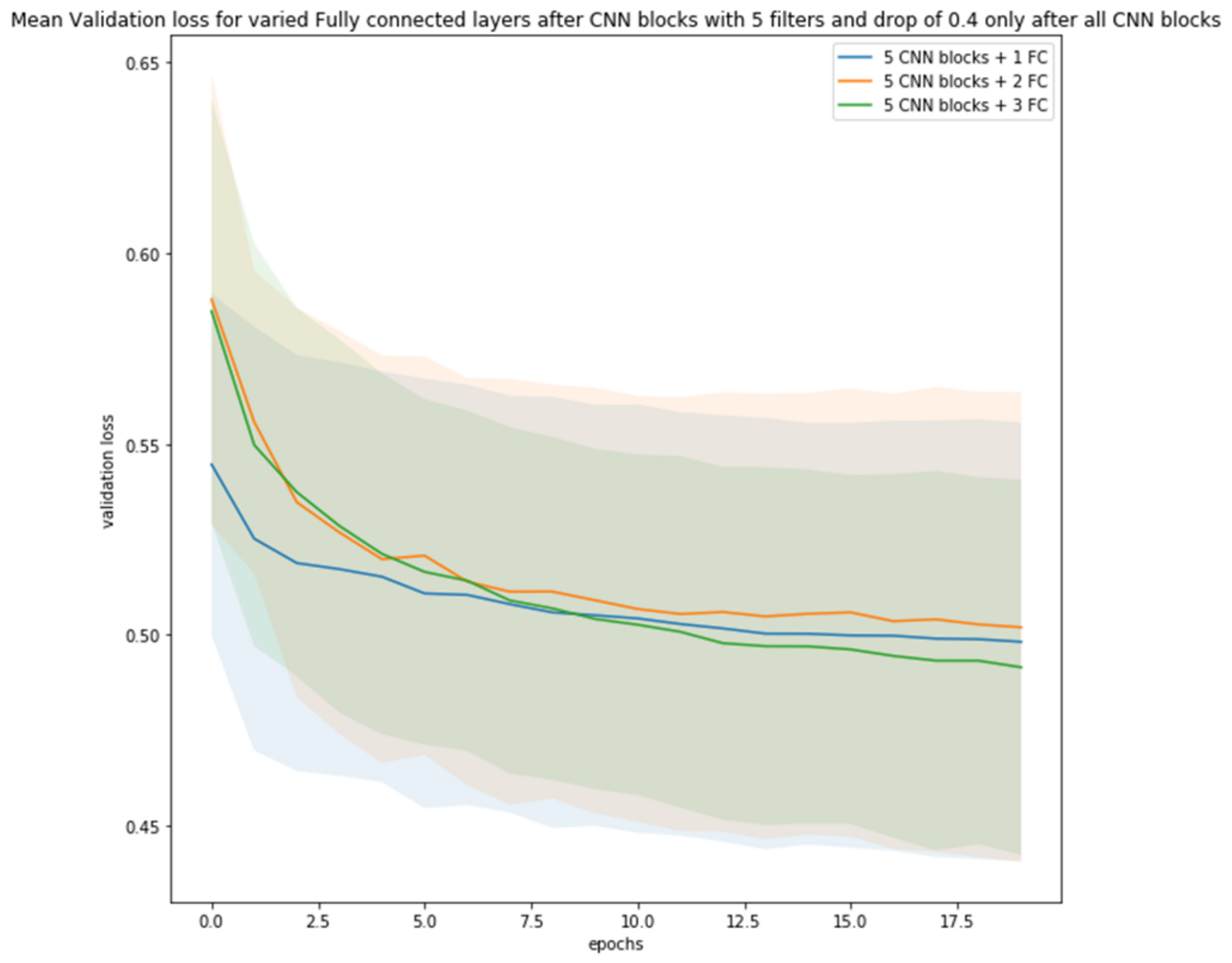}
    \label{fig:exp-fc-layers}
    \caption{Mean Validation accuracy and validation loss across all the 10 folds for varying number of fully connected layers; Standard deviation indicated by the shaded region}
\end{figure}
\end{landscape}

In the 3 CNN layer blocks, when an additional FC layer is added, the model got overfitted. 
Thus, we experiment by varying the fully connected layers (1,2 and 3) added after the 5 CNN block model, each block consisting of 5 convolutional filters and with a dropout of 0.4 only after all the convolutional blocks. The number of neurons in the subsequent FC layers is equal to the neurons resulting from the flattening operation. For eg. if the number of neurons after flattening is 20, then the additional FC layers used will have 20 neurons and they work as a simplified multi-layer perceptron (MLP).\\ Models are trained for 20 epochs with varying CNN block depth, and the corresponding mean validation accuracies and validation losses across all 10 folds are plotted as shown in Figure \ref{fig:exp-fc-layers} and the model accuracies are shown in Table \ref{tab:Table Vary FC}. Model structure is summarised in Appendix A3.

\begin{table}[!htbp]
\caption{Mean accuracy of model trained with varying FC layers to the 5 CNN block model}
\begin{tabular}{l l l l l} \hline
\multicolumn{1}{c}{\multirow{1}{*}{}} & \multicolumn{4}{c}{Accuracy / AUC $\pm$ SD averaged across all 10 folds} \\
\cline{2-5}
\multicolumn{1}{c}{\multirow{1}{*}{}} & \multicolumn{4}{c}{AUC} \\
\cline{3-5}
Model & Mean Acc & All test data & MCI vs CN & AD vs CN\\  \hline
\multirow{1}{10em}{5 CNN blocks + 1 FC} & $74.9 \pm7.5$ & $\textbf{0.836} \pm0.053$ & $\textbf{0.756} \pm0.076$ & $0.930 \pm0.057$\\
\multirow{1}{10em}{5 CNN blocks + 2 FC} & $75.7 \pm5.6$ & $0.820 \pm0.061$ & $0.736 \pm0.085$ & $0.917 \pm0.053$ \\
\multirow{1}{10em}{5 CNN blocks + 3 FC} & $\textbf{76.0} \pm6.0$ & $0.832 \pm0.054$ & $0.738 \pm0.078$ & $\textbf{0.941} \pm0.044$ \\ \hline
 \end{tabular}
    \label{tab:Table Vary FC}
\end{table}

\section{Applying best hyper parameters on whole brain input scans}

From all the experiments performed, the best hyper parameters observed for the optimization and training hyper parameters are batch size of 8/64, learning rate and decay as lr = 0.001, decay = 0.01 and dropout =0.4 after all the convolutional blocks. Two of the best models are

\begin{enumerate}
    \item 3 CNN block with a dropout layer of 0.4 after all CNN blocks and with 5 filters in each convolution block
    \item  5 CNN block with a dropout layer of 0.4 after all CNN blocks and with 20 filters in each convolution block
\end{enumerate}

These parameters were then used to train the model on the whole brain input scans and the results are shown in table 5.7.

\begin{table}[h]
\caption{Mean accuracy of model trained on whole brain model with the best hyper parameters}
\begin{tabular}{l l l l l} \hline
\multicolumn{1}{c}{\multirow{1}{*}{}} & \multicolumn{4}{c}{Accuracy / AUC $\pm$ SD averaged across all 10 folds} \\
\cline{2-5}
\multicolumn{1}{c}{\multirow{1}{*}{}} & \multicolumn{4}{c}{AUC} \\
\cline{3-5}
Model & Mean Acc & All test data & MCI vs CN & AD vs CN\\  \hline
\multirow{1}{10em}{3 CNN blocks } & $73.4 \pm6.1$ & $0.856 \pm0.062$ & $0.769 \pm0.107$ & $0.930 \pm0.057$\\
\multirow{1}{10em}{5 CNN blocks} & $76.0 \pm6.0$ & $0.832 \pm0.054$ & $0.738 \pm0.078$ & $0.941 \pm0.044$ \\ \hline
 \end{tabular}
    \label{tab:Table-wb-models}
\end{table}
\FloatBarrier
\chapter{Evaluation}
\section{LRP visualizations on the whole brain model}

\begin{figure}
    \centering
    \includegraphics[width = 5.7in]{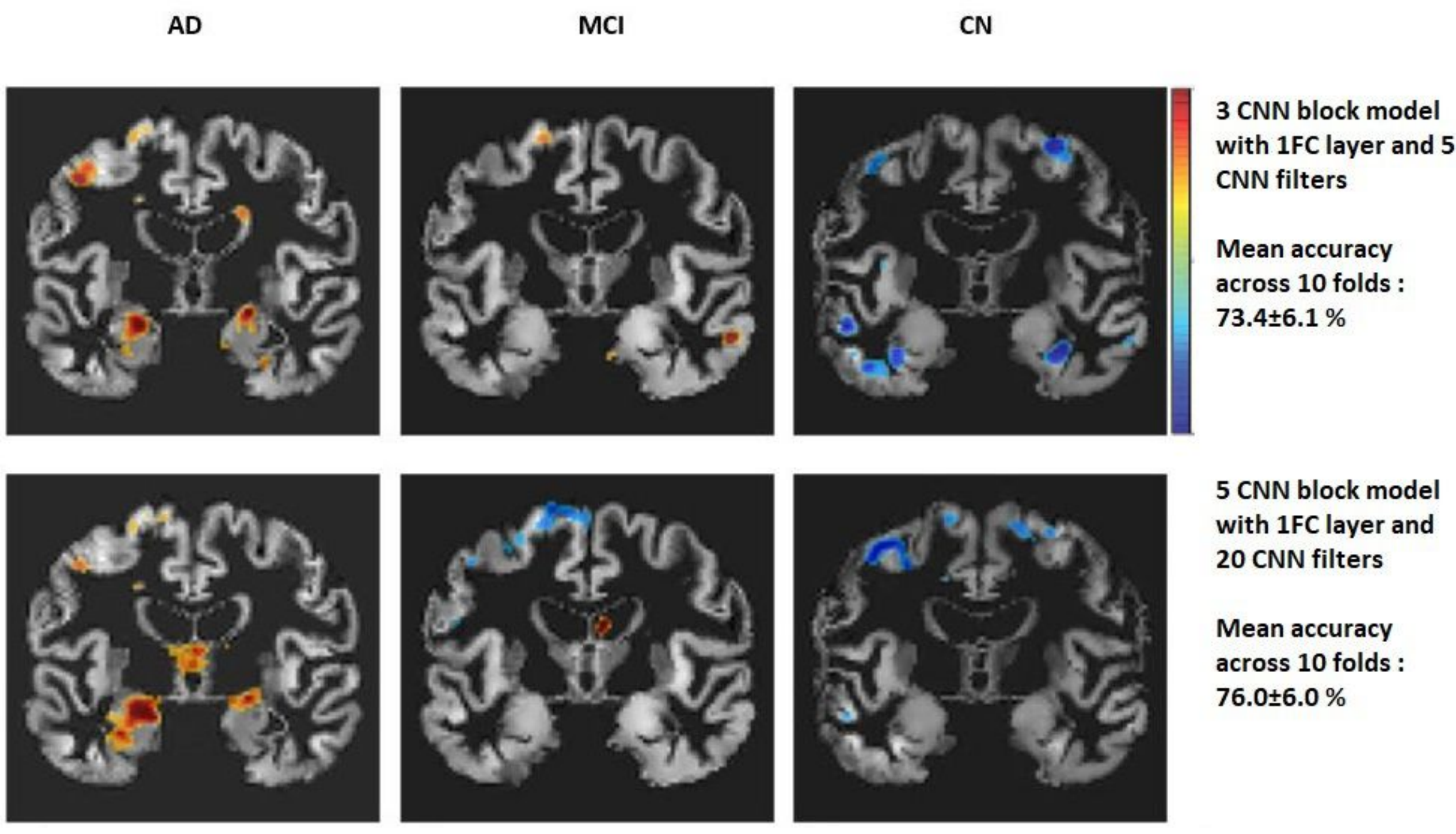}
    \caption{LRP visualizations for the 3 CNN and 5 CNN block models for the whole brain model}
    \label{fig:wb_lrp_viz}
\end{figure}

In this section, the visualizations are presented on the models that were trained using the whole brain volume scans, as reported in section 5.9.
The iNNvestigate library \cite{innvestigate} was used for this implementation. As the intensity range of the relevance maps varied significantly between different subjects and the different models, they were scaled linearly to a fixed range allowing for an excellent visual comparison. The raw relevance maps were overlaid on the original input data with 50\% transparency, and a smoothed and thresholded visualization is obtained by displaying the most prominent clusters, i.e., the top 30 percentile of intensity values.\\
This allows us for qualitative evaluation, and it was observed that for AD subjects, the relevance maps were concentrated in the hippocampal region, as shown in figure \ref{fig:wb_lrp_viz}. 

This also corresponds with the medical literature, in which they analyze the volume atrophy in the hippocampal region, which is considered as a reliable biomarker associated with AD since the atrophy in this region has been linked to cognitive impairment suggestive of AD \cite{hippo-atrophy}.\\
In the case of scans that were predicted as MCI it was observed that there was very little relevance around the hippocampal region and no positive relevance for subjects predicted as CN.

\section{Dice Similarity coefficient (DSC)}

The similarity between the relevance maps obtained from the different models using LRP is explored in this section.
DSC is used as a statistical validation metric to evaluate the performance of image segmentation of MRI images \cite{DSC}. It compares two binary sets of data and checks how close these sets are related. The value of a DSC ranges from 0 to 1, 0 indicating no spatial overlap between two sets of binary segmentation results, to 1, indicating complete overlap. DSC score for two sets, X and Y, is defined as 
\begin{equation}
    DSC = \frac{2|X \cap Y|}{|X| + |Y|}
\end{equation}
In order to obtain the binary relevance maps, the relevance maps are thresholded by considering only the positive relevance scores.
We get the DSC by comparing the heat relevance maps of two different models for a single subject predicted as AD. These are analyzed in two ways, firstly by considering only the relevance in the hippocampal region, obtained by segmenting out the hippocampal region alone from the relevance map using a binary template of the hippocampal region and secondly, the relevance produced for the full input volume.\\ The binarized relevance maps are shown in figures 6.2 to 6.4. They are analyzed for robustness, where the 3 CNN block model was trained for 3 times, and the relevance maps are checked for how close the relevance maps are produced during each run, as presented in Table \ref{tab:dice-relevance-robustness}. Comparison between different trained models using the 32 coronal slices as well as the whole brain volume as input were also done as depicted in Tables \ref{tab:dice-relevance-diff-models} and \ref{tab:dice-relevance-wb}. In Figure 6.2, the number tags (1,2,3,4) indicate the four different models compared as presented in Table 6.2 in the respective order.\\ The volume of relevance map with respect to the hippocampal area was also attained by considering the ratio between the number of pixels occupied in the hippocampal region in the relevance map against the number of pixels occupied in the hippocampal template. Heat maps of the DSC were also obtained by taking each pair of the different models trained using the subvolume of the scan (32 coronal slices) for both the relevance comparison against the hippocampal region alone and the full input volume as shown in Figure 6.5.

From the heatmaps shown in Figure 6.5, it can be observed that the relevance maps when comparing the relevance map in the hippocampal region for all the variations of 3 CNN and 5 CNN block models obtain a high similarity score except the model trained without the residualization process of the input scans which show less relevance in the hippocampal region with the scores ranging from 0.76 to 0.81. When the relevance heat maps are compared against the full input space, they show less similarity between the models with the scores ranging between 0.51 to 0.61 in the region of the heatmap, as mentioned above.\\
This indicates that the relevance map for all the different models trained except the model trained without the residualized input scans shows high relevance in the hippocampal region.

Although this comparison was illustrated only for a single AD subject , it cannot be a representative for all AD subjects as the relevance maps vary for each subject. Hence a group statistical analysis    comparing the correlation between the relevance maps for all subjects is required. For this purpose the Pearson correlation analysis as discussed in the next section is done where the relevance in the hippocampal region is correlated against the hippocampal volume of the subjects.

\begin{table}
\caption{Evaluating the robustness of relevance map trained on 3 CNN block model}
\begin{tabular}{l l l l l } \hline
\multicolumn{3}{c}{\multirow{3}{*}{}} & \multicolumn{2}{c}{Dice coefficient} \\
\cline{4-5}
Model & Accuracy & Vol of relevance wrt & Relevance in & Overall relevance\\&& hippocampal template & hippocampal region \\\hline
\multirow{1}{10em}{Run 1} & 87.88\% & 61.77\% & 1  & 1 \\
\multirow{1}{10em}{Run 2} & 86.32\% & 58.73\% & 0.804  & 0.532   \\
\multirow{1}{10em}{Run 3} & 86.14\% & 45.27\% & 0.776  & 0.614 \\ \hline
 \end{tabular}
    \label{tab:dice-relevance-robustness}

\vspace{1cm}

\caption{Evaluating the relevance map trained on different models}
\begin{tabular}{l l l l l } \hline
\multicolumn{3}{c}{\multirow{3}{*}{}} & \multicolumn{2}{c}{Dice coefficient} \\
\cline{4-5}
Model & Accuracy(\%) & Vol of relevance wrt & Relevance in & Overall relevance \\&& hippocampal template & hippocampal region \\\hline
\multirow{1}{10em}{3 CNN block} & 87.88\% & 61.77\% & 1  & 1 \\
\multirow{1}{10em}{5 CNN block} & 88.02\% & 72.32\% & 0.840  & 0.481   \\
\multirow{1}{10em}{3 CNN block without} & 86.36\% & 60.53\% & 0.797  & 0.564   \\
\multirow{1}{10em}{data augmentation} & & & &   \\
\multirow{1}{10em}{3 CNN block without} & 80.62\% & 38.42\% & 0.497  & 0.233 \\
\multirow{1}{10em}{residual input} & & && \\ \hline
 \end{tabular}
    \label{tab:dice-relevance-diff-models}

\vspace{1cm}

\caption{Evaluating the relevance map trained on whole brain model model}
\begin{tabular}{l l l l l } \hline
\multicolumn{3}{c}{\multirow{3}{*}{}} & \multicolumn{2}{c}{Dice coefficient} \\
\cline{4-5}
Model & Accuracy(\%) & Vol of relevance wrt & Relevance in & Overall relevance \\&& hippocampal template & hippocampal region \\\hline
\multirow{1}{10em}{3 CNN block} & 78.78\% & 44.69\% & 1  & 1 \\
\multirow{1}{10em}{5 CNN block} & 83.33\% & 75.03\% & 0.725  & 0.521 \\ \hline
 \end{tabular}
    \label{tab:dice-relevance-wb}

\end{table}

\begin{figure}
    \centering
    \includegraphics[height=2.4in,width=5.6in]{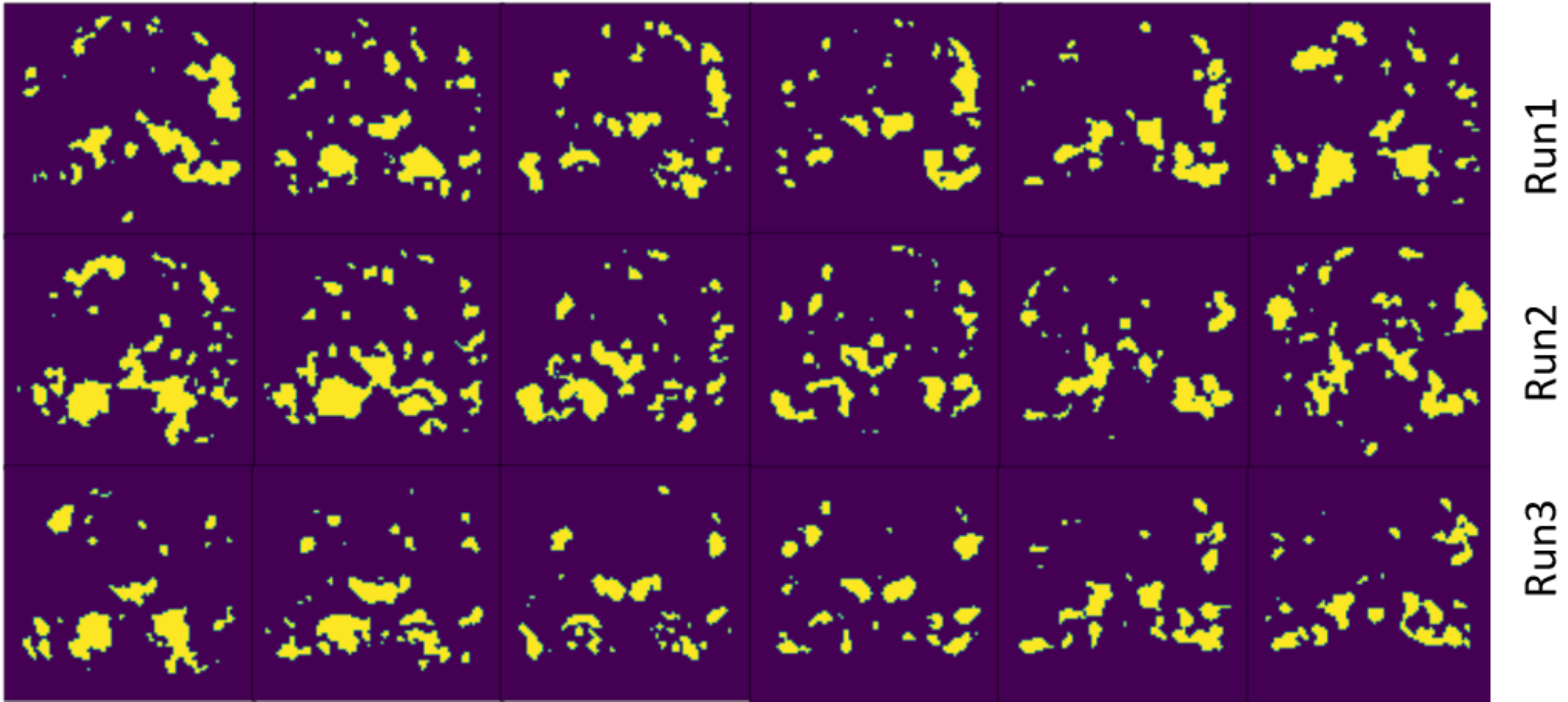}
    \caption{Binarized relevance maps after thresholding to check the visual robustness of relevance map trained on 3 CNN block model}
    \label{fig:thresh-robustness-3runs}
    
    \centering
    \includegraphics[height=2.4in,width=5.7in]{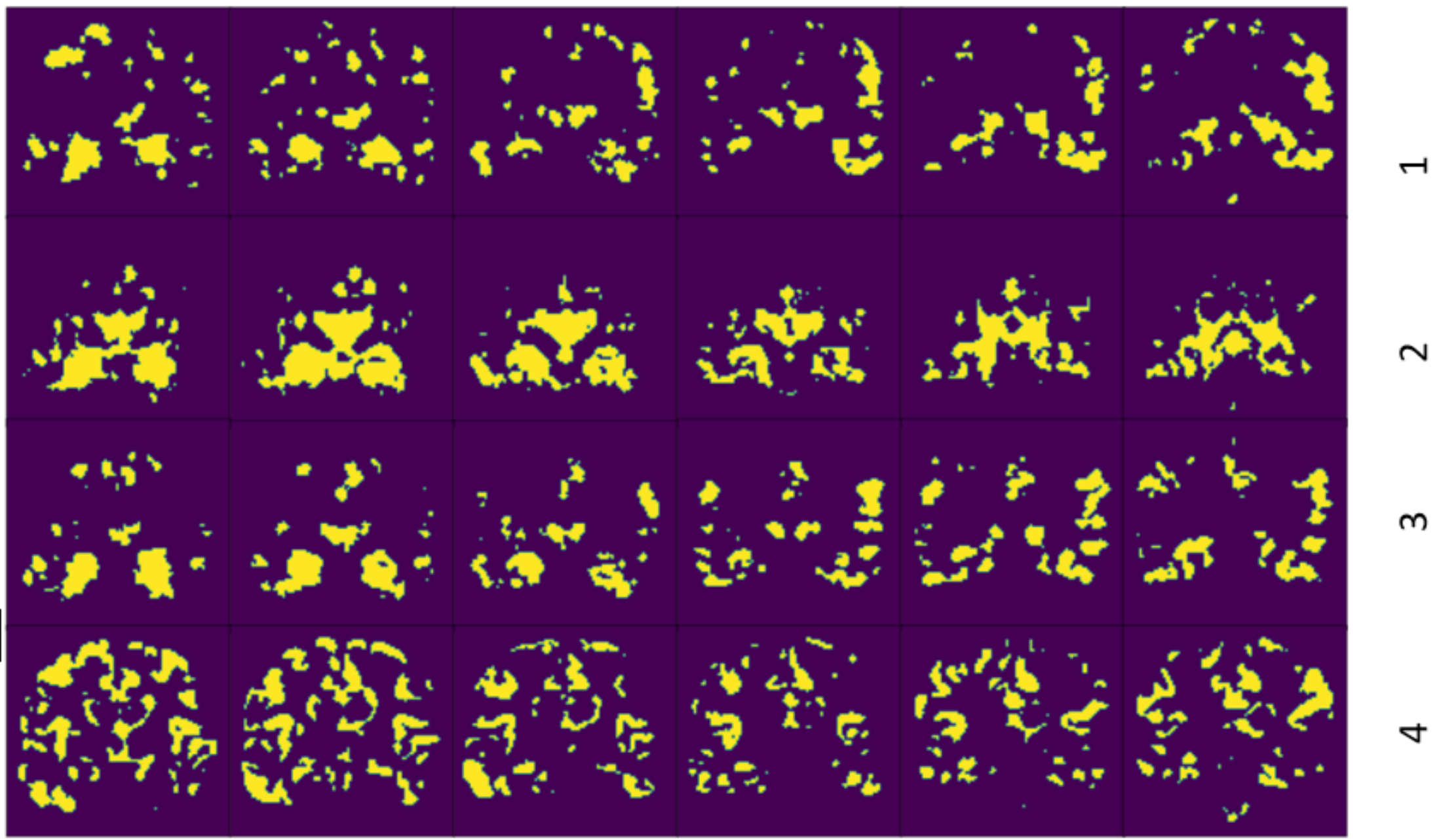}
    \caption{Binarized relevance maps after thresholding to check the difference in relevance map between different models}
    \label{fig:thresh-diff-models}

    \centering
    \includegraphics[height=1.5in,width=6in]{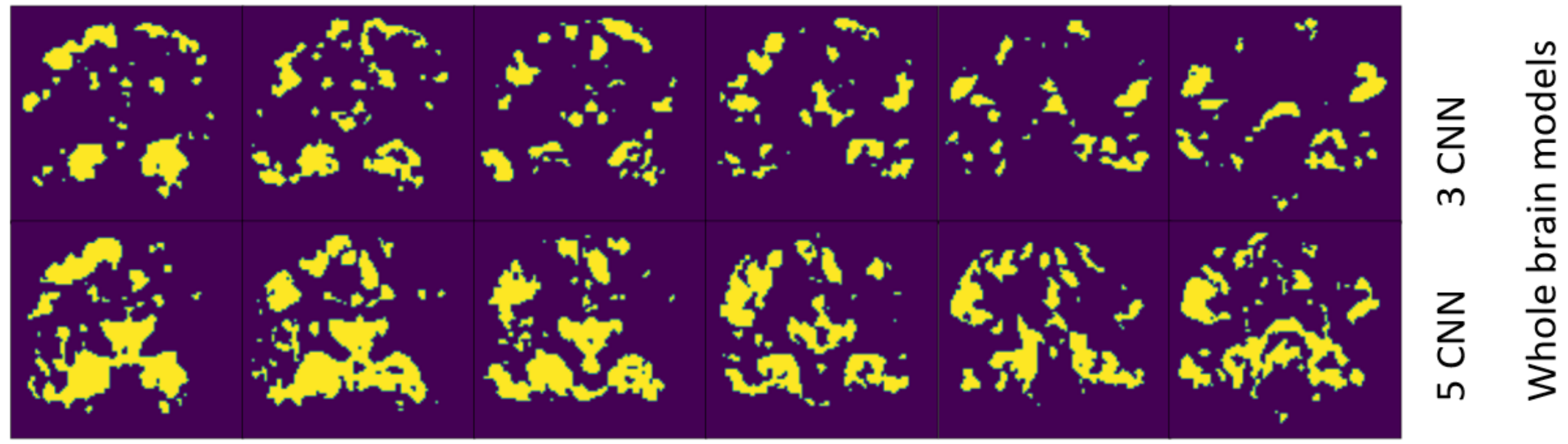}
    \caption{Binarized relevance maps after thresholding to check the difference in relevance map between different whole brain models }
    \label{fig:thresh-wb}
\end{figure}

\begin{figure}
    \includegraphics[height=4in,width=4.5in]{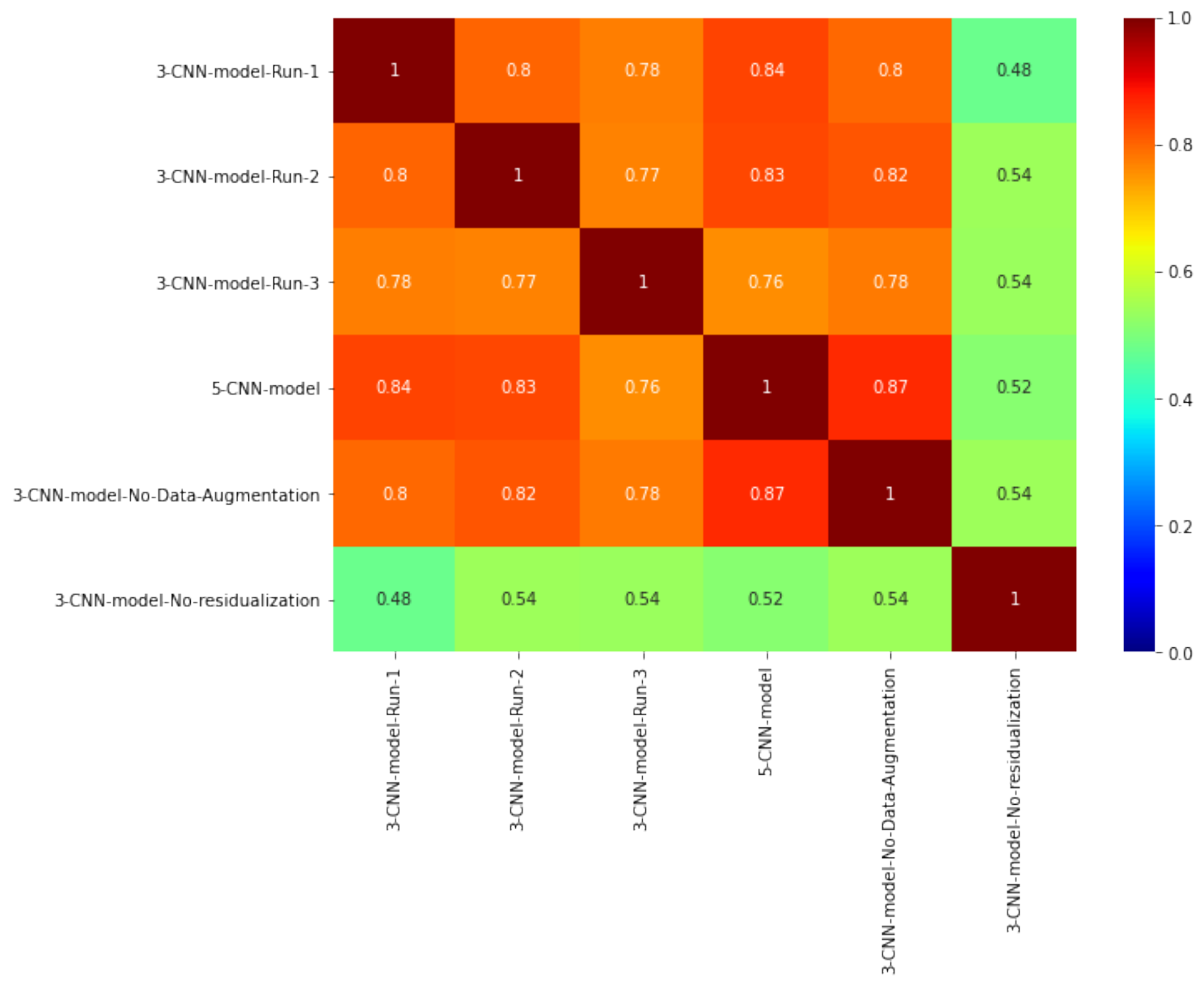}
    \label{fig:DSC-hippo}
    \newline
    \includegraphics[height=4in,width=4.5in]{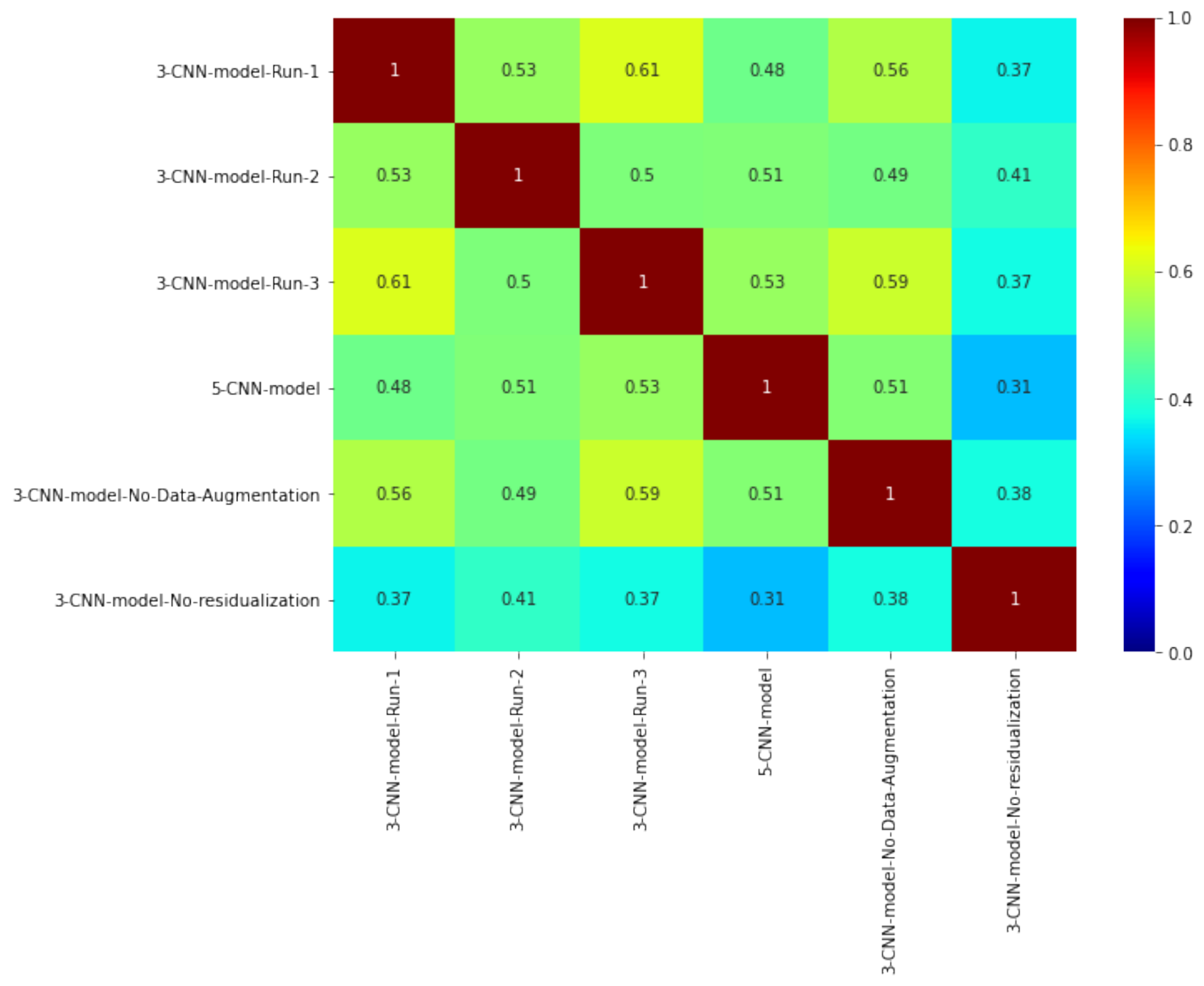}
    \caption{Heat map of DSC scores comparing each pair of models by considering the relevance scores in segmented hippocampal region and the whole input region as shown from top to bottom}
    \label{fig:DSC-wb}
\end{figure}
\newpage
\section{Pearson Correlation of CNN relevance heatmaps and hippocampus volume}
Correlation is a statistical measure used to assess the linear association between two continuous variables. In this section, we perform a group statistic analysis in which the relevance scores in the hippocampal region of all the subjects in the ADNI dataset are correlated with the hippocampus volume per subject. We expect that we get a negative correlation between the CNN relevance score and hippocampus volume, i.e., a low hippocampus region with a high relevance score for an AD subject and vice-versa for a CN subject. This is because the volume in the hippocampus volume is less for an AD subject due to the atrophy in the region and since the relevance maps show the regions in the brain that the model predicts for the classification between AD and CN, we expect the model to show relevance in this region as was verified in the previous section. We use Pearson correlation \cite{pearson-coeff} and obtain the scatter plots showing the correlation between the aggregated relevance of hippocampus voxels against the hippocampus volume. Pearson's correlation coefficient $\rho$ of two variable X and Y is given by the formula:

\begin{equation}
    \rho_{X,Y} = \frac{cov(X,Y)}{\sigma_X \sigma_Y}
\end{equation}

where cov(X,Y) is the covariance of the two variables X and Y and $\sigma_{X}$, $\sigma_{Y}$ are the standard deviations of the variables X and Y. It has a value between +1 and -1. A value of +1 is total positive linear correlation, 0 is no linear correlation, and -1 is total negative linear correlation.

\subsection{Correlation plots on 32 slice model}
The scatter plots and the Pearson's correlation generated for the subvolume of input scans (89 * 32 * 94), used for the tuning of the hyper parameters for the 3 CNN and 5 CNN block models are shown in figures 6.6 and 6.7 respectively. The 3 CNN block model has a higher coefficient score of -0.71 compared with the latter model with a score of -0.68.
\begin{figure}
    \centering
    \includegraphics[width = 5.7in]{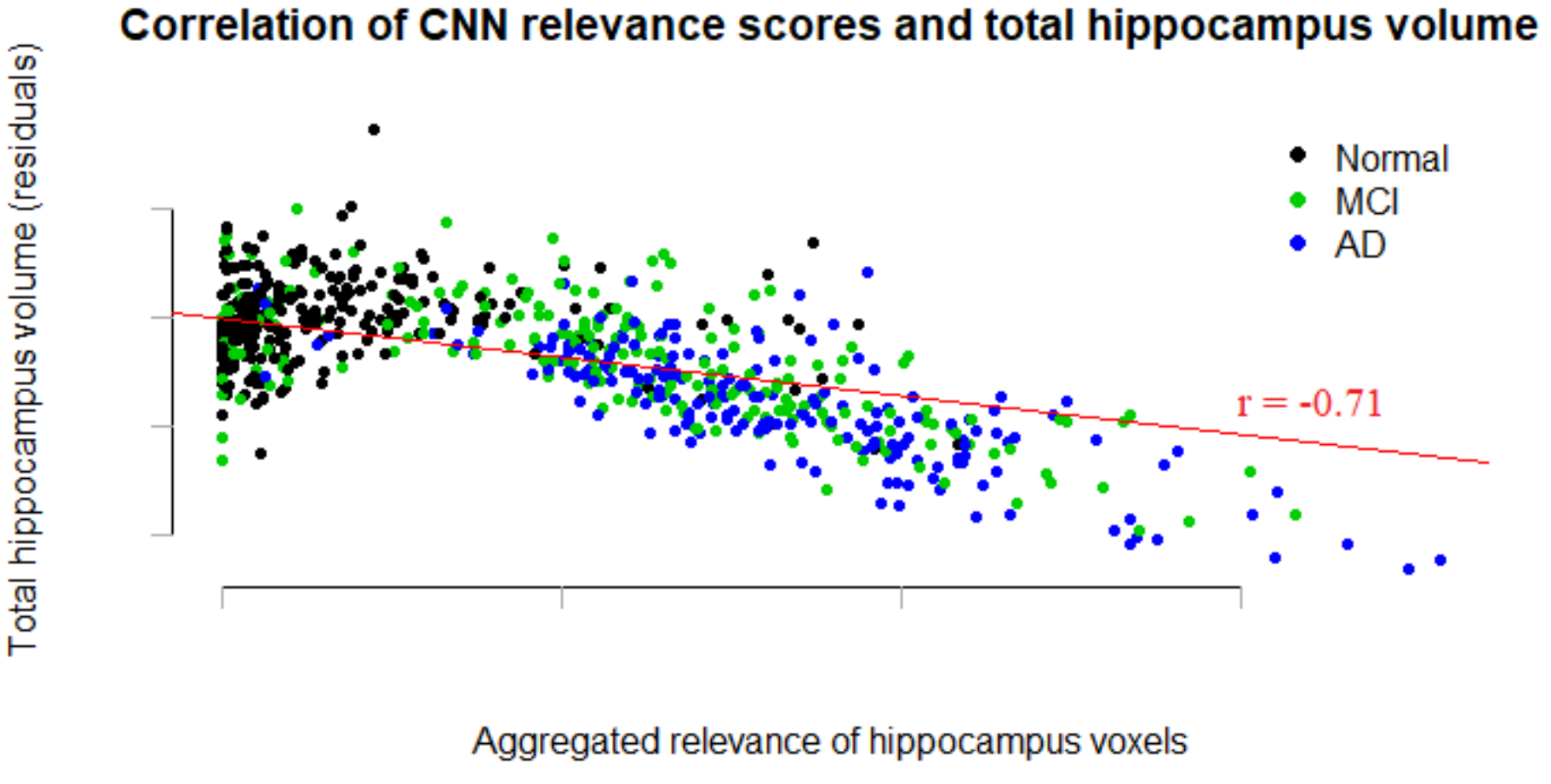}
    \caption{Correlation plot of relevance heatmap vs total hippocampus volume for 3 CNN block model}
    \label{fig:3cnn-32slice-corr-plot}
    \vspace{1cm}
    \centering
    \includegraphics[width =5.7in]{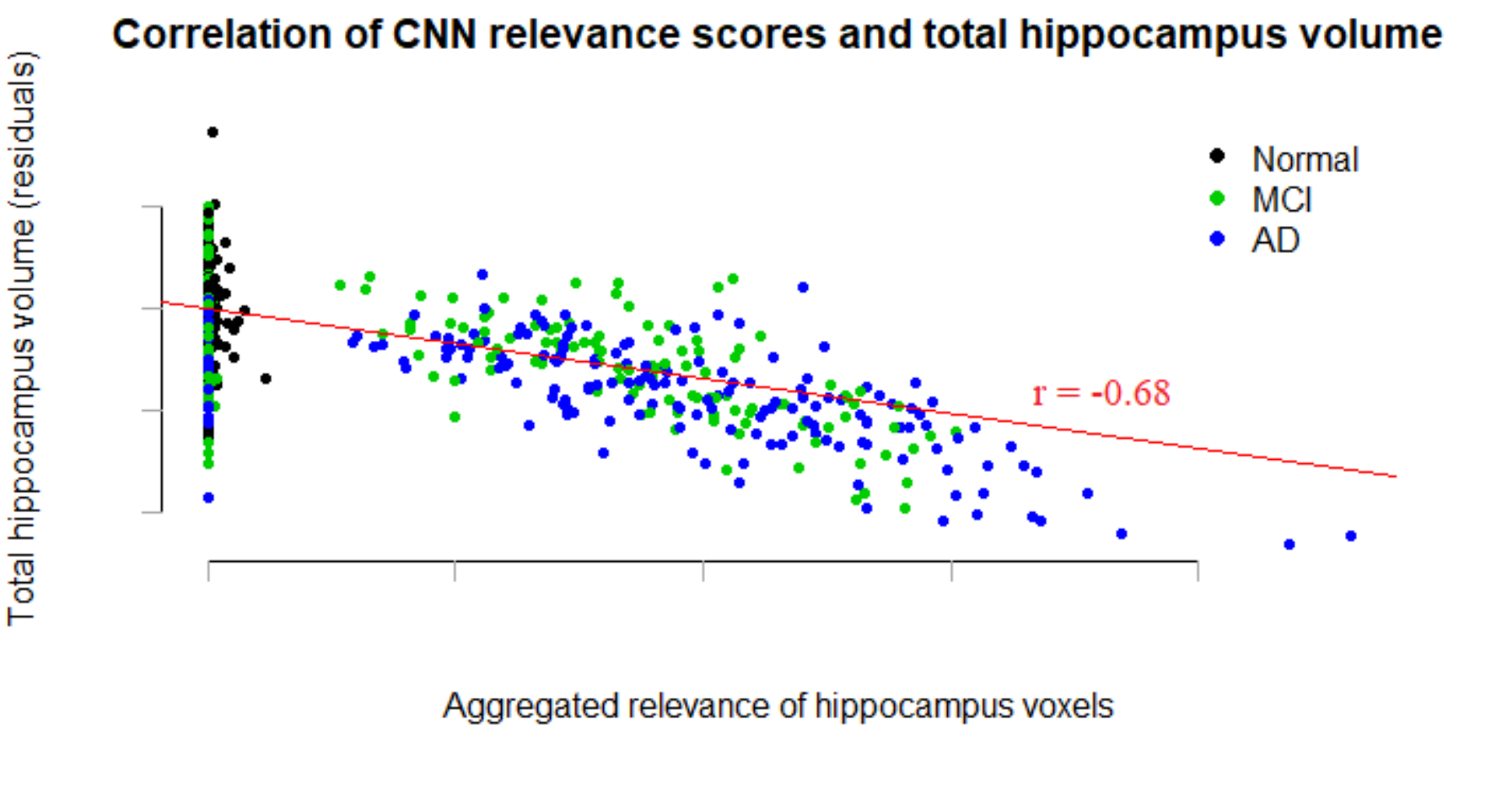}
    \caption{Correlation plot of relevance heatmap vs total hippocampus volume for 5 CNN block model}
    \label{fig:5cnn-32slice-corr-plot}
\end{figure}

\subsection{Correlation plots on Whole brain model}
 The scatter plots and the pearson's correlation generated for the whole brain volume of input scans (89 * 111 * 94) for the 3 CNN and 5 CNN block models are shown in figures 6.8 and 6.9 respectively. The 3 CNN block model has a higher coefficient score of -0.69 compared with the latter model with a score of -0.56.
\begin{figure}
    \centering
    \includegraphics[height=3.5in,width=5.7in]{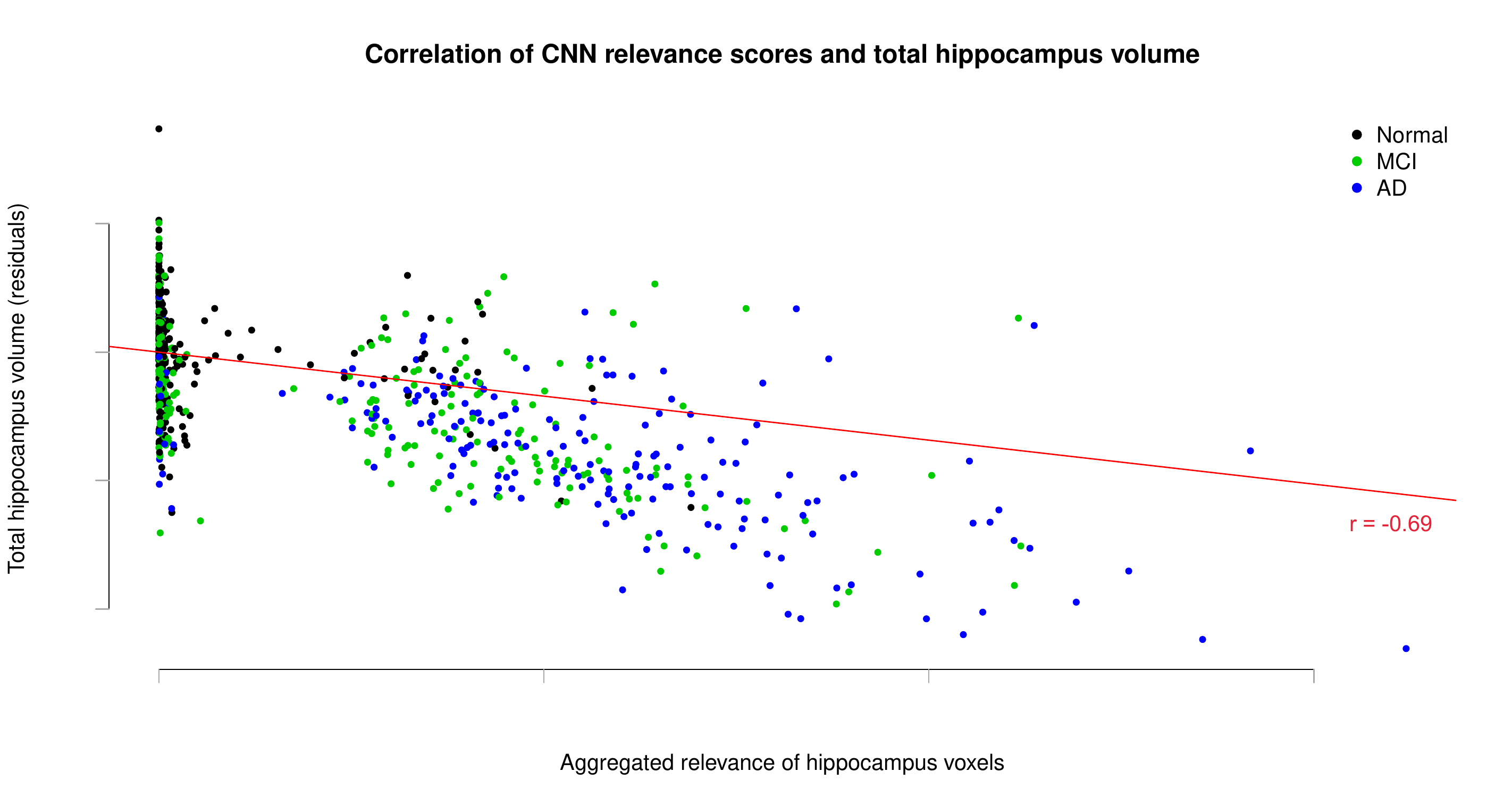}
    \caption{Correlation plot of relevance heatmap vs total hippocampus volume for 3 CNN block model using whole brain volume as input}
    \label{fig:3cnn-wb-corr-plot}
    \vspace{1cm}
    \centering
    \includegraphics[height=3.5in,width=5.7in]{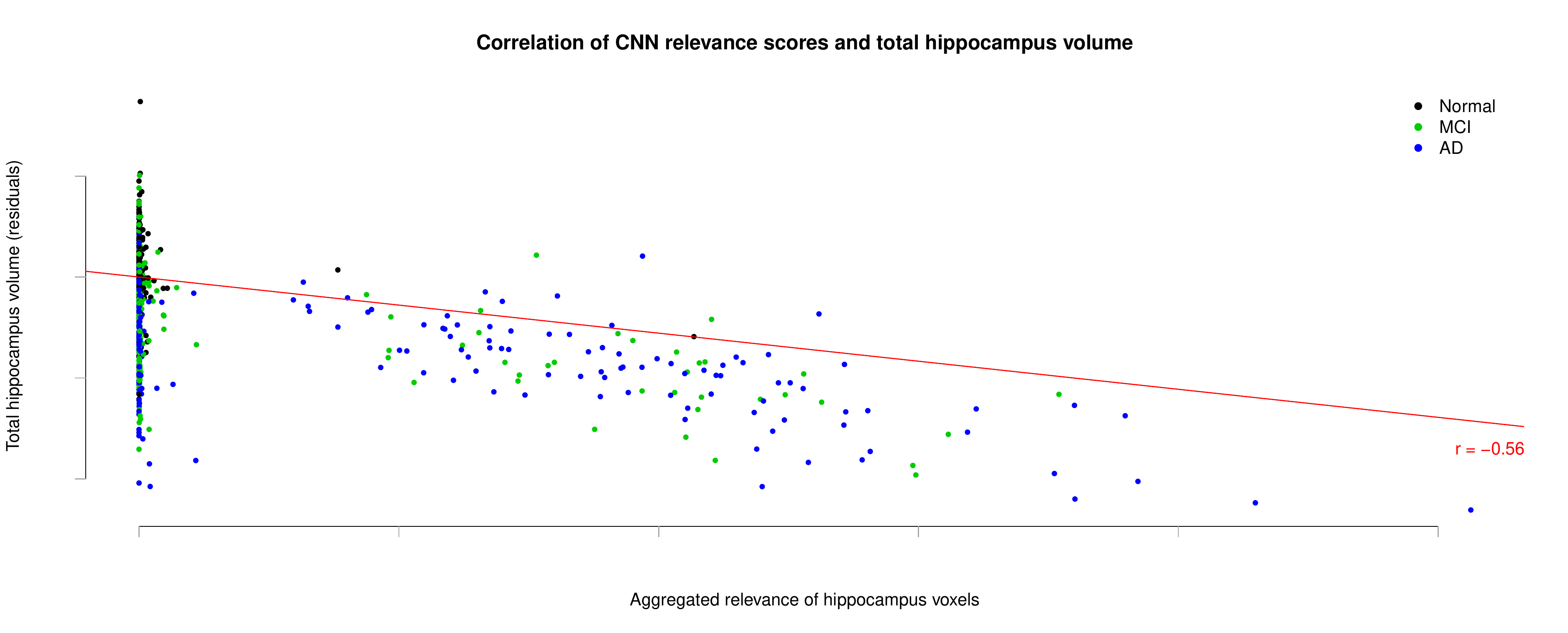}
    \caption{Correlation plot of relevance heatmap vs total hippocampus volume for 5 CNN block model using whole brain volume as input}
    \label{fig:5cnn-wb-corr-plot}
\end{figure}

It can be observed from these results that the 3 CNN block model generalizes well and gives a higher correlation value when compared with the 5 CNN block model in the ADNI data when the whole volume of input scan is considered.

\section{Testing on AIBL dataset}
Results from the two previous sections indicate that the simple model of 3 CNN block architecture gives high relevance on the hippocampal area with high accuracy in the classification of AD vs. CN and generalizes well with a high correlation score on the ADNI dataset.
Thus the fold with the highest accuracy obtained by training the 3 CNN block model on the full input volume of the scan was applied to the AIBL dataset and tested.

By applying the model on the AIBL data, we get an accuracy of 82 \% with an AUC score of 0.82 and an f1 score of 0.88 and 0.62 for the classification of control (NC) and MCI/AD respectively. We get a high recall / True positive rate (TPR) of 0.91 for the prediction of NC subjects, however the recall of prediction of AD/MCI is low with a value of 0.56. This low value is attributed due to the fact that in the case of the trained model using ADNI data, the LMCI was combined along with AD to be categorized as one class. In contrast, in the AIBL dataset, there are only MCI instead of LMCI samples. As discussed in the introduction, the difference between the two is that MCI is a combination of Early MCI (EMCI) and Late MCI (LMCI), with the latter being the stage closest for a subject to have AD. Therefore the MCI subjects in the AIBL dataset could have many subjects that are in the early MCI stage with less atrophy in the hippocampus region. This aspect is not explored further due to the lack of medical knowledge for segregating the late MCI scans from the MCI subjects as this process requires the help of a radiologist. On comparing the prediction results against the ground truth labels of AD and MCI, it was verified that the AD subjects are correctly classified with high TPR of 0.91 against the MCI. Thus the model correctly classifies between AD and NC subjects and finds it difficult to correctly classify MCI subjects against the binary classes of NC and MCI/AD. We get a Pearson's correlation score of -0.66 for the correlation of relevance scores and the hippocampal volume for this dataset, and this result is comparable with the result in the previous section where we got a score of -0.69. From these results, we can conclude that the model generalized well as the test set results are comparable with the results in the training set. The LRP visualization, confusion matrix, ROC curve and the correlation plots are shown in figures 6.10 to 6.13, respectively, and the classification report is shown in Table 6.4. 

\begin{center}
\begin{table}[!htbp]
    \centering
     \caption{Classification Report for AIBL data}
\begin{tabular}{l l l l } \hline
& Precision  & Recall & f1-score \\ \hline
 \multirow{1}{10em}{NC(n=455)}& 0.85 & 0.91 & 0.88 \\
 \multirow{1}{10em}{MCI/AD(n=166)}& 0.70 & 0.56 & 0.62 \\ 
 \\
 \multirow{1}{10em}{accuracy} &&& 0.82 \\ \hline
 \end{tabular}
\label{tab:AIBL-report}
\end{table}
\end{center}

\begin{figure}[!htbp]
    \centering
    \includegraphics[width = 5in]{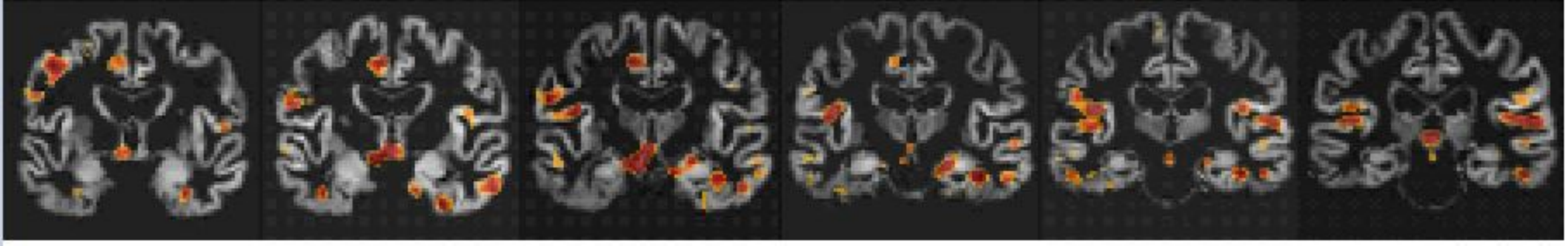}
    \caption{Lrp visualization on model trained with 3 CNN blocks using whole brain residualized input on AIBL data for an AD subject}
    \label{fig:AIBL-lrp}
    \centering
    \includegraphics[height=3in,width=5in]{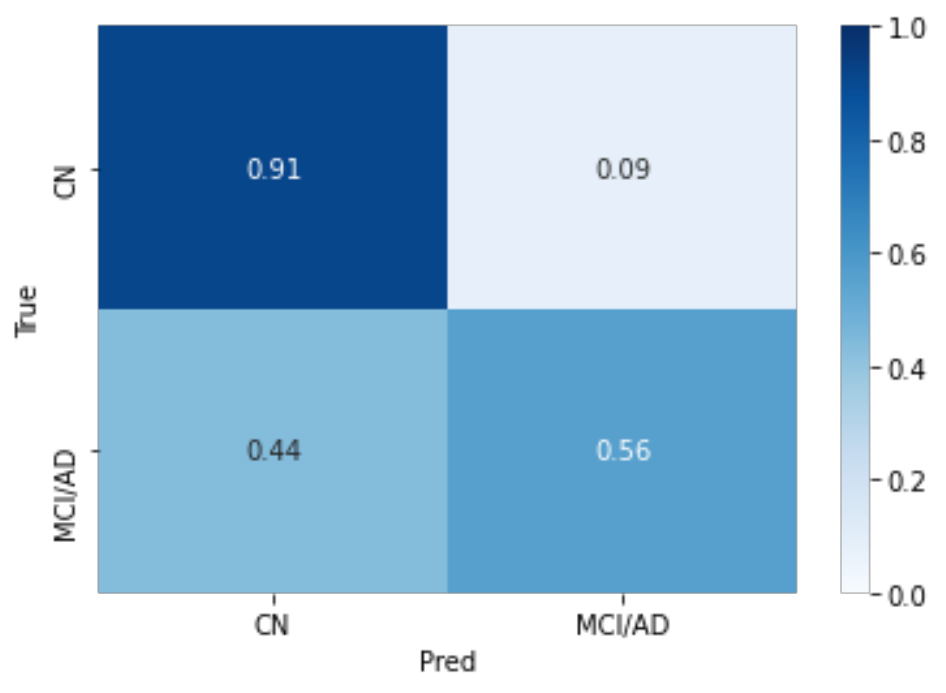}
    \caption{Confusion Matrix for AIBL data}
    \label{fig:AIBL-confusion-matrix}
\end{figure}

\begin{figure}
    \centering
    \includegraphics[height=3in,width=5in]{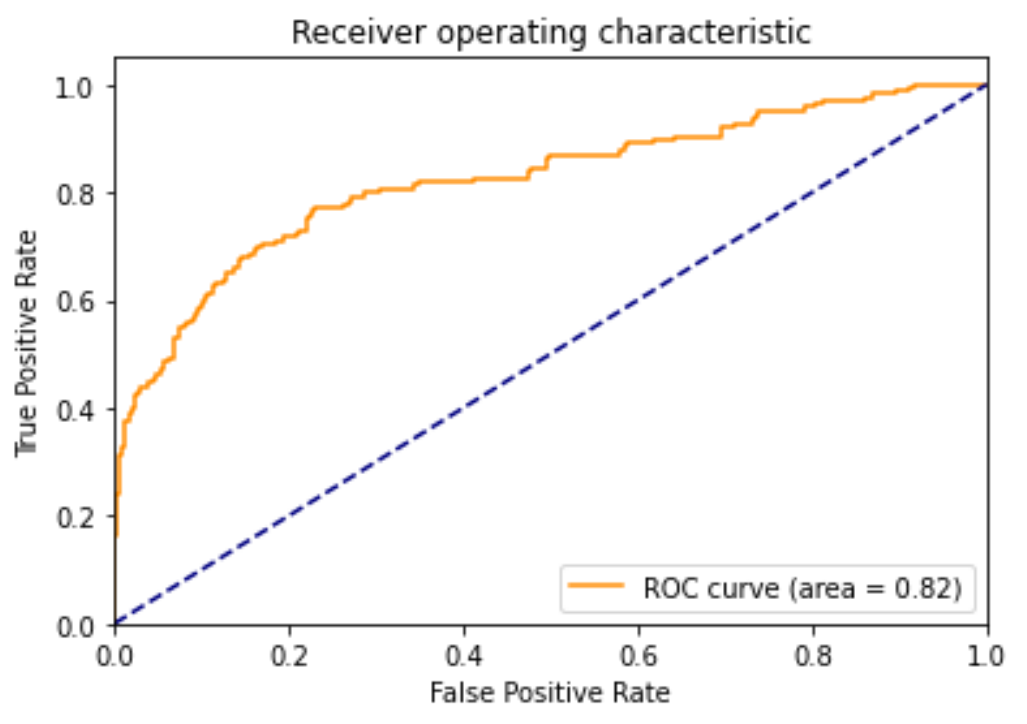}
    \caption{ROC curve for AIBL data}
    \label{fig:AIBL-ROC-curve}  
    \centering
    \includegraphics[width = 5in]{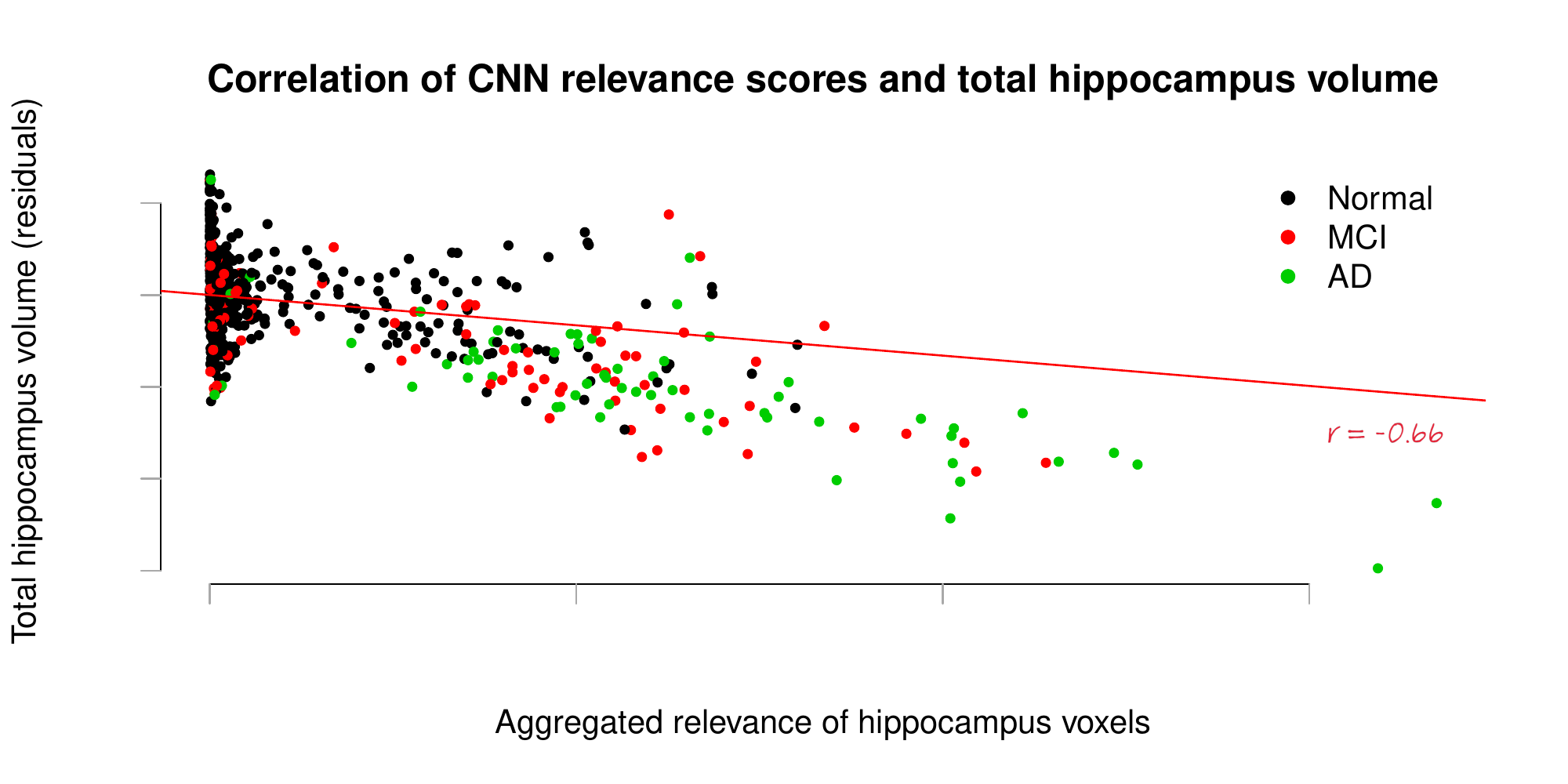}
    \caption{Correlation plot of relevance heatmap vs total hippocampus volume for 3 CNN block model using whole brain volume as input on AIBL dataset}
    \label{fig:AIBL-3cnn-corr}
\end{figure}

\chapter{Interactive web based Application}
A web application is developed that connects the front-end user interface events to real-time running Python code using Bokeh \footnote{https://docs.bokeh.org}. Bokeh is an interactive visualization library for web browsers that can be used to create interactive plots, dashboards, and data applications and can be run using Python or Javascript. In this application, we need to provide the MRI scans that need to be tested after preprocessing them using the SPM8 software as well as the residualized scans, as the best models are trained using the residual input. We can then select the subject and the coronal slice which the user wishes to be displayed, and the threshold range can be adjusted to show only regions of high relevance in the LRP visualization.

Further, the transparency value of the relevance map overlaid on top of the input image can be adjusted. We can also adjust the cluster size by filtering out only regions of high relevance density and hiding the regions that are activated along only a few pixels. An aid to view the best coronal slice having the highest relevance per subject is also provided by comparing the histogram of the overall relevance for each coronal slice. A bokeh server can be initiated to run in the local machine and the server uses the application code to create a session in the browser of the local machine to view the visualization. 
\begin{figure}
    \centering
    \includegraphics[]{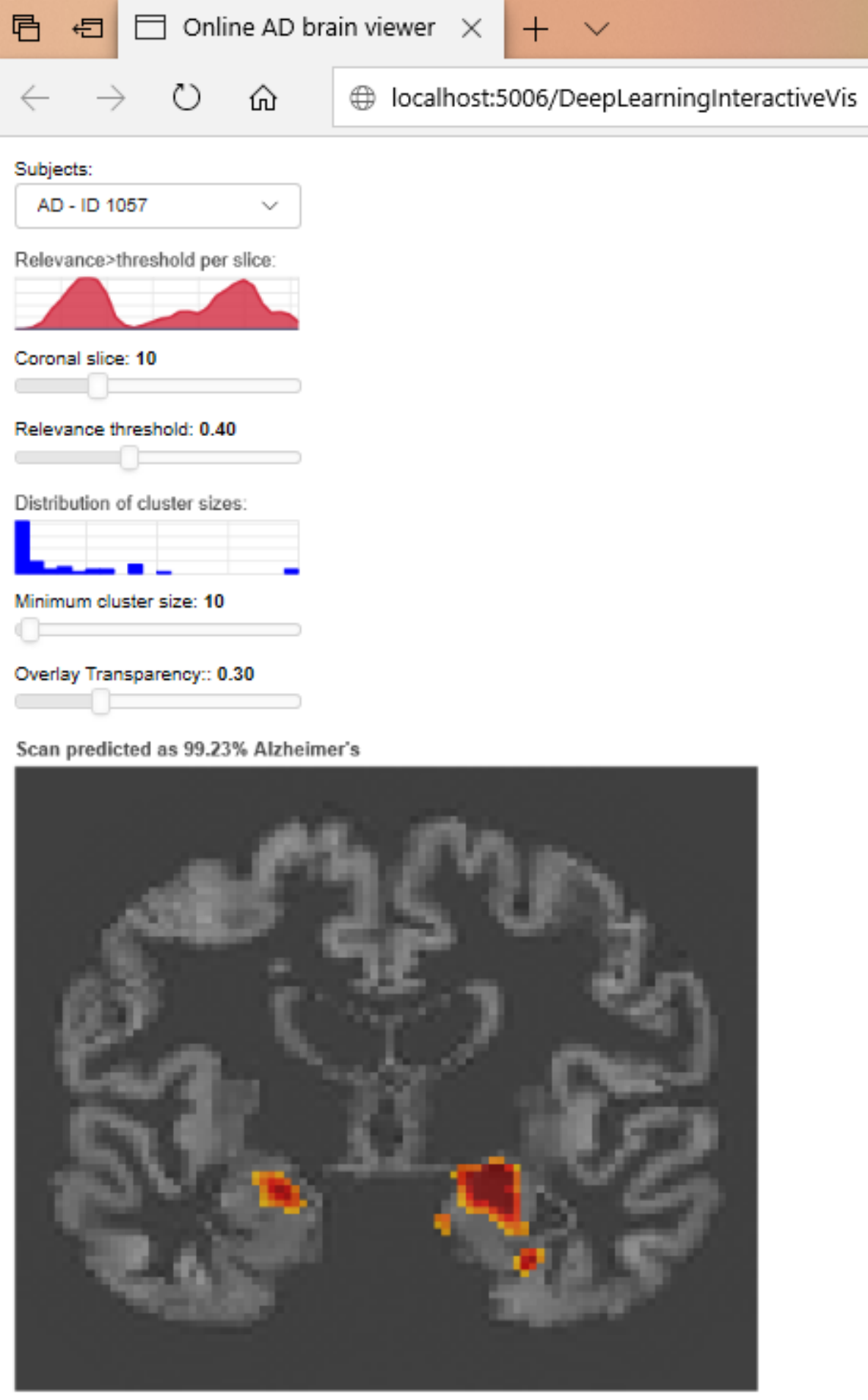}
    \caption{Interactive bokeh web app displaying neural network relevance maps for individual subjects overlaid on the normalized gray matter maps}
    \label{fig:bokeh}
\end{figure}

\chapter{Conclusion and Outlook}
From the various experimentation performed with varying the different hyperparameters, we could conclude that the results for mean accuracy across all the 10 folds resulted in almost the same value. We observed that by varying the amount of data augmentation did not result in better accuracy results, but had to be trained for a longer duration when only a few samples are considered for training the model. It was also noted that we get a  boost in the accuracy and the LRP visualization results when the input scans are trained using the residuals, i.e., with the removal of covariates such as age, gender, TIV and magnetic field strength from the input scans. Overall it was found that two models, the 3 CNN block model using 5 convolution filters per block and the 5 CNN block model using 20 convolutional filters per block gave slightly better accuracy results. As expected, the accuracy results of the model trained on the subvolume of MRI scans consisting of 32 coronal slices as input gives a higher result when compared to the model trained on the whole brain input volume. This is due to the increased computational complexity of dealing with high dimensional data and dealing with a higher number of parameters to be considered, which leads to overfitting. The number of parameters that the 3 CNN block model considers for the subvolume and whole-brain as input is 6402 and 17,292 and in the case of 5 CNN block models, the values are 11,402 and 44,402. The relevance maps obtained using LRP visualizations highlighted regions in the hippocampus region for AD subjects which is considered as the biomarker for AD. The relevance maps were then evaluated for similarity between different models using Dice Similarity coefficient and the hippocampal region relevances were correlated using Pearsons correlations.    
It was found that the 5 CNN block model gave a higher accuracy and higher relevance scores in the hippocampal activation when compared to the 3 CNN block model. However, when considering the Pearson's correlation compared with the whole dataset for the correlation between the aggregate sum of relevance in the hippocampus against the hippocampal volume, we found that the 3 CNN block model with a the negative correlation score of -0.69 is more closely correlated than 5 CNN block model with a score of -0.56. This result indicates that the 3 CNN block model with fewer parameters generalizes well for all the various subjects. This is further validated when the correlation score is tested on the independent AIBL test dataset, where a correlation score of -0.66 was obtained.

Thus we can conclude that a simpler model architecture with fewer parameters to train performs best without overfitting. It is to be noted that the common models such as VGG net and Alex net were also tested, but the models overfitted due to the higher number of parameters that the model needed to train.

It was also observed in this study that the classification of MCI subjects against AD or CN is difficult. Hence further research is required by using a multi-modal approach of including PET scans along with MRI scans and including features such as the build-up of amyloid-beta and tau proteins in the cerebrospinal fluid (CSF) as a biomarker along with the hippocampus volume.

\newpage
\section*{Appendix}
This section shows the architecture of the models attaining the high accuracy while tuning the hyper parameters of the CNN model using the subvolume of the input scans (89 * 32 * 94) as explained in sections 5.6 - 5.8.
Abbreviations used in the model:
BN: batch normalization layer; Conv3D: 3D convolutional layer; FC: fully connected layer.
\section*{A1 - 3 CNN block model with 1 FC layer and 5 convolution filters per CNN block}
\begin{figure}[!htbp]
    \centering
    \includegraphics{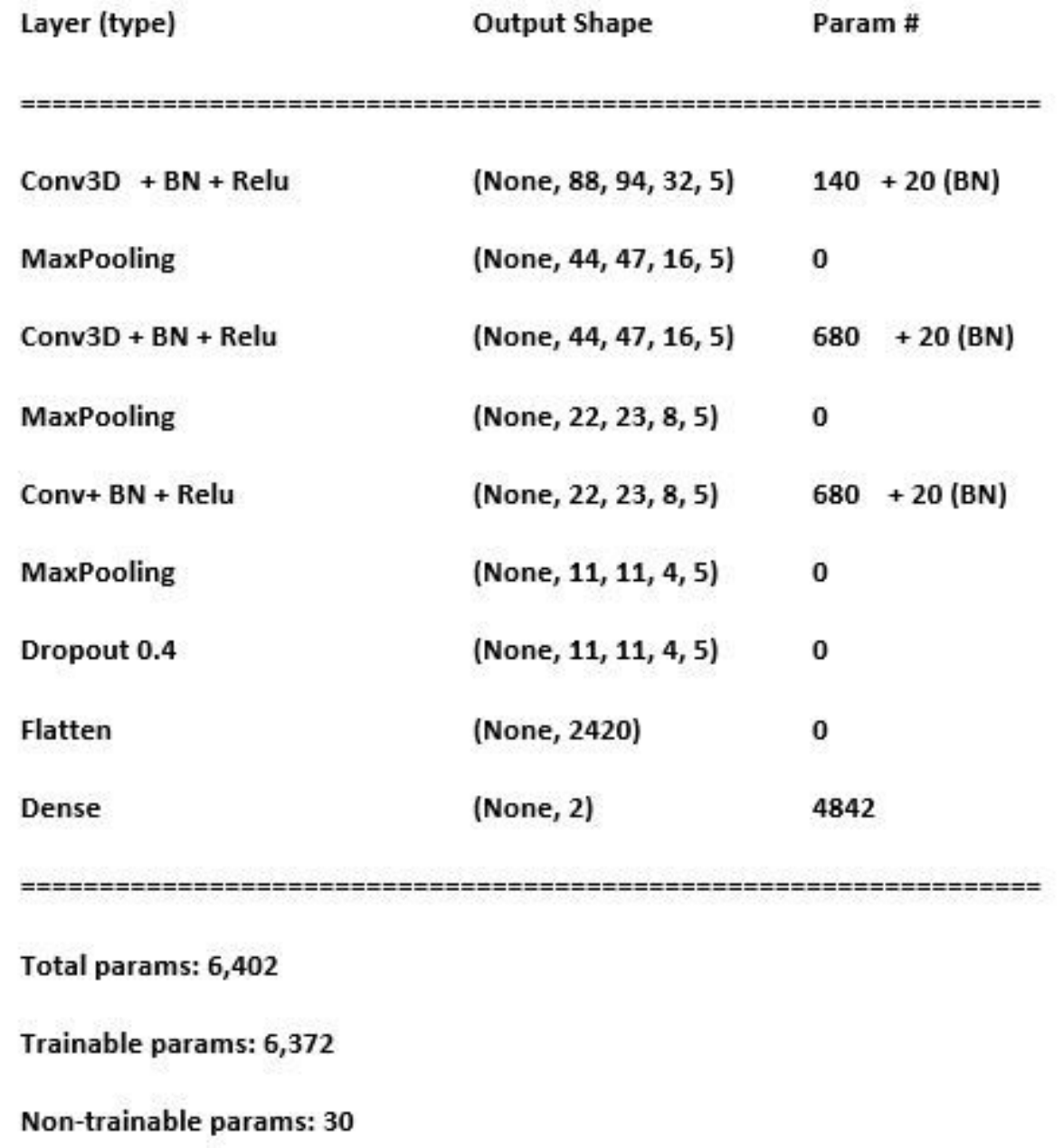}
\end{figure}
\FloatBarrier
\newpage
\section*{A2 - 5 CNN block model with 1 FC layer and 20 convolution filters per CNN block}
\begin{figure}[!htbp]
    \centering
    \includegraphics{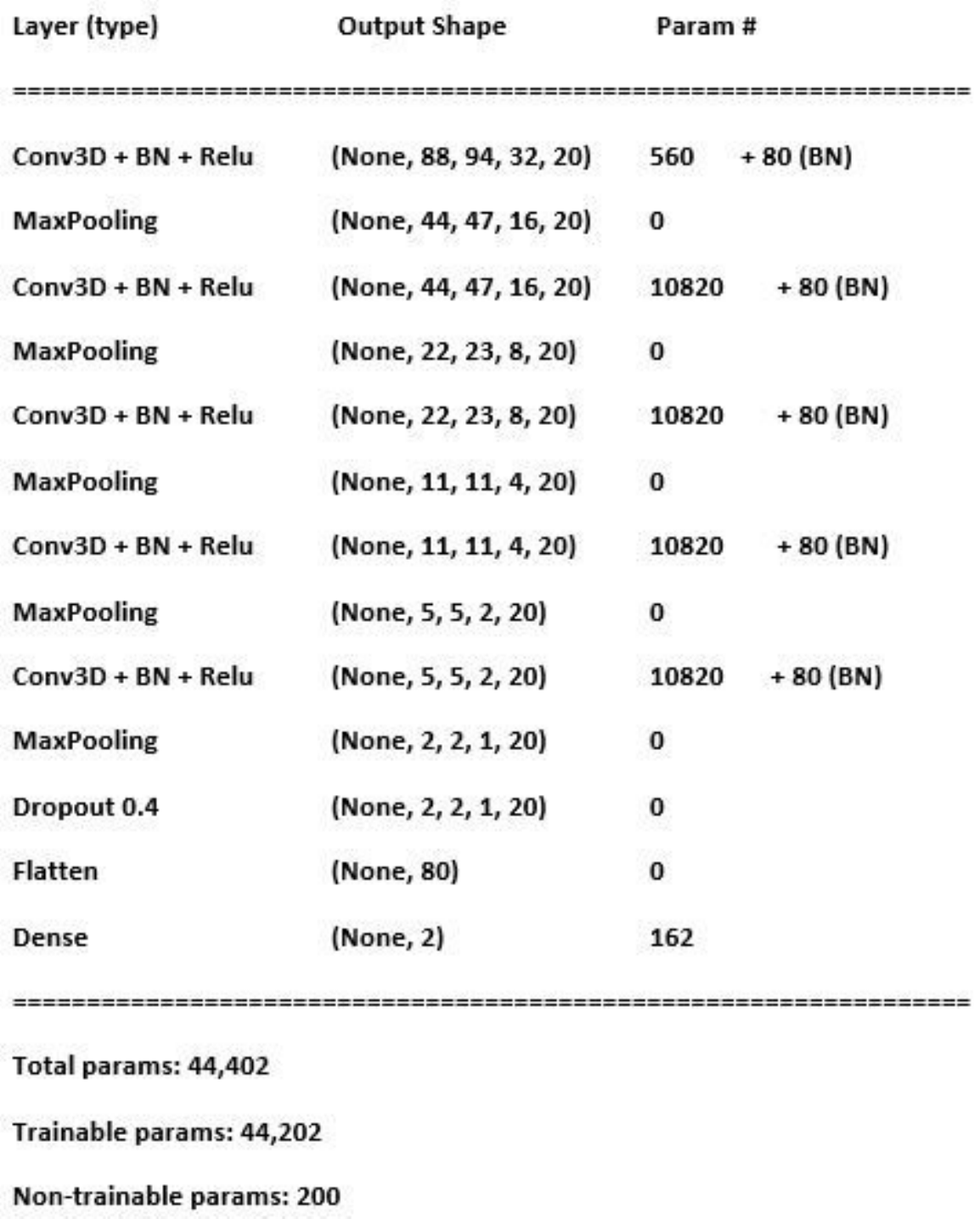}
\end{figure}
\FloatBarrier
\newpage
\section*{A3 - 5 CNN block model with 3 FC layer and 5 convolution filters per CNN block}
\begin{figure}[!htbp]
    \centering
    \includegraphics{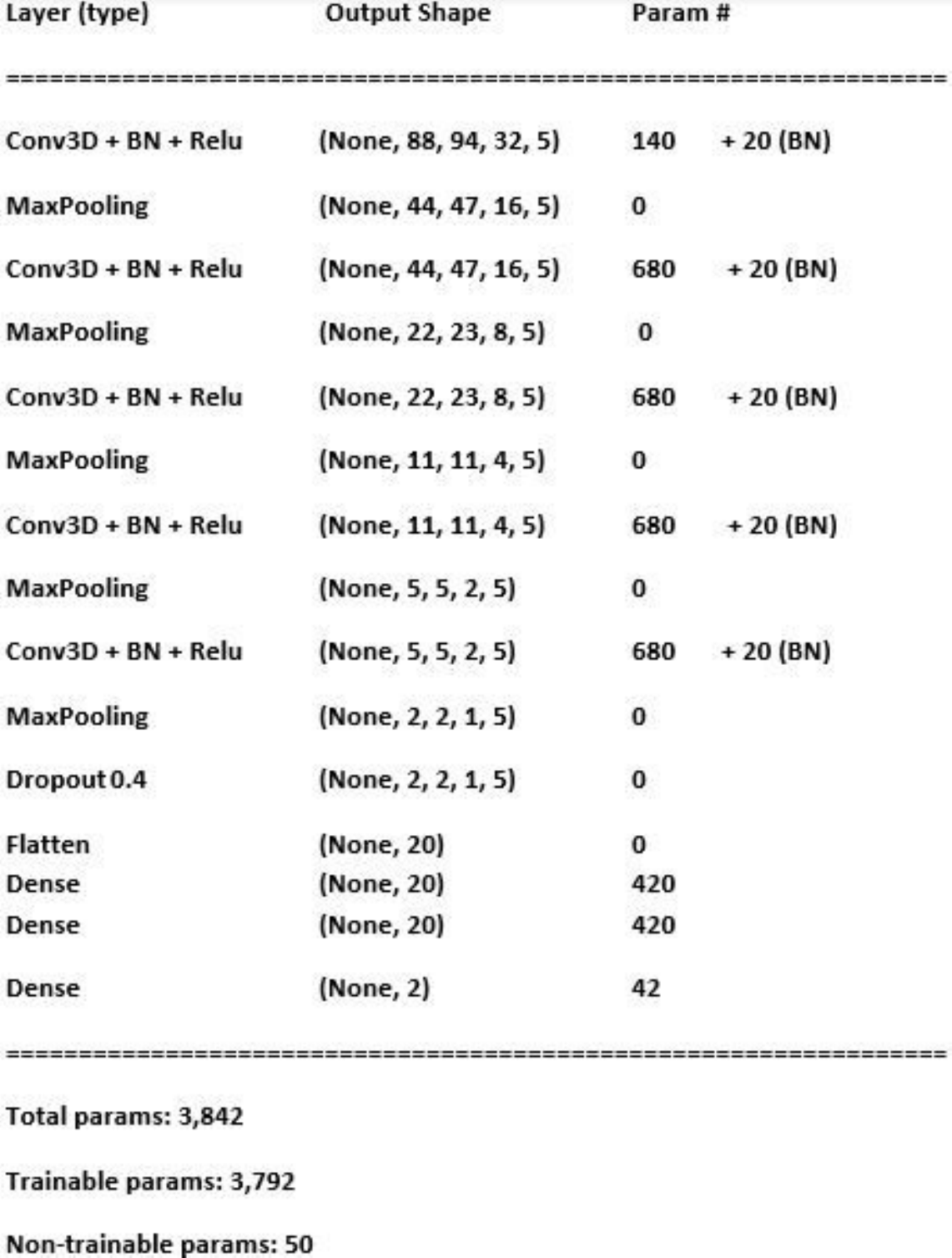}
\end{figure}
\FloatBarrier

The total number of parameters when the whole input volume of dimension 89 * 111 * 94 for the 3 CNN and 5 CNN block models are 17,292 and 44,722.

\clearpage

\appendix
\printbibliography
\pagenumbering{gobble}
\chapter*{Statutory Declaration}
I hereby declare that I have written this master's thesis myself and that I have not used any sources or resources other than those specified. Sentences or parts of sentences quoted literally are marked as such; other references with regard to the statement and scope are indicated by full details of the publications concerned. The thesis in the same or similar form has not been submitted to any examination body and has not been published. This thesis was not yet, even in part, used in another examination or as a course performance.

Rostock, \today

	\rule{6cm}{0.5pt}\\
	\parbox[l][1cm][c]{6cm}{Arjun Haridas Pallath}

\end{document}